\documentclass[a4paper,11pt]{article}
\pdfoutput=1
\usepackage{jcappub}

\usepackage{tikz,xcolor,hyperref}
\usepackage{amsmath, amssymb, amsthm, graphicx, epsfig, fancyhdr,epsfig, slashed}
\usepackage[normalem]{ulem}
\usepackage{tikzsymbols}
\usepackage{natbib}
\usepackage{float}
\usepackage{bm}
\usepackage{orcidlink}
\usepackage{latexsym}
\usepackage{booktabs}
\usepackage{subfig}
\usepackage{amsfonts}
\usepackage{bigints}
\usepackage{xcolor}
\usepackage{braket}
\usepackage{indent first}
\usepackage{slashed}
\usepackage{multirow}
\usepackage{blindtext}
\usepackage{titlesec}
\useunder{\uline}{\ul}{}
\usepackage{environ}
\usepackage{mathtools}
\usepackage{subcaption}
\usepackage{float} 

\usepackage{relsize}
\usepackage{tcolorbox}

\newcommand{\gs}{g_\star}
\newcommand{\gss}{g_{\star s}}
\newcommand{\Trh}{T_\text{rh}}

\newcommand{\arh}{a_\text{rh}}
\newcommand{\Tmax}{T_\text{max}}

\newcommand{\rR}{\rho_R}
\newcommand{\rp}{\rho_\phi}

\newcommand{\Gp}{\Gamma_\phi}

\newcommand{\mueff}{\mu_\text{eff}}
\newcommand{\yeff}{y_\text{eff}}

\renewcommand{\ast}{a_{\star}}
\newcommand{\Tst}{T_{\star}}

\newcommand{\frh}{f_\text{rh}}

\newcommand{\afo}{a_\text{fo}}
\newcommand{\Tfo}{T_\text{fo}}
\newcommand{\asph}{a_{\rm sph}}
\newcommand{\yphn}{y_{\phi N}}
\newcommand{\Gsp}{\Gamma_{\rm sph}}
\newcommand{\GeV}{\text{GeV}}
\newcommand{\amd}{a_{\rm md}}
\newcommand{\Tsph}{T_{\rm sph}}

\begin{document}
\title{Leptogenesis with sub-electroweak-\\scale reheating temperature}
\author[a]{Basabendu Barman,}
\author[a,b]{Arindam Basu,}
\author[c]{Debasish Borah,}
\author[c]{and Nayan Das}
\affiliation[a]{\,\,Department of Physics, School of Engineering and Sciences, SRM University AP, Amaravati 522240, India}
\affiliation[b]{\,\,School of Physical Sciences, Indian Association for the Cultivation of Science, Jadavpur, Kolkata 700032, India}
\affiliation[c]{\,\,Department of Physics, Indian Institute of Technology Guwahati, Assam 781039, India}
\emailAdd{basabendu.b@srmap.edu.in}
\emailAdd{arindam.basu.phy@gmail.com}
\emailAdd{dborah@iitg.ac.in}
\emailAdd{nayan.das@iitg.ac.in }
\abstract{We study the generation of the baryon asymmetry of the Universe via leptogenesis during the post-inflationary reheating epoch, considering reheating temperatures below the temperature of sphaleron freeze-out. Within the framework of a monomial inflaton potential during reheating, we analyze three perturbative reheating scenarios in which the inflaton decays into (i) a pair of Standard Model (SM)-like bosons, (ii) a pair of SM-like fermions, or (iii) exclusively into a pair of heavy right-handed neutrinos, which eventually decays into the SM final states after briefly dominating the energy density of the Universe. For each case, we identify the regions of parameter space that successfully reproduce the observed baryon asymmetry consistently tracking the sphaleron interaction rate during reheating, while satisfying existing cosmological constraints. We also highlight the potential of future primordial gravitational wave observations to probe this class of scenarios.}
\maketitle
\section{Introduction}
\label{sec:intro}
Several astrophysical and cosmological observations~\cite{Planck:2018vyg, ParticleDataGroup:2020ssz} indicate that the visible matter content of the present Universe is highly asymmetric. Although such an asymmetry can be generated dynamically if the Sakharov conditions are satisfied~\cite{Sakharov:1967dj}, the Standard Model (SM) does not provide sufficient CP violation and departure from thermal equilibrium to account for the observed baryon asymmetry of the Universe (BAU). This shortcoming has motivated a wide range of beyond-the-SM scenarios. Among them, leptogenesis~\cite{Fukugita:1986hr} is one of the most attractive and extensively studied mechanisms. In this framework, a lepton asymmetry is generated first and is subsequently converted into a baryon asymmetry through electroweak sphaleron processes that violate $(B+L)$~\cite{Kuzmin:1985mm}. A notable feature of leptogenesis is its intimate connection to the origin of neutrino masses, another phenomenon unexplained within the SM. In the simplest realization, namely the Type-I seesaw mechanism~\cite{Minkowski:1977sc,Yanagida:1979as,Gell-Mann:1979ijt, GellMann:1980vs,Mohapatra:1979ia,Schechter:1980gr,Schechter:1981cv}, heavy right-handed neutrinos (RHNs) generate small neutrino masses, while their CP-violating out-of-equilibrium decays produce the required lepton asymmetry. In conventional thermal leptogenesis with hierarchical RHN masses, successful baryogenesis typically requires RHN masses above $10^9$ GeV, known as the Davidson--Ibarra bound~\cite{Davidson:2002qv}. Consequently, the reheating temperature after inflation must satisfy $\Trh \gtrsim 10^{10}$ GeV to ensure efficient thermal production of RHNs. These stringent requirements can be relaxed in several ways, including non-thermal production of RHNs during reheating after inflation~\cite{Lazarides:1991wu,Murayama:1992ua,Kolb:1996jt,Giudice:1999fb,Asaka:1999yd,Asaka:1999jb,Hamaguchi:2001gw,Hahn-Woernle:2008tsk,Hamada:2018epb,Eijima:2019hey,Maleknejad:2020pec,Barman:2021tgt,Barman:2021ost,Barman:2022gjo,Lazarides:2022ezc,Datta:2022jic,Datta:2023pav,Barman:2024jqh,Barman:2024ujh,Chowdhury:2026zox}, resonant enhancement of the CP asymmetry through quasi-degenerate RHN masses~\cite{Pilaftsis:2003gt}, or the recently proposed wash-in leptogenesis scenario~\cite{Domcke:2020quw}.

In conventional scenarios, where RHNs are produced during the post-inflationary reheating era, the reheating temperature, $\Trh$, is typically assumed to be higher than the sphaleron freeze-out temperature, $T_{\rm fo}\simeq 131$ GeV~\cite{DOnofrio:2014rug}, ensuring the efficient conversion of the generated lepton asymmetry into the observed baryon asymmetry through sphaleron processes. On the other hand, the lower bound on the reheating temperature is only about $\sim$ 4 MeV, as required by successful big bang nucleosynthesis (BBN), which lies far below the sphaleron freeze-out temperature. Motivated by this temperature window, we explore the possibility of achieving successful leptogenesis even when $\Trh$ is below the electroweak scale. While the sphaleron decoupling temperature remains unchanged since we do not modify the electroweak symmetry-breaking history of the SM\footnote{See~\cite{Bhandari:2025ufp} for a scenario involving a first-order electroweak phase transition.}, we exploit the fact that during reheating the thermal bath can temporarily attain temperatures much higher than $\Trh$, modifying the high temperature behaviour of the sphaelron rate with respect to standard radiation-dominated cosmology. We particularly focus on the reheating epoch, during which the inflaton oscillates around the minimum of a generic monomial potential and decays perturbatively into (i) into SM particles, together with heavy RHNs, and (ii) heavy RHNs exclusively, that subsequently decay into the SM final states. Although leptogenesis during reheating has been studied previously~\cite{Hahn-Woernle:2008tsk,Datta:2022jic}, primarily for a quadratic inflaton potential, and explicit inflaton-RHN couplings have been considered in Refs.~\cite{Hahn-Woernle:2008tsk,Barman:2021tgt,Datta:2023pav}, here we focus on the low-reheating-temperature regime, particularly $\Trh<T_{\rm fo}$. We further investigate the observational prospects of this framework through its imprint on the primordial gravitational wave (GW) spectrum, with an inflationary origin. The prolonged reheating phase associated with either of these couplings can significantly modify the shape of the primordial GW spectrum, potentially bringing the resulting signal within the reach of future GW observatories. 

This paper is organized as follows. In Sec.~\ref{sec:setup} we outline the particle physics setup as well as the reheating dynamics. In Sec.~\ref{sec:lepto} we discuss the details of leptogenesis with sub-EW-scale reheating temperatures, and finally we conclude in Sec.~\ref{sec:concl}.
\section{The set-up}
\label{sec:setup}
\subsection{Particle physics framework}
The interaction Lagrangian relevant for leptogenesis is given by,
\begin{align}
-\mathcal{L}\supset\overline{\ell}_{L_\alpha}\,(y_{\nu})_{\alpha i}\,\widetilde{H}\,N_i
+\frac{1}{2}\,\overline{N_i^c}\,(M)_{ii}\,N_i
+\text{h.c.},\qquad (i=1,2,3)\,,
\end{align}
where the SM particle content is extended by three generations of right-handed neutrinos (RHNs), $N_i$, which are singlets under the SM gauge symmetry. Here, $\ell_{L_{\alpha=e,\mu,\tau}}$ denote the SM left-handed lepton doublets, while $\widetilde{H}=i\sigma_2 H^*$ is the conjugate Higgs doublet, with $H$ being the SM Higgs field and $\sigma_i$ the Pauli matrices. The Yukawa matrix can be conveniently expressed using the Casas--Ibarra (CI) parametrization~\cite{Casas:2001sr},
\begin{align}\label{eq:Ynu}
y_{\nu}=-i\,\frac{\sqrt{2}}{v}\,\mathcal{U}\,
D_{\sqrt{m}}\,\mathcal{R}^{T}\,D_{\sqrt{M}}\,,
\end{align}
where $\mathcal{U}$ is the Pontecorvo--Maki--Nakagawa--Sakata (PMNS) matrix~\cite{Zyla:2020zbs}, which relates the flavour and mass eigenstates of the light neutrinos. The diagonal matrices
\begin{align}
D_{\sqrt{m}}
&=\text{diag}\!\left(\sqrt{m_1},\,\sqrt{m_2},\,\sqrt{m_3}\right),\\
D_{\sqrt{M}}
&=\text{diag}\!\left(\sqrt{M_1},\,\sqrt{M_2},\,\sqrt{M_3}\right),
\end{align}
contain the square roots of the light-neutrino and RHN masses, respectively, where the RHNs are assumed to be in a mass-diagonal basis. The matrix $\mathcal{R}$ is a complex orthogonal matrix satisfying $\mathcal{R}^{T}\mathcal{R}=\mathbb{I}$. 

Ae assume a hierarchical RHN mass spectrum: $M_1 \ll M_{2,3}$. Furthermore, the lepton number violating interactions of the lightest RHN, $N_1$, are assumed to be sufficiently rapid to wash out any asymmetry generated from the decays of $N_{2,3}$. As a result, the final lepton asymmetry is entirely determined by the CP-violating decays of $N_1$. In this case, the CP asymmetry originates from the interference between the tree-level and one-loop decay amplitudes of the RHNs into lepton (or anti-lepton) and Higgs doublets. After summing over the lepton flavours, the resulting asymmetry can be written as~\cite{Davidson:2008bu}
\begin{align}
\epsilon_{\Delta L}=\sum_\alpha \epsilon_{1\alpha}^{\ell}=\frac{1}{8\pi (y_\nu^\dagger\,y_\nu)_{11}}
\sum_{j\neq 1}\text{Im}\!\left[
\left((y_\nu^\dagger\,y_\nu)_{1j}\right)^2\right]\,\mathcal{F}\!\left(\frac{M_j^2}{M_1^2}\right)\,,
\end{align}
where the loop function is given by
\begin{align}
\mathcal{F}(x)=\sqrt{x}\,\left[
\frac{1}{1-x}+1-(1+x)\ln\!\left(\frac{1+x}{x}\right)\right]\,.
\end{align}

We further introduce a singlet scalar field $\phi$ beyond the SM particle content, which serves as the inflaton in the present framework. The inflationary dynamics associated with $\phi$ will be discussed in the next section. Unless forbidden by an additional symmetry, the inflaton will interact with the RHNs via
\begin{align}\label{eq:LphiNN}
& \mathcal{L}\supset-\left(\yphn\right)_i\,\phi\,\overline{N_i^c}\,N_i+\text{h.c.}\,.
\end{align}
This interaction allows for the production of RHNs through the two-body decay of the inflaton. 
\begin{figure}[htb!]
\centering
\includegraphics[scale=0.14]{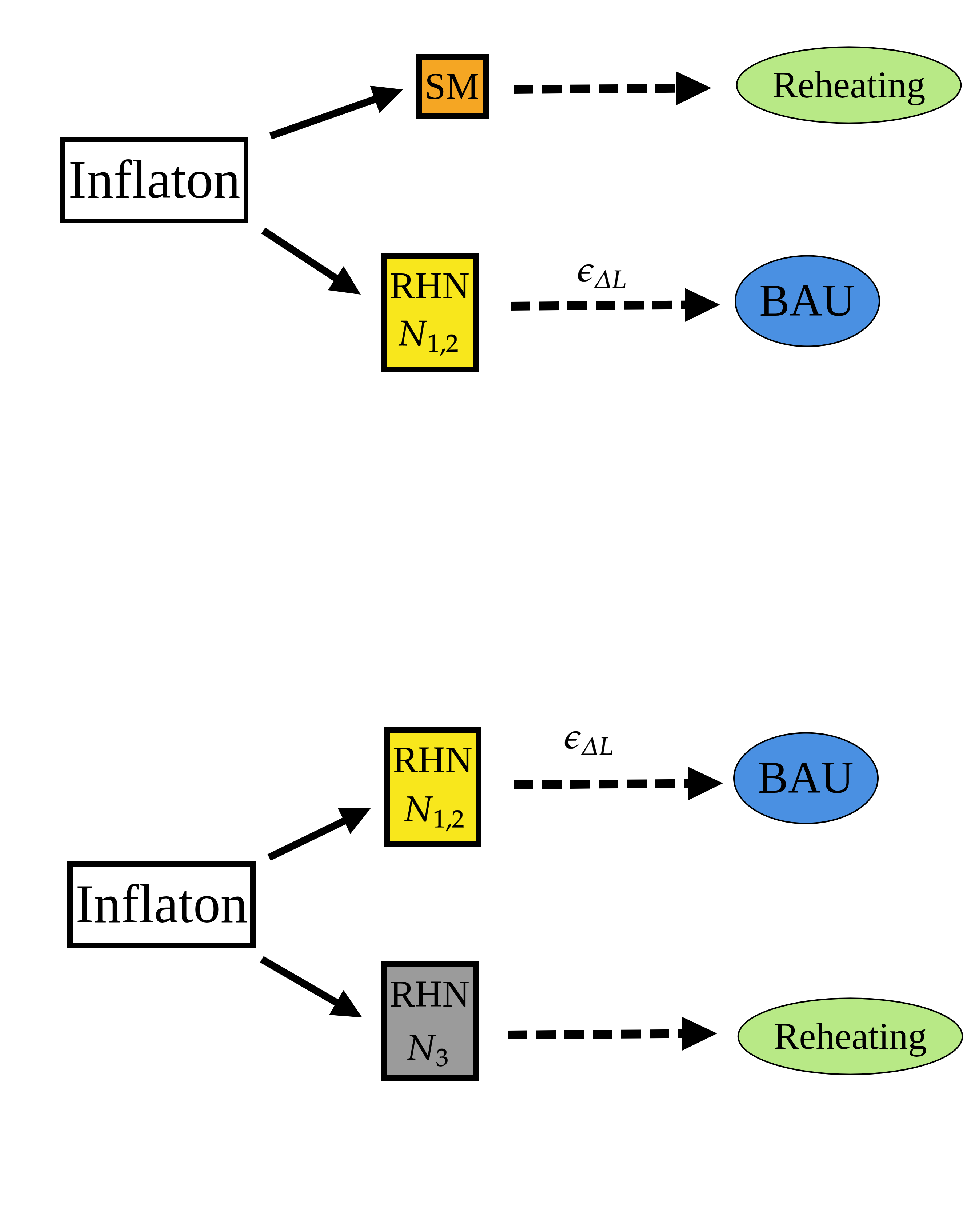}
\caption{A schematic diagram showing two different reheating scenarios in the present set-up. The top panel shows {\it Scenario-A}, while the bottom shows {\it Scenario-B}. This figure is generated using online open source latex editor {\texttt mathcha.io}.}
\label{fig:schema}
\end{figure}
\subsection{Reheating dynamics}
We study the post-inflationary oscillations of the inflaton $\phi$ around the minimum of a monomial potential
\begin{equation}\label{eq:inf-pot}
V(\phi) = \lambda\, \frac{\phi^n}{\Lambda^{n - 4}}\,,
\end{equation}
where the dimensionless coupling $\lambda$ decides the overall scale of the potential and the dimensionful parameter $\Lambda$ is another scale that depends on the form of the full inflationary potential (we provide further details in Appendix.~\ref{sec:inflation}). Potential of the form in Eq.~\eqref{eq:inf-pot} arise in several inflationary frameworks, such as $\alpha$-attractor T- or E-models~\cite{Kallosh:2013hoa, Kallosh:2013yoa, Kallosh:2013maa, Kallosh:2015lwa} and the Starobinsky model~\cite{Starobinsky:1980te, Starobinsky:1981vz, Starobinsky:1983zz, Kofman:1985aw}. The oscillating inflaton condensate obeys the equation of motion (EoM)~\cite{Turner:1983he}
\begin{equation} \label{eq:eom0}
\ddot\phi + (3\, \mathcal{H} + \Gp)\, \dot\phi + V'(\phi) = 0\,,
\end{equation} 
where $\mathcal{H}$ is the Hubble rate, $\Gp$ the dissipation rate, dots denote time derivatives, and primes field derivatives. During reheating one can approximate $\mathcal{H}\simeq\sqrt{\rp/(3M_P^2)}$. Defining $\rp \equiv \frac12\, \dot\phi^2+ V(\phi)$ and $p_\phi \equiv \frac12\, \dot\phi^2 - V(\phi)$, with EoS $w \equiv p_\phi/\rp = (n - 2) / (n + 2)$~\cite{Turner:1983he}, the inflaton energy density evolves as
\begin{equation} \label{eq:drhodt}
\frac{d\rp}{dt} + \frac{6\, n}{2 + n}\, \mathcal{H}\, \rp = - \frac{2\, n}{2 + n}\, \Gp\, \rp\,.
\end{equation}
For $a_I \ll a \ll \arh$, the expansion term dominates and Eq.~\eqref{eq:drhodt} yields
\begin{equation} \label{eq:rpsol}
\rp(a) \simeq \rp (\arh) \left(\frac{\arh}{a}\right)^\frac{6\, n}{2 + n}\,.
\end{equation}
Here $a_I$ and $\arh$ denote the scale factors at the end of inflation and reheating. Since reheating is inflaton-dominated,
\begin{equation} \label{eq:Hubble}
\mathcal{H}(a) \simeq \mathcal{H}(\arh) \times
\begin{dcases}
\left(\frac{\arh}{a}\right)^\frac{3\, n}{n + 2} &\text{ for } a \leq \arh\,,\\[10pt]
\left(\frac{\arh}{a}\right)^2 &\text{ for } \arh \leq a\,.
\end{dcases}
\end{equation}
At $a=\arh$, $\rR(\arh)=\rp(\arh)=3\, M_P^2\, \mathcal{H}(\arh)^2$. Successful BBN requires $\Trh > T_\text{BBN} \simeq 4$~MeV~\cite{Sarkar:1995dd, Kawasaki:2000en, Hannestad:2004px, DeBernardis:2008zz, deSalas:2015glj,Hasegawa:2019jsa}. The radiation density satisfies~\cite{Garcia:2020wiy}
\begin{equation} \label{eq:rR}
\frac{d\rR}{dt} + 4\, \mathcal{H}\, \rR = + \frac{2\, n}{2 + n}\, \Gp\, \rp\,,
\end{equation}
which, using Eq.~\eqref{eq:rpsol}, gives
\begin{equation} \label{eq:rR_int}
\rR(a) \simeq \frac{2\, \sqrt{3}\, n}{2 + n}\, \frac{M_P}{a^4} \int_{a_I}^a \Gp(a')\, \sqrt{\rp(a')}\, a'^3\, da'\,.
\end{equation}
Allowing for scale-dependent $\Gp$, the effective inflaton mass follows from Eq.~\eqref{eq:inf-pot},
\begin{equation}\label{eq:inf-mass1}
m_\phi(a)^2 \equiv \frac{d^2V}{d\phi^2} = n\, (n - 1)\, \lambda\, \frac{\phi^{n - 2}}{\Lambda^{n - 4}}
\simeq n\, (n-1)\, \lambda^\frac{2}{n}\, \Lambda^\frac{2\, (4 - n)}{n} \rp(a)^{\frac{n-2}{n}}\,,
\end{equation}
or
\begin{equation}\label{eq:inf-mass}
m_\phi(a) \simeq m_I \left(\frac{a_I}{a}\right)^\frac{3 (n-2)}{n+2}\,,
\end{equation}
with $m_I\equiv m_\phi(a_I)$; for $n \neq 2$, $m_\phi$ is field-dependent and induces time-dependent dissipation. Finally, note that, the interaction in Eq.~\eqref{eq:LphiNN} allows RHN production through inflaton decay, $\phi\to NN$, whenever $m_\phi(a)>2M_N$, $M_N$ being mass of the relevant RHN. For $n>2$, this channel closes beyond the critical scale factor,  
\begin{align}\label{eq:ast}
a_\star= a_I\,\left(\frac{2 \,M_N}{m_I}\right)^\frac{n+2}{3(2-n)},
\end{align}
after which inflaton decay into RHNs becomes kinematically forbidden. For $n=2$, the inflaton mass remains independent of the scale factor, and hence the above equation becomes invalid. 

With this setup, we investigate leptogenesis with reheating temperature below the electroweak scale, in the backdrop of two distinct reheating scenarios, as described in Tab.~\ref{tab:scenarios}, with a schematic diagram depicted in Fig.~\ref{fig:schema}. Further, for clarity of reading, in box~\ref{box:notation} we have listed a few notations that we have frequently used in the paper. As we will show in the following sections, these two scenarios lead to markedly different mechanisms for asymmetry generation during reheating at temperatures below the electroweak scale.  
\begin{table}[htb!]
\centering
\begin{tabular}{p{2.2cm} p{10.5cm}}
\toprule
\textbf{Reheating scenarios} & \textbf{Description} \\
\midrule
\textit{Scenario-A} &
The inflaton reheats the Universe through its \textbf{direct} perturbative decay into a pair of SM bosons or fermions. 
\\[10pt]
\textit{Scenario-B} &
The inflaton decays \textbf{exclusively} into the three generations of RHNs. One of the RHNs is sufficiently long-lived to dominate the energy density of the Universe and subsequently reheat it through its decay into the SM final states. \\
\bottomrule
\end{tabular}
\caption{Summary of the reheating scenarios considered in this work.}
\label{tab:scenarios}
\end{table}
\begin{tcolorbox}
[colback=white!5!white,colframe=green!50!black,colbacktitle=green!75!black,title=Guide to notations, label={box:notation}]
\begin{itemize}
\item [] $\star_I:$ any quantity ($\star$) with the subscript $``I"$ is defined at the beginning of reheating (end of inflation)
\item [] $\arh:$ Scale factor at the end of reheating
\item [] $\afo:$ Scale factor at which sphalerons freeze-out ($\Tfo\equiv T(\afo)\simeq 130$ GeV)
\item [] $\asph:$ Scale factor at which sphalerons enter equilibrium during reheating ($T_{\rm sph}\equiv T(\asph)$)
\end{itemize}
\label{tempDefinitions}
\end{tcolorbox}
\subsubsection{{\it Scenario-A:} direct reheating}
\label{sec:caseA}
Here we consider reheating proceeds via inflaton dissipation into bosonic final states through $\mu\,\phi\,|H|^2$, giving
\begin{equation} \label{eq:bos_gamma}
\Gp(a) = \frac{\mueff^2}{8\pi\, m_\phi(a)}\,,
\end{equation}
where $\mueff \ne \mu$ (for $n\neq2$) denotes the oscillation-averaged coupling~\cite{Shtanov:1994ce, Ichikawa:2008ne, Garcia:2020wiy}. The resulting radiation density is
\begin{equation} \label{eq:rR_bos}
\rR(a) \simeq \frac{3\, n}{1 + 2\, n}\, M_P^2\, \Gp(\arh)\, \mathcal{H}(\arh) \left(\frac{\arh}{a}\right)^\frac{6}{2 + n} \left[1 - \left(\frac{a_I}{a}\right)^\frac{2\, (1 + 2 n)}{2 + n}\right]\,,
\end{equation}
considering instantaneous thermalization of the decay products. The SM temperature evolves as
\begin{align} \label{eq:Tevol}
& T(a) \simeq \Trh \left(\frac{\arh}{a}\right)^\alpha\,, & \alpha = \frac32\, \frac{1}{n + 2}\,.
\end{align}

Alternatively, reheating may occur via perturbative decay of the inflaton condensate into a pair of SM-like fermions through $y_\psi\,\bar{\psi}\,\psi\,\phi$, with a dissipation rate
\begin{equation} \label{eq:fer_gamma}
    \Gp(a) = \frac{\yeff^2}{8\pi}\, m_\phi(a)\,,
\end{equation}
where the effective coupling $\yeff \ne y_\psi$ (for $n \neq 2$) is obtained after averaging over several oscillations~\cite{Shtanov:1994ce, Ichikawa:2008ne, Garcia:2020wiy}. The evolution of the SM energy density (Eq.~\eqref{eq:rR_int}) in this case becomes~\cite{Bernal:2022wck}
\begin{equation} \label{eq:rR_fer}
    \rR(a) \simeq \frac{3\, n}{7 - n}\, M_P^2\, \Gp(\arh)\, \mathcal{H}(\arh) \left(\frac{\arh}{a}\right)^\frac{6 (n - 1)}{2 + n} \left[1 - \left(\frac{a_I}{a}\right)^\frac{2 (7 - n)}{2 + n}\right]\,,
\end{equation}
for $n\neq7$, while for $n=7$ 
\begin{align}
\rR(a)\simeq \frac{14}{3}\,M_P^2 \,\alpha_\phi(\arh)\,\mathcal{H}(\arh)\,\left(\frac{\arh}{a}\right)^4\,\log\left(\frac{a}{a_I}\right)\,.  
\end{align}
The temperature of the SM bath evolves as
\begin{align} \label{eq:Tevol2}
& T(a) \simeq \Trh \left(\frac{\arh}{a}\right)^\alpha\,, &\alpha=\frac32\, \frac{n - 1}{n + 2}\,.
\end{align}

\begin{figure}[htb!]
\centering
\includegraphics[scale=0.55]{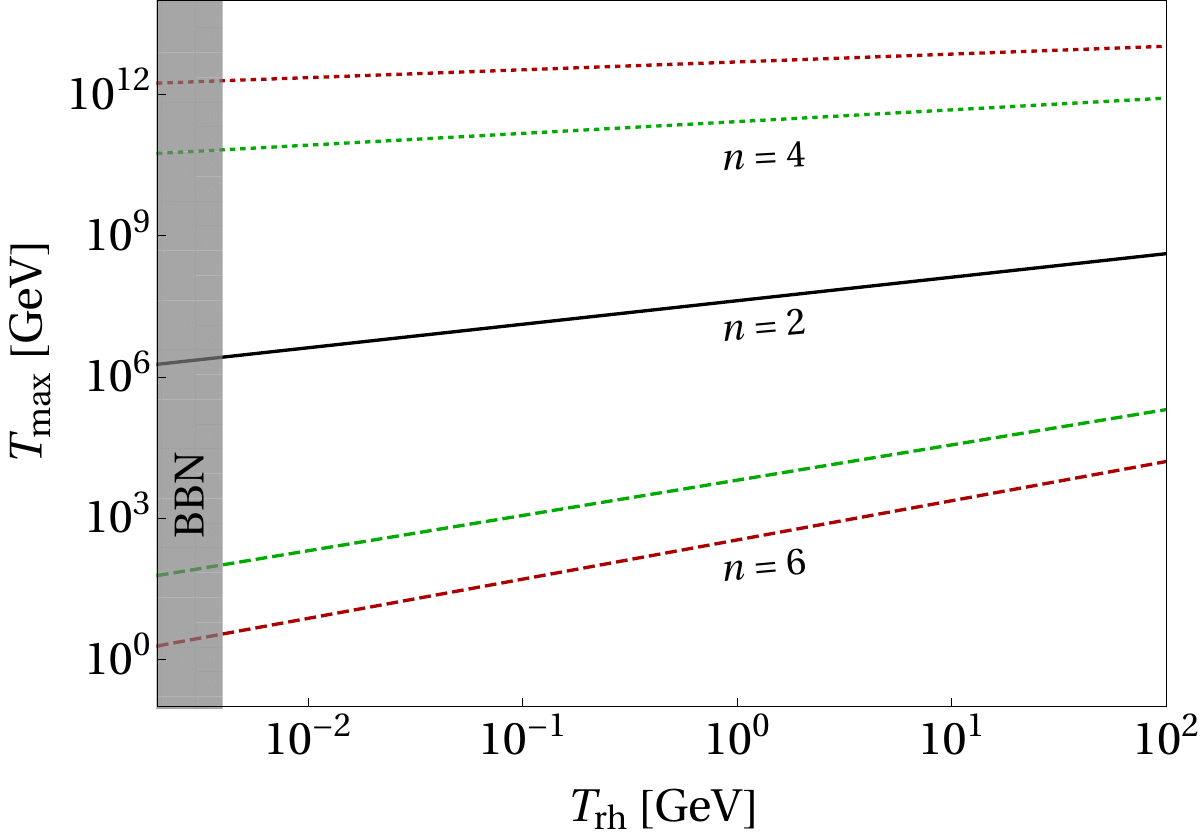}
\caption{{\it Scenario-A}: Maximum temperature $\Tmax$ during reheating, as a function of reheating temperature $\Trh$, for different choices of $n$. The dashed and dotted lines correspond to bosonic and fermionic reheating scenarios, respectively. The gray shaded region corresponds to $\Trh<4$ MeV.}
\label{fig:Tmax}
\end{figure}

Finally, the maximum temperature of the bath during reheating reads,
\begin{align}\label{eq:Tmax1}
\Tmax\simeq \Trh
\begin{cases}
\left(\arh/a_I\right)^\frac{3\,(n-1)}{2\,(n+2)} & \text{fermionic reheating}\,,
\\[10pt]
\left(\arh/a_I\right)^\frac{3}{2\,(n+2)} & \text{bosonic reheating}\,,
\end{cases}
\end{align}
which can be several orders of magnitude larger than $\Trh$, for $\arh\gg a_I$. Note that, for the same $\Trh$, $\Tmax$ corresponding to bosonic reheating is much lower compared to the fermionic scenario, for $n>2$. This can be seen from Fig.~\ref{fig:Tmax}. This behavior arises because the inflaton decay rate into fermions is proportional to $m_\phi(a)$, whereas the decay rate into bosons scales as $1/m_\phi(a)$. Since the inflaton mass decreases with time, decays into bosons become increasingly efficient, while decays into fermions become less efficient. As a result, reheating proceeds more effectively for bosonic final states than for fermionic ones at later times. Consequently, the temperature evolves differently with the scale factor in the two cases, leading to a steeper $T$-versus-$a$ slope for fermionic final states than for bosonic final states. Roughly speaking, a larger decay rate injects more energy into the radiation bath and therefore sustains a higher temperature.  
\begin{figure}[htb!]
\centering
\includegraphics[scale=0.375]{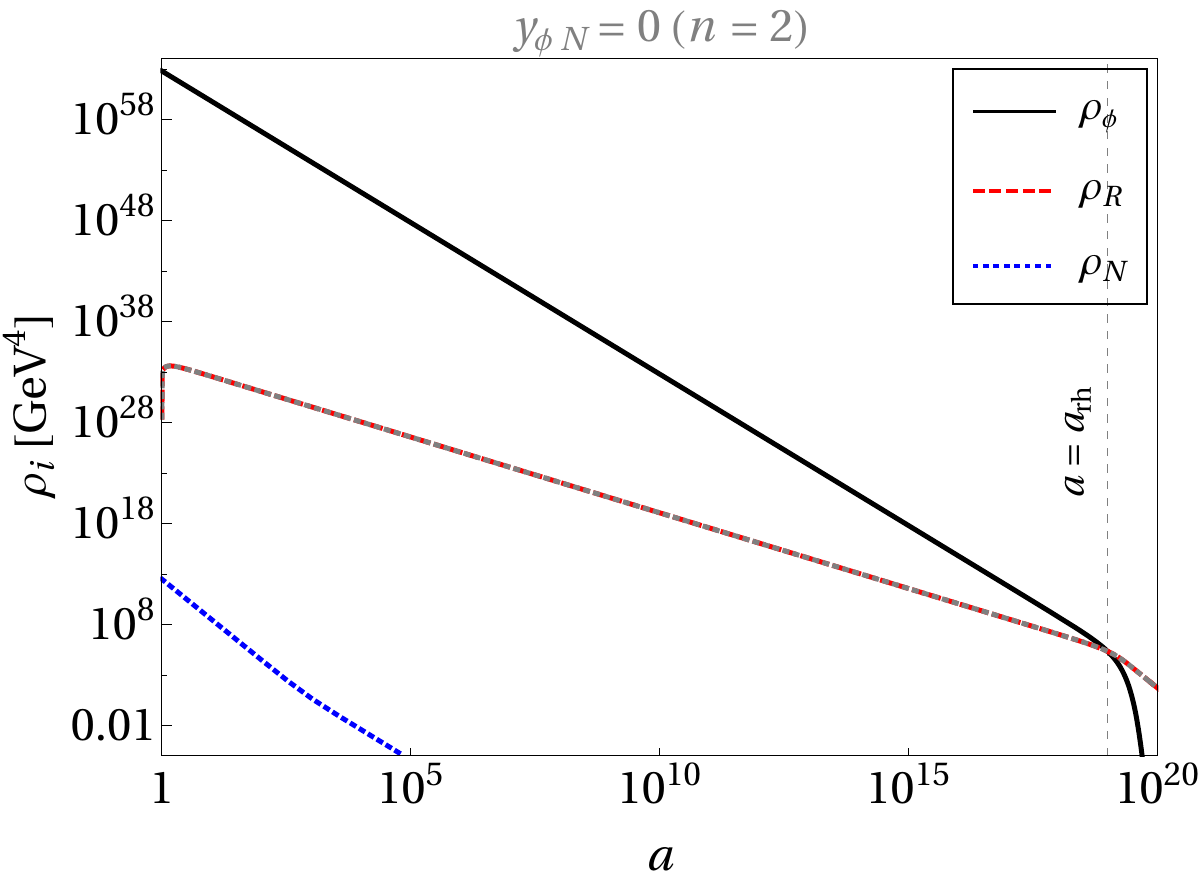}~\includegraphics[scale=0.375]{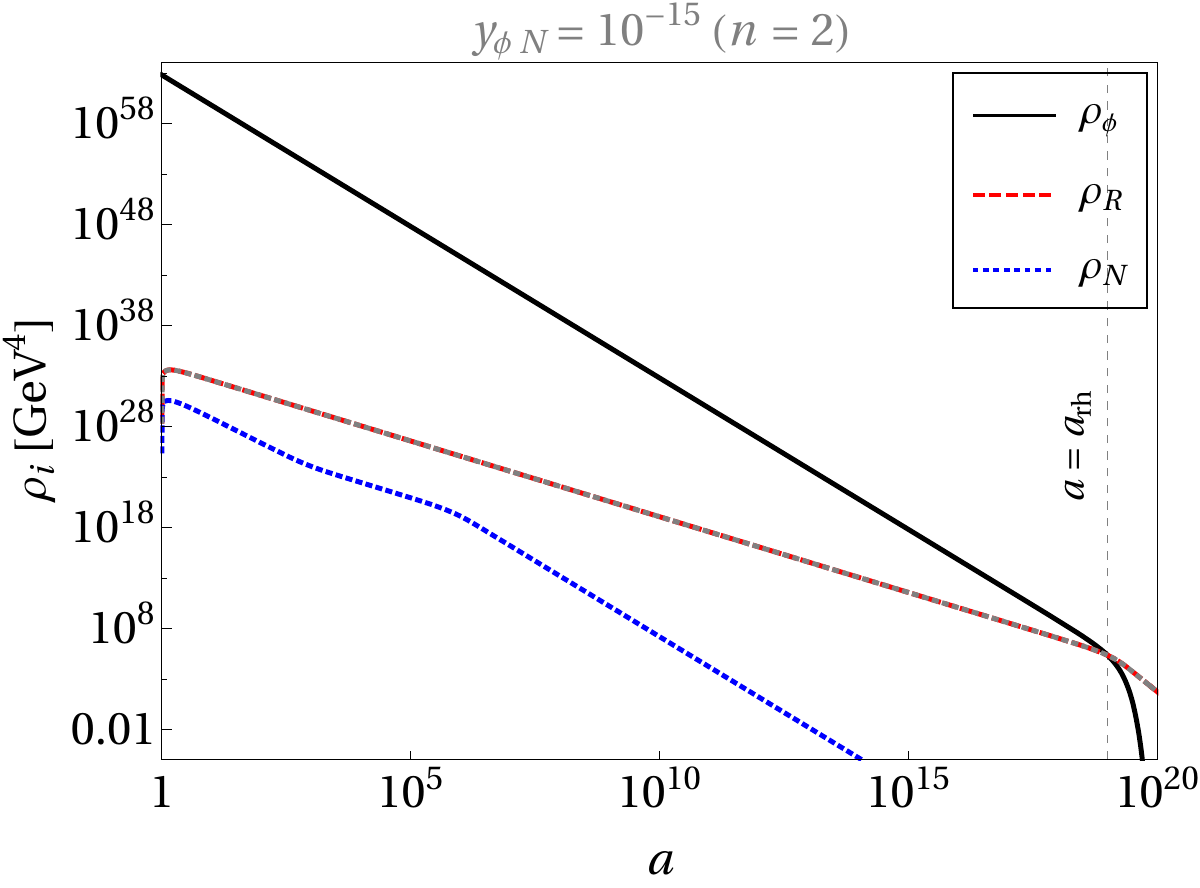}
\\[10pt]
\includegraphics[scale=0.42]{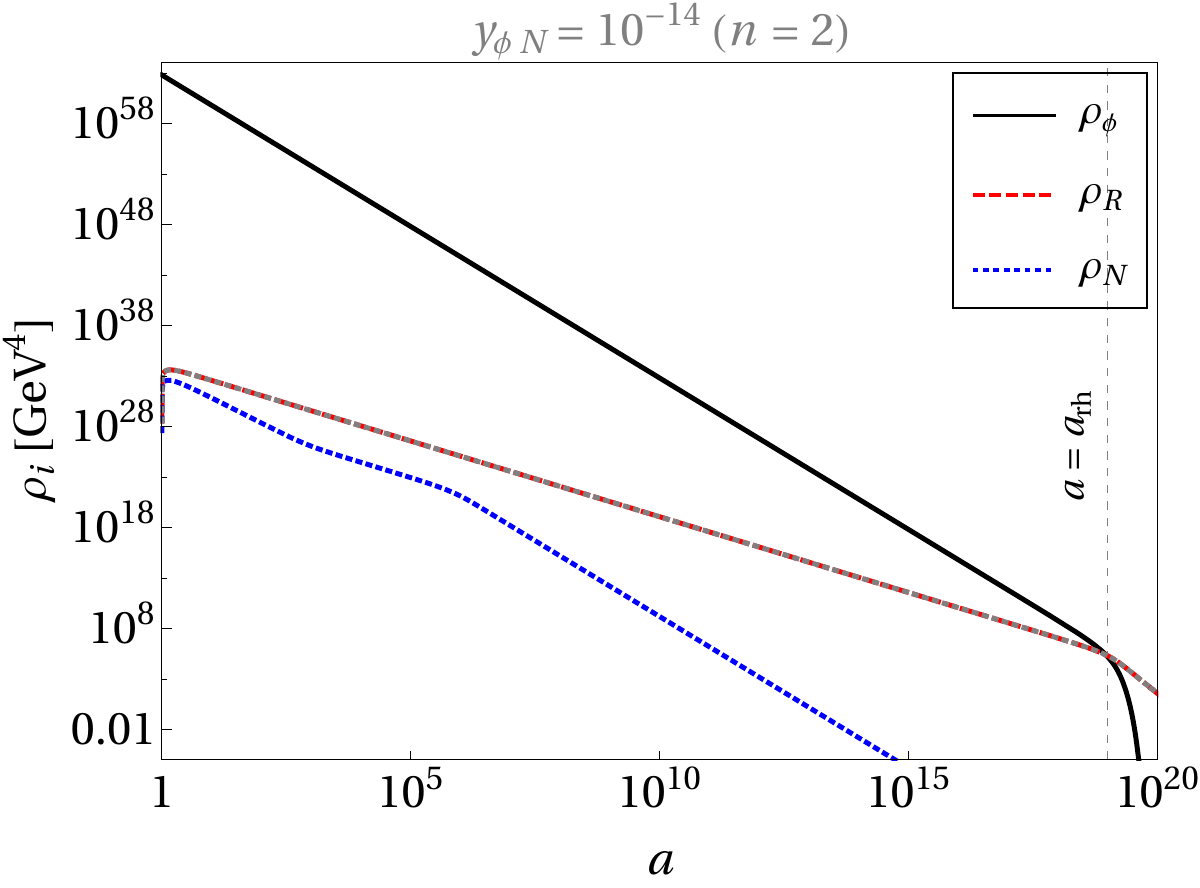}~
\caption{Evolution of inflaton (black solid), radiation (red dashed) and RHN (blue dotted) energy densities, as a function of the scale factor for $n=2$. In all cases we have fixed $\Trh=10$ GeV and $M_{N_1}=10^{10}$ GeV. In all cases the gray dotted curve shows the radiation energy density entirely from inflaton decay.}
\label{fig:rhon2}
\end{figure}
\begin{figure}[htb!]
\centering
\includegraphics[scale=0.375]{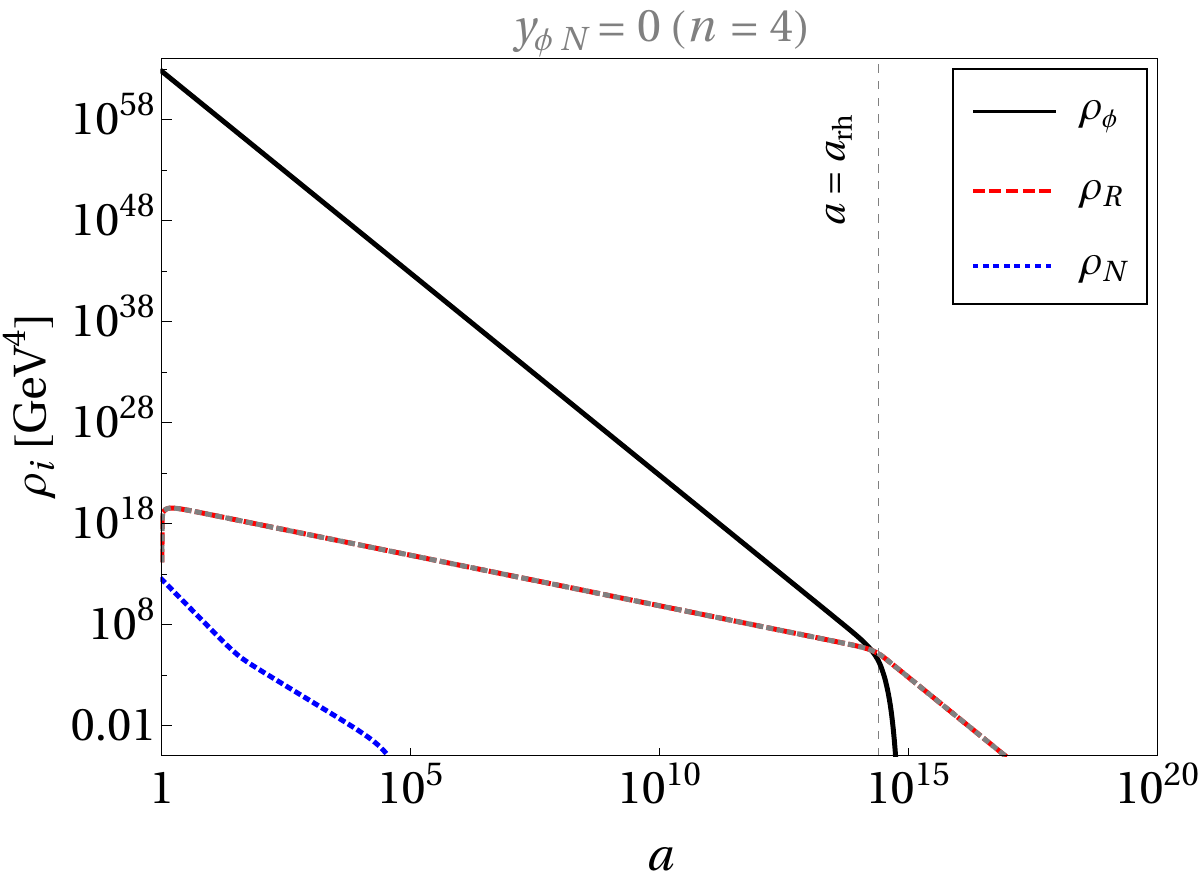}~\includegraphics[scale=0.375]{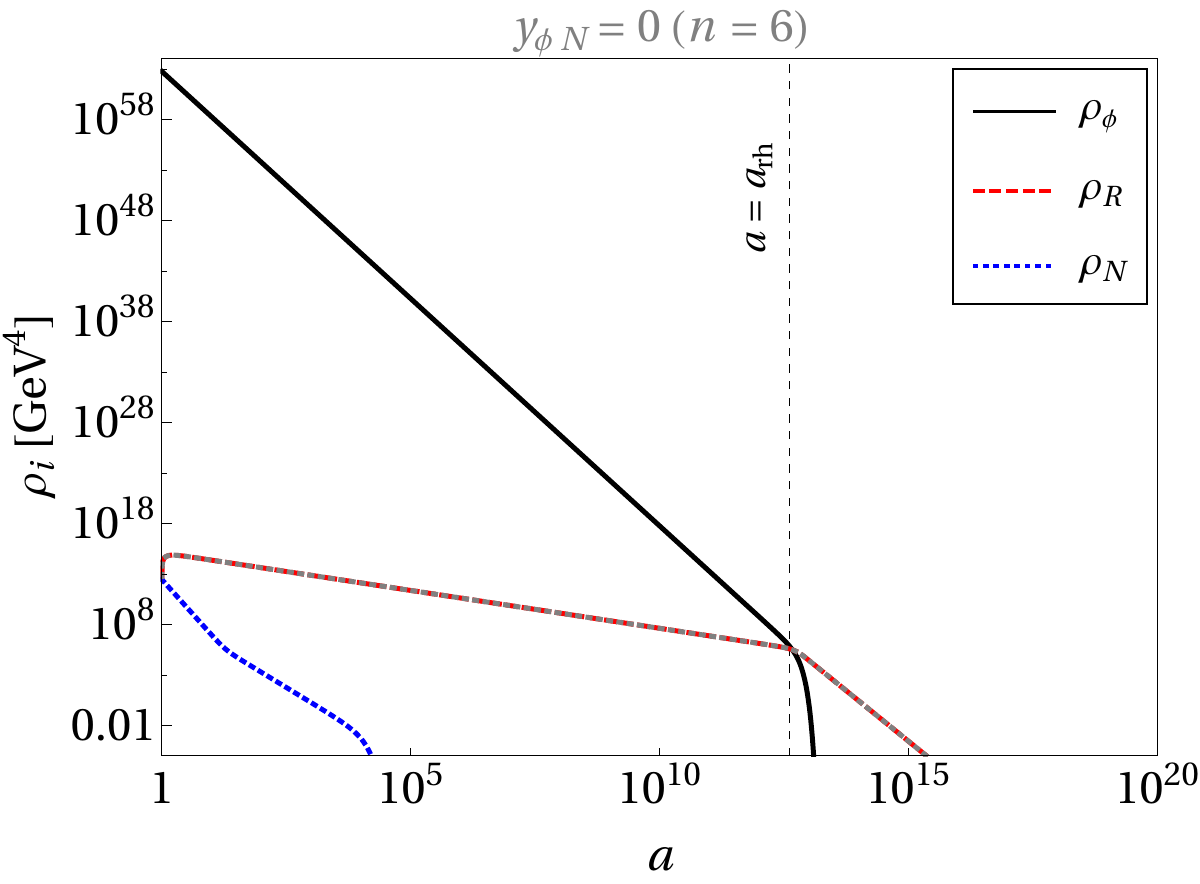}
\\[10pt]
\includegraphics[scale=0.375]{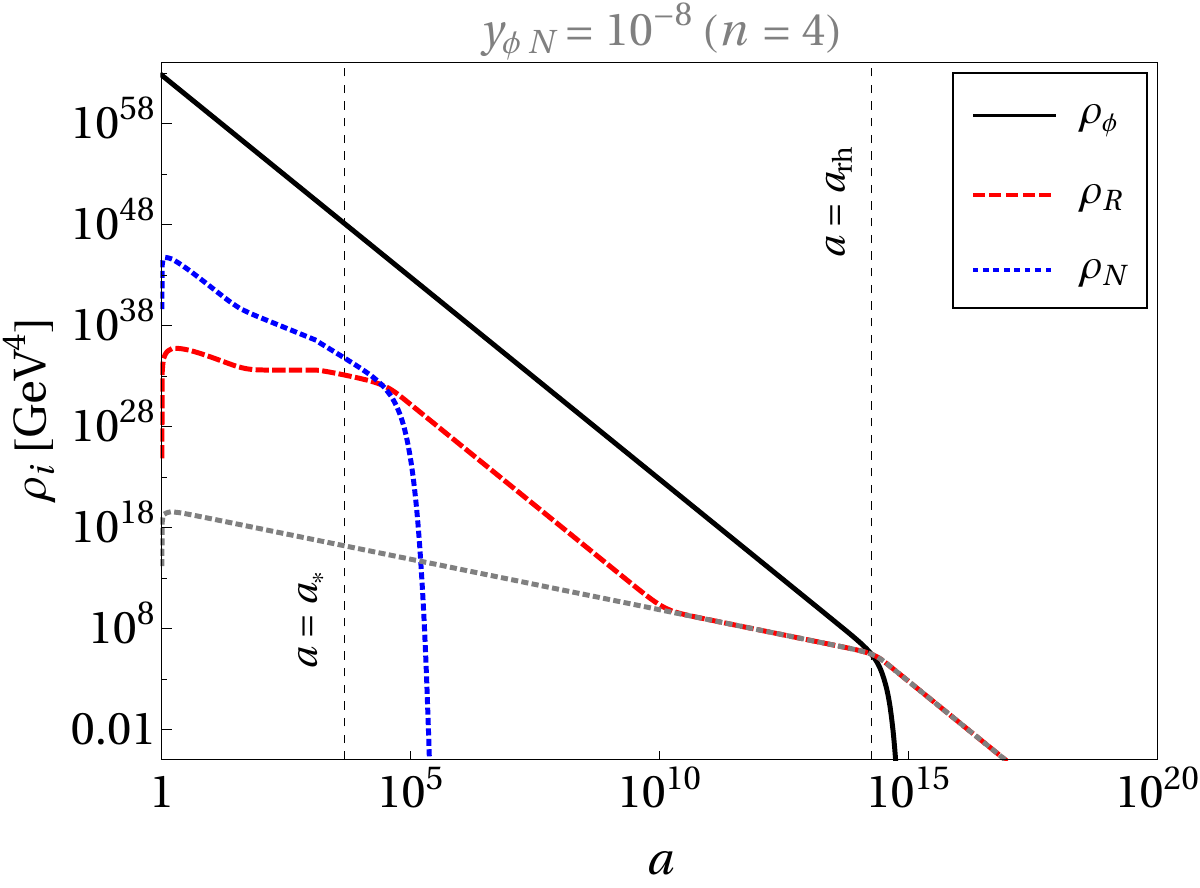}~\includegraphics[scale=0.375]{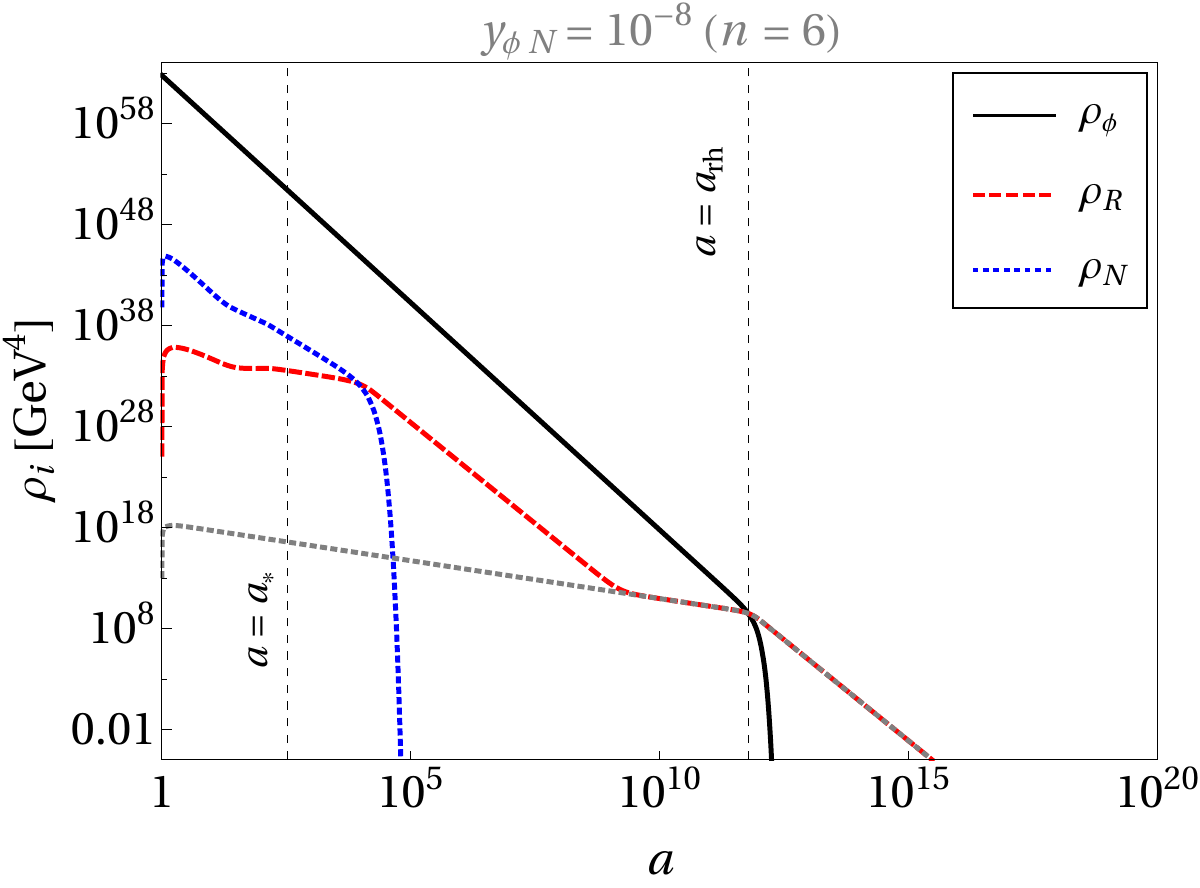}
\\[10pt]
\includegraphics[scale=0.375]{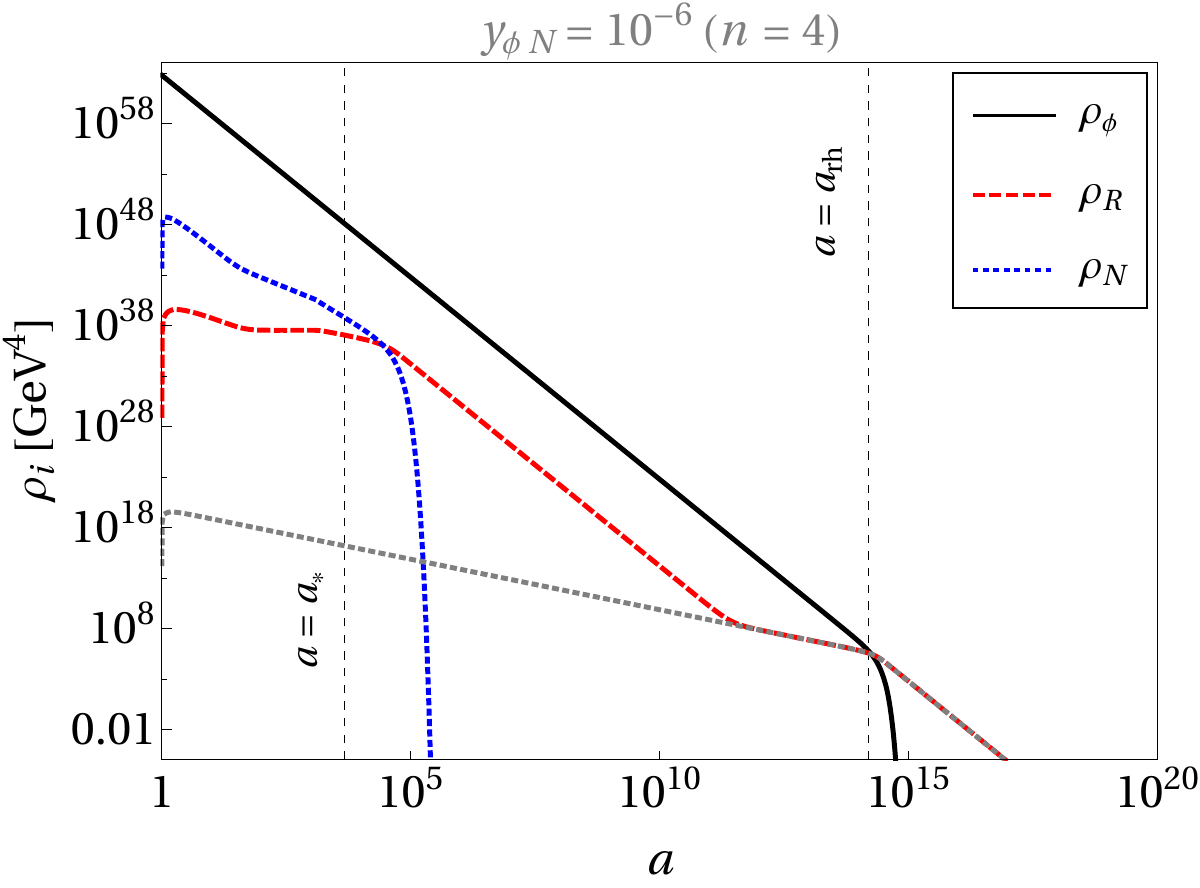}~\includegraphics[scale=0.375]{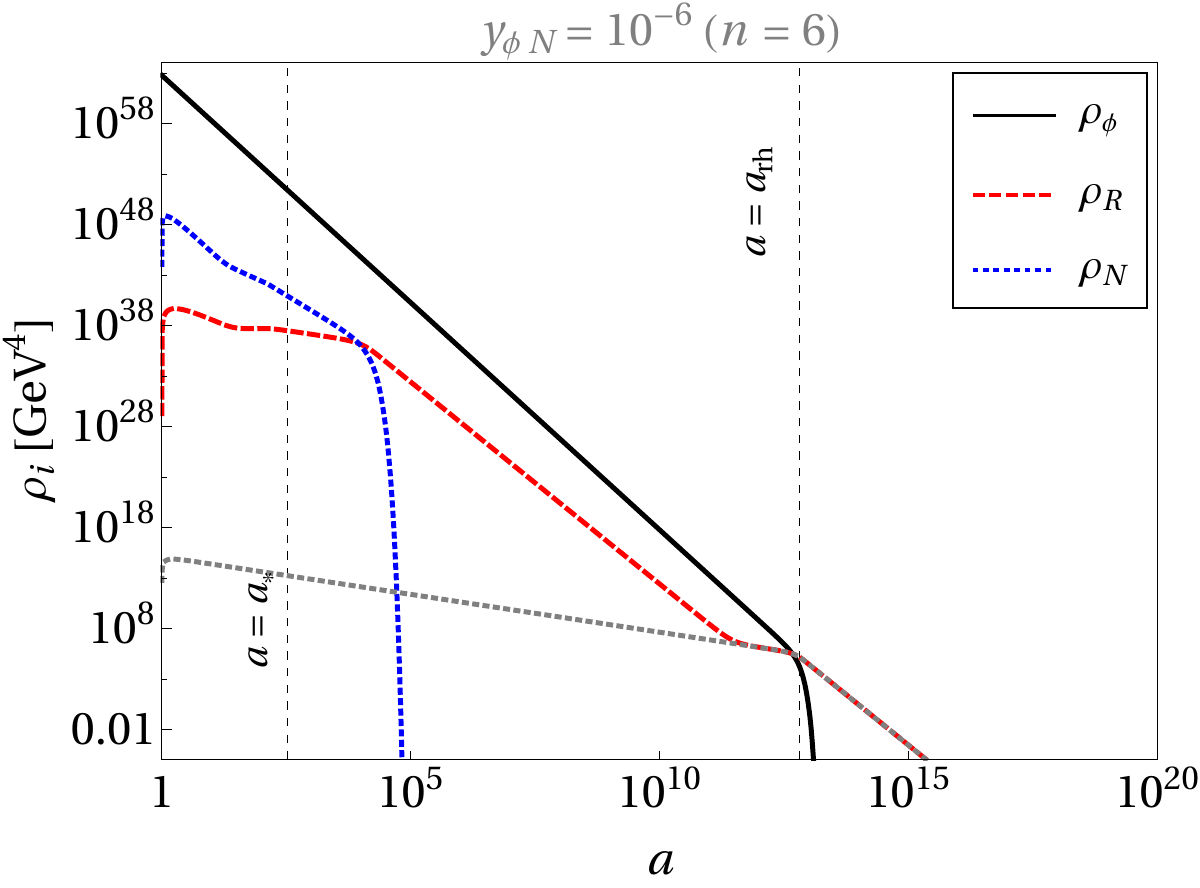}
\caption{{\it Scenario-A:} Bosonic reheating. evolution of inflaton (black solid), radiation (red dashed) and RHN (blue dotted) energy densities, as a function of the scale factor for $n=4$ (left column) and $n=6$ (right column). We have fixed $\Trh=10$ GeV. In all cases the gray dotted curve shows the radiation energy density entirely from inflaton decay.}
\label{fig:rho-bos}
\end{figure}
\begin{figure}[htb!]
\centering
\includegraphics[scale=0.375]{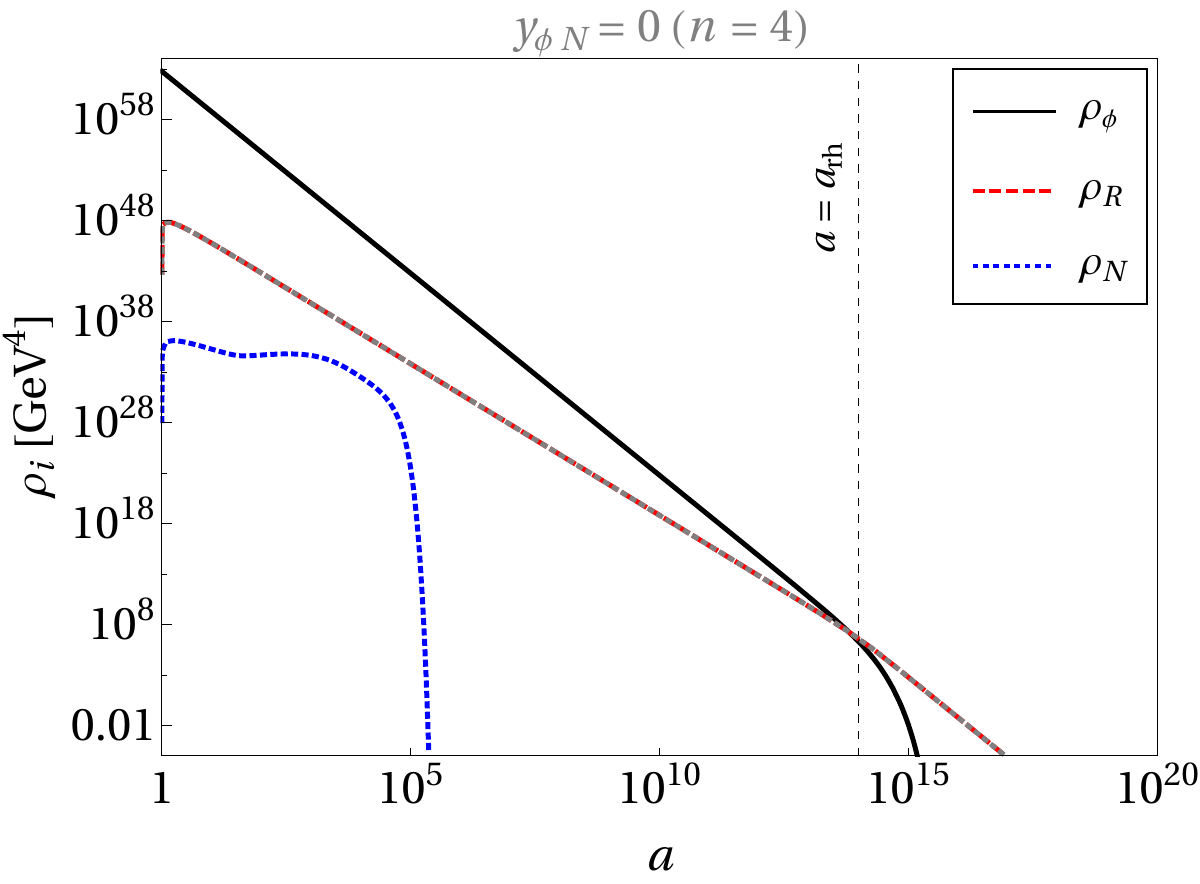}~\includegraphics[scale=0.375]{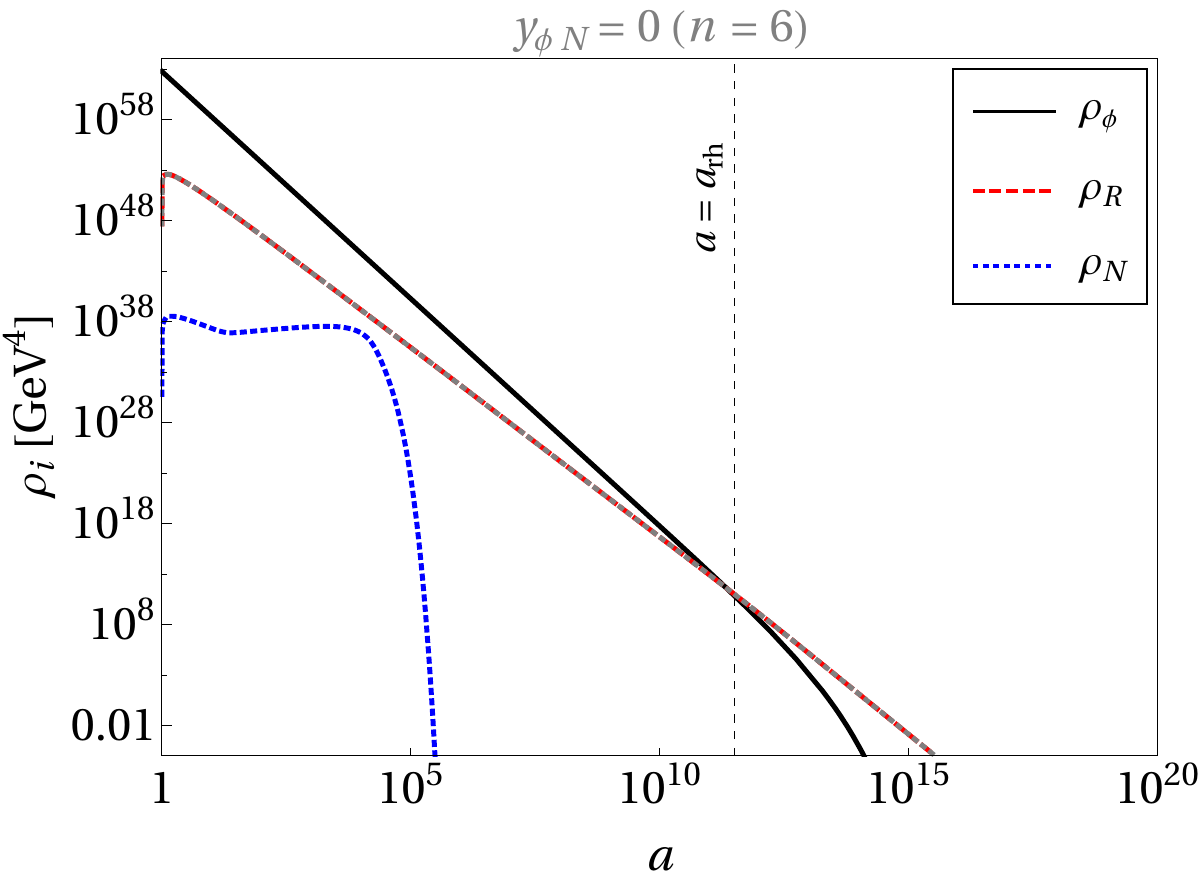}
\\[10pt]
\includegraphics[scale=0.375]{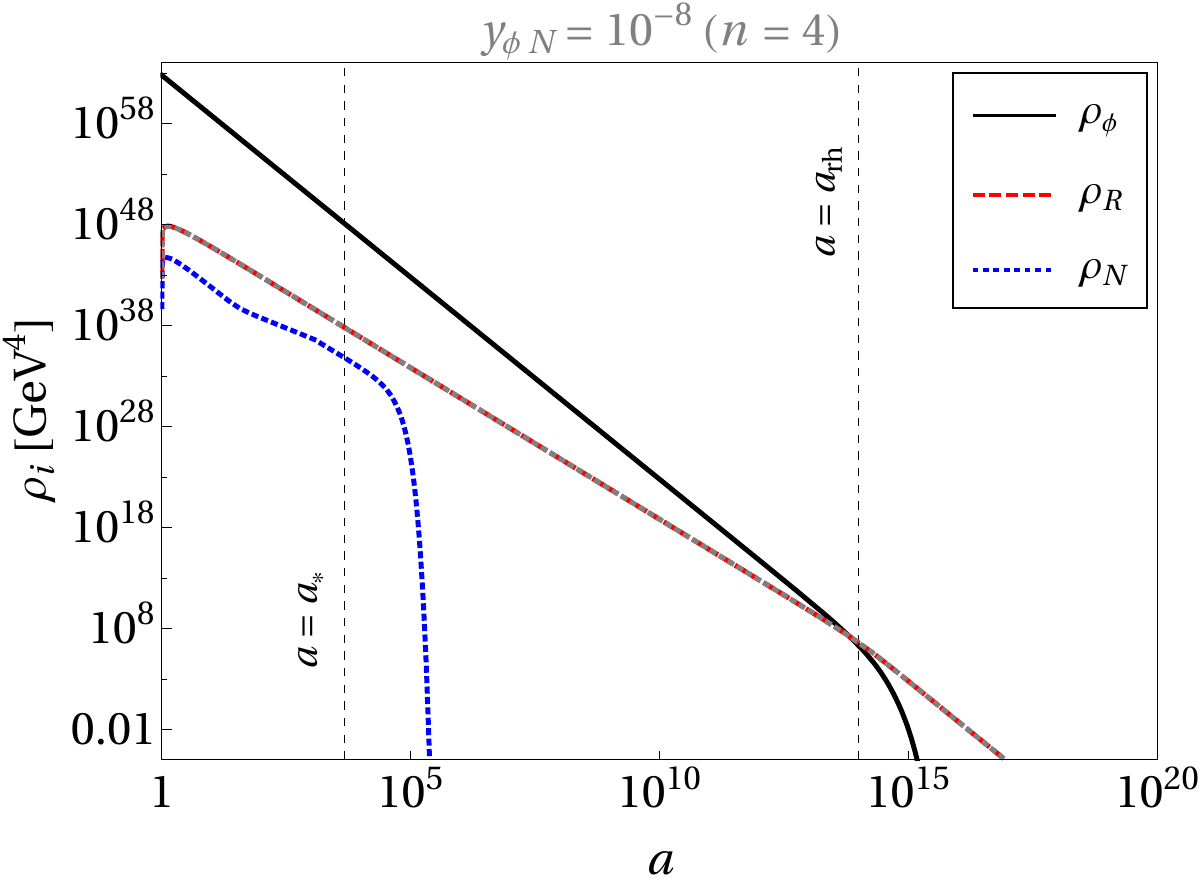}~\includegraphics[scale=0.375]{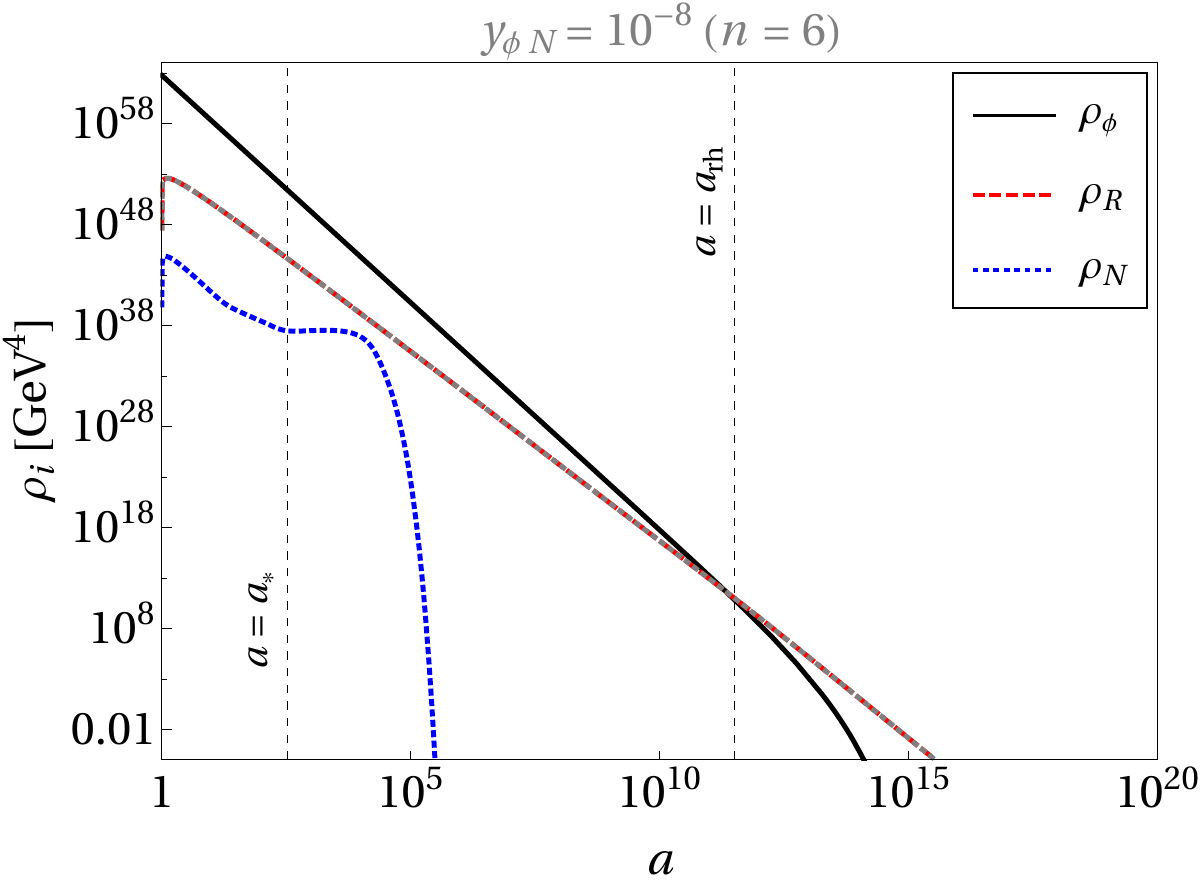}
\\[10pt]
\includegraphics[scale=0.375]{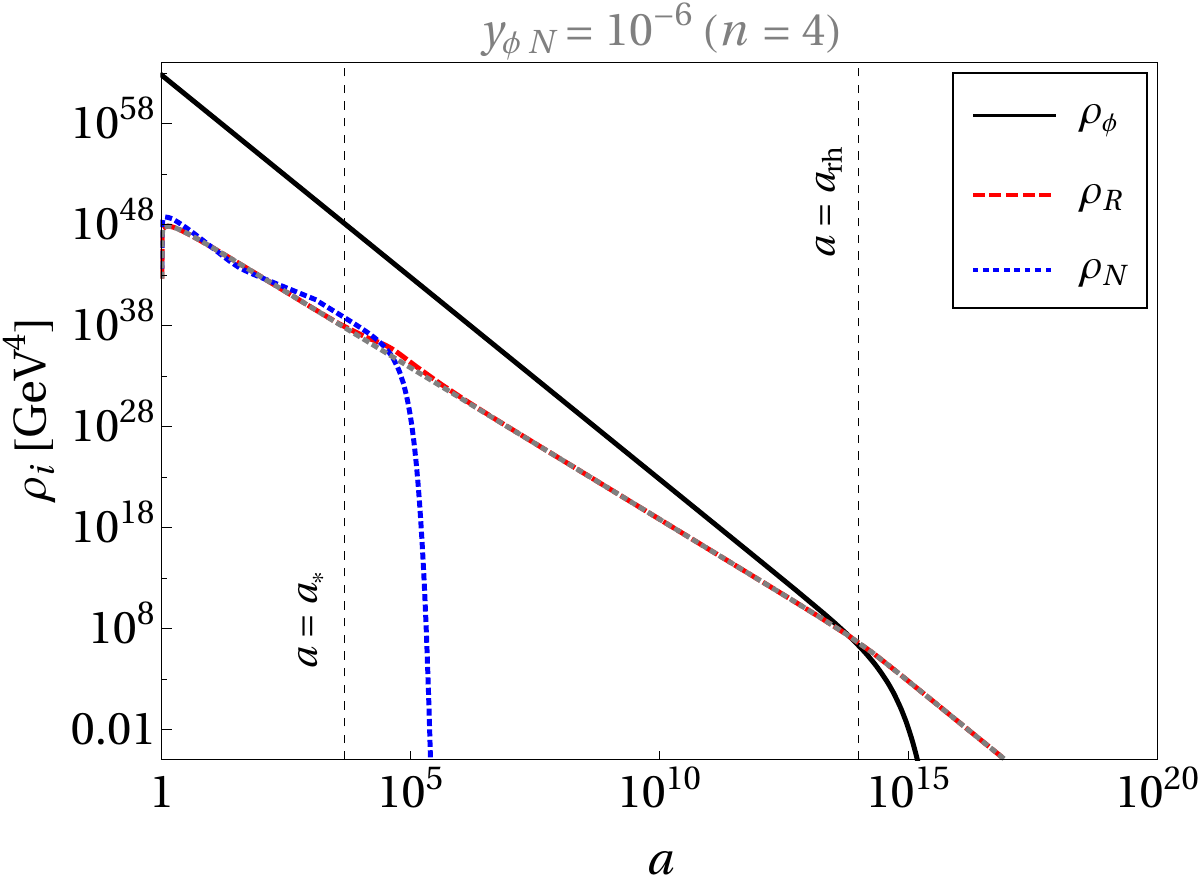}~\includegraphics[scale=0.375]{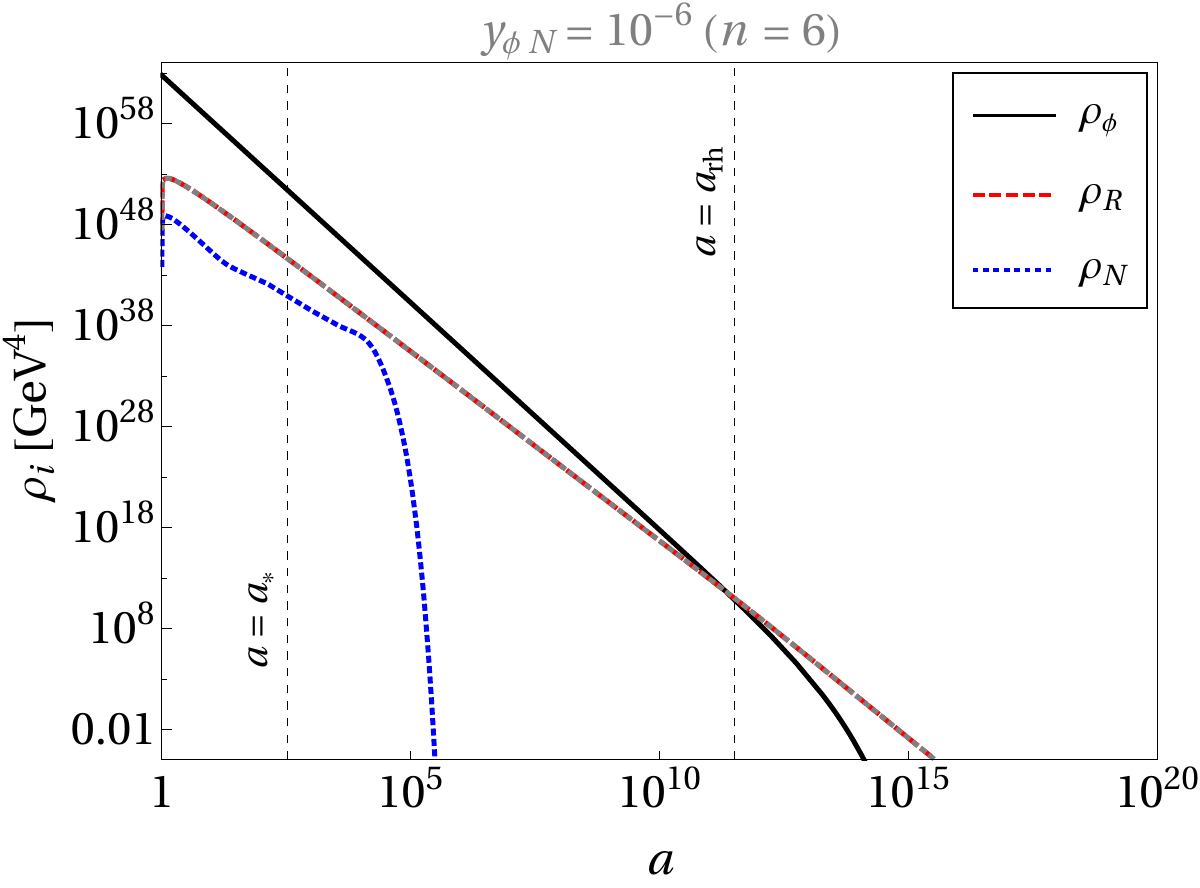}
\caption{{\it Scenario-A:} Fermionic reheating. evolution of inflaton (black solid), radiation (red dashed) and RHN (blue dotted) energy densities, as a function of the scale factor for $n=4$ (left column) and $n=6$ (right column). We have fixed $\Trh=10$ GeV and $M_{N_1}=10^{10}$ GeV. In all cases the gray dotted curve (overlapped with the red dashed curve) shows the radiation energy density entirely from inflaton decay.}
\label{fig:rho-fer}
\end{figure}
\begin{figure}[htb!]
\centering
\includegraphics[scale=0.45]{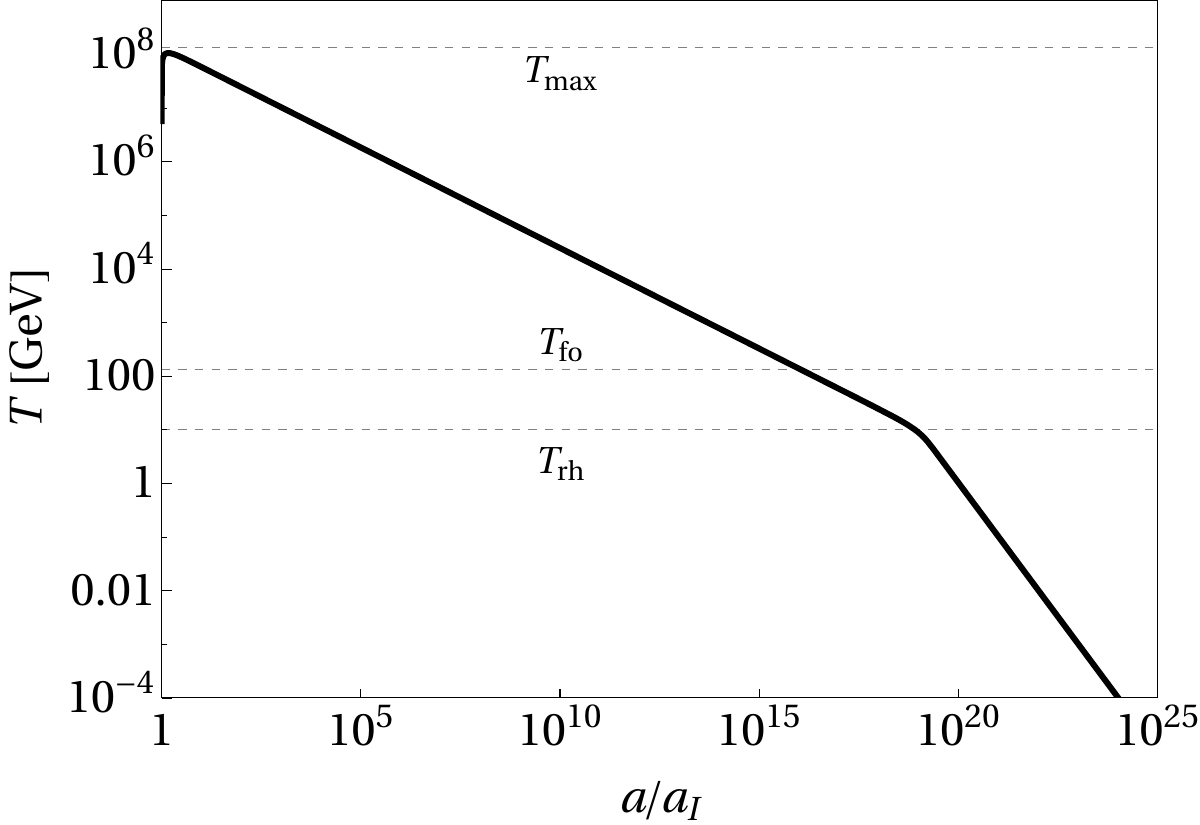}\\[10pt]
\includegraphics[scale=0.375]{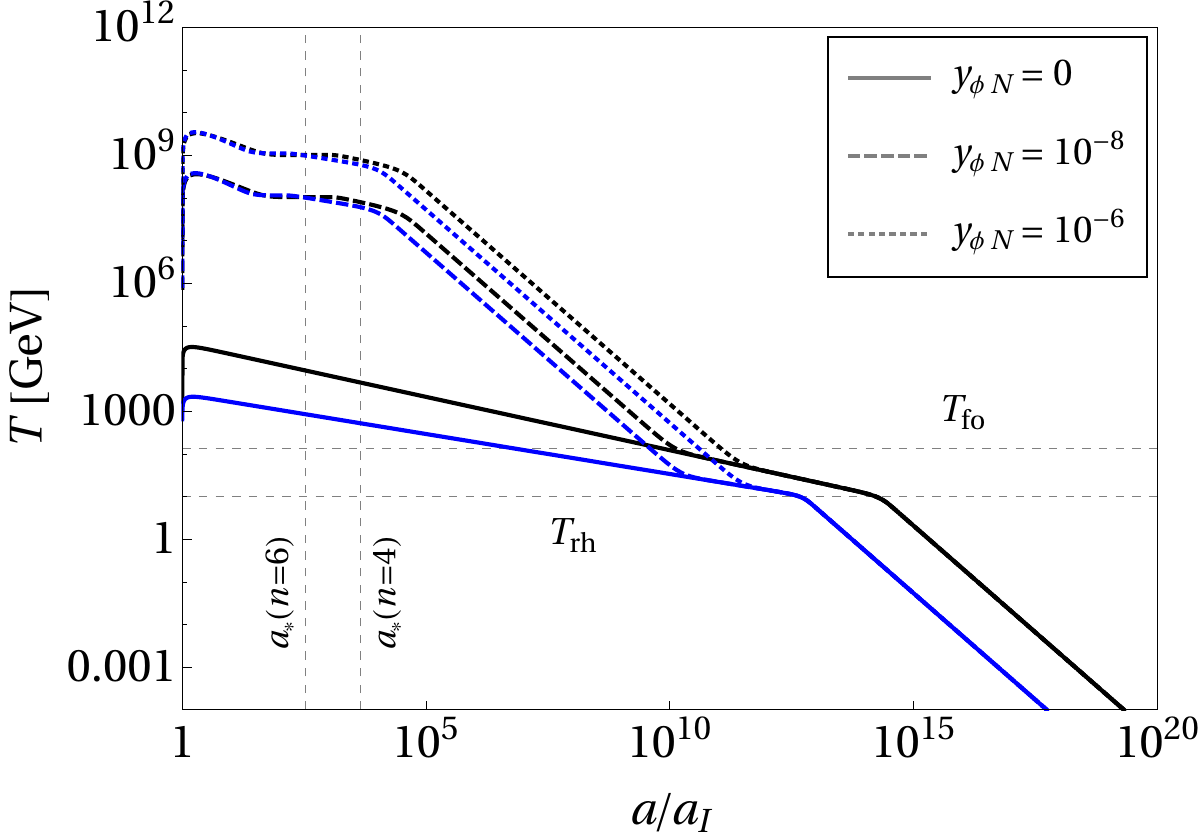}~\includegraphics[scale=0.375]{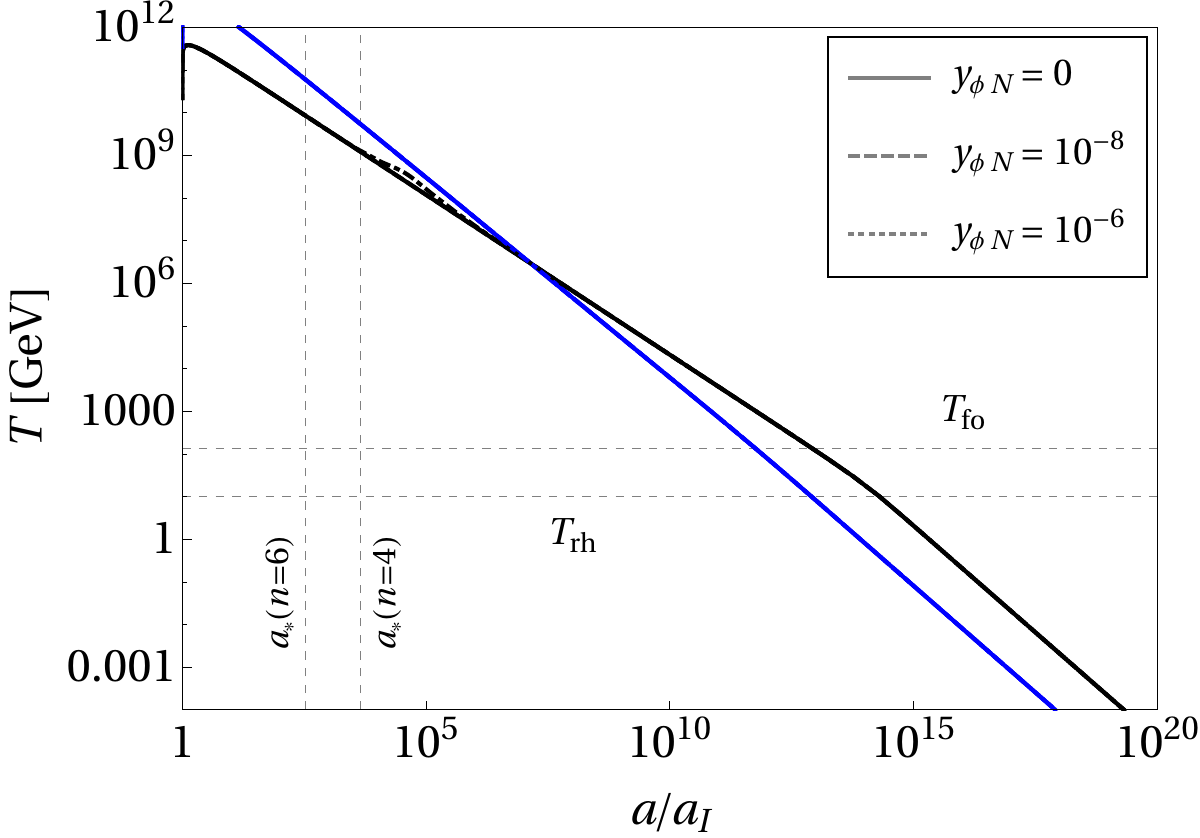}
\caption{Temperature evolution for {\it Scenario-A}, for different choices of $\yphn$, as denoted by different curves, considering bosonic (bottom left) and fermionc (bottom right) reheating scenarios. The blue and black curves correspond to $n=4$ and $n=6$, respectively. The top panel shows $n=2$ case.}
\label{fig:T-evol}
\end{figure}
The evolution of the energy densities of the inflaton ($\rp$), SM radiation ($\rR$), and RHNs ($\rho_N$) is shown in Fig.~\ref{fig:rhon2} for $n=2$. In this case, the bosonic and fermionic reheating scenarios coincide. When the inflaton–RHN coupling is switched on, the radiation bath receives contributions both from RHN decays as well as from the decay of the inflaton condensate. We choose $\yphn$ in such a way that, the reheating is completed entirely via the SM channel. With increase in $\yphn$, the RHN energy density starts rising as the branching of the inflaton into RHN final state rises. Now, as the physical three-momentum of any particle (both massive and massless) decays with the expansion of the Universe, it is possible to determine the epoch $a_{\rm NR}$ at which $N_1$ become non-relativistic,
\begin{align}\label{eq:aNR}
& a_{\rm NR}\simeq
a_I\,\left(\frac{m_I} {2\,M_1}\right)^{\frac{n+2}{4(n-1)}}\,,   
\end{align}
considering initial RHN momenta $p(a_I)\simeq m_\phi/2$ and $p(a_{\rm NR})\simeq M_1$. For an RHN of mass $10^{10}$ GeV, we find, $a_{\rm NR}/a_I\sim \{780,\,34,\,20\}$, for $n=\{2,\,4,\,6\}$ respectively, implying the RHN becomes non-relativistic at an early stage of reheating. Finally, reheating is completed once the radiation energy density (black curve) overtakes the inflaton energy density (red curve) at a scale factor,
\begin{align}\label{eq:arh}
\arh\simeq a_I\,\left(\frac{2}{3}\,\frac{\rp(a_I)}{M_P^2\,\mathcal{H}(\arh)^2}\right)^\frac{n+2}{6n}\,. 
\end{align}
For $\Trh=10$ GeV, we find, $\arh/a_I\sim\{10^{19},\,1.7\times 10^{14},\,4.5\times 10^{12}\}$, for $n=\{2,\,4,\,6\}$, respectively.

Beyond $n=2$, the evolution of the energy densities exhibits non-trivial features. For $n=\{4,\,6\}$, the inflaton energy density redshifts as $\rho_\phi\propto\{a^{-4},\,a^{-4.5}\}$, similar to that of free radiation. In the bosonic reheating scenario, the inflaton dissipation rate scales as $\Gamma_\phi\propto 1/m_\phi$. Since $m_\phi(a)$ decreases with time for $n>2$, the dissipation rate gradually increases, making bosonic reheating progressively more efficient than fermionic reheating, for which the dissipation rate scales as $\Gamma_\phi\propto m_\phi$. Consequently, during the inflaton-dominated era, the radiation energy density (or equivalently, the temperature) decreases much more slowly with the scale factor in the bosonic case than in the fermionic case, as illustrated in Figs.~\ref{fig:rho-bos} and \ref{fig:rho-fer}. Specifically, for $n=\{4,\,6\}$, the radiation energy density scales as $\rho_R\propto\{a^{-1},\,a^{-0.75}\}$ in the bosonic reheating scenario, whereas it scales as $\rho_R\propto\{a^{-3},\,a^{-3.75}\}$ for fermionic reheating, in case of radiation production during pure inflaton domination. Since the maximum thermal bath energy density is lower in the bosonic reheating case [cf. Fig.~\ref{fig:Tmax}], while the RHN production rate $\Gamma_{\phi\to NN}\propto m_\phi(a)$, the RHN energy density can temporarily dominate over the radiation energy density, depending on the choice of $\yphn$. Once $a>\ast$, the inflaton no longer has sufficient mass to decay into a pair of RHNs. The RHN population then simply redshifts and decays into SM particles, causing its energy density to decrease. Nevertheless, the radiation produced from RHN decays continues to dominate over the radiation directly generated by inflaton decays until approximately $a\sim10^{10}$, when bosonic reheating becomes sufficiently efficient. The reheating is completed shortly thereafter. In contrast, for fermionic reheating (Fig.~\ref{fig:rho-fer}), the radiation produced directly from inflaton decay is practically indistinguishable from that originating from intermediate RHN decays. This is because both $\phi\to NN$ and $\phi\to\psi\psi$ decay widths exhibit the same $\propto m_\phi$ dependence, and therefore evolve identically with time.

The temperature evolution for $n=2$ is shown in the top panel of Fig.~\ref{fig:T-evol}. As expected, the temperature follows the standard reheating behavior, $T\propto a^{-3/8}$, confirming that reheating proceeds through the SM channel. For bosonic reheating, shown in the bottom left panel, the temperature scales as $T\propto\{a^{-4},\,a^{-3/16}\}$ for $n=\{4,\,6\}$, respectively, when $\yphn=0$. Once $\yphn$ is turned on, the temperature rises much earlier because the radiation bath receives an additional contribution from the decay of RHNs produced directly from inflaton decay. Nevertheless, the final reheating is still driven by the bosonic channel, since RHN production from the inflaton ends at $a=\ast\ll\arh$. A similar behavior is observed for fermionic reheating, as shown in the bottom right panel. In this case, however, the temperature evolution is nearly identical for both $n=4$ and $n=6$, causing the corresponding curves to overlap. 
\subsubsection{{\it Scenario B:} reheating via RHN decay}
\label{sec:caseB}
In the preceding section, we examined leptogenesis with low reheating temperatures in scenarios where the inflaton reheats the Universe by decaying into SM-like bosonic or fermionic states. We now turn to an even more economical setup in which the inflaton condensate decays exclusively through two-body channels into right-handed neutrino states alone. In this construction, the inflaton directly populates all three RHN species, without any branching fraction explicitly into the SM final state. Two of them, $N_{1,2}$, can account for the observed baryon asymmetry of the Universe, while the third state, $N_3$, may be arranged to be sufficiently long-lived\footnote{In~\cite{Haque:2023zhb,Haque:2024zdq}, a similar set-up has been studied, where the RHNs are produced entirely via graviton-mediation during reheating.}. In such a scenario, the eventual decay of $N_3$ governs the onset and duration of the reheating epoch itself. We thus assume inflaton coupling to the SM sector to be absent by construction. A particularly intriguing consequence of this framework is that, if $N_3$ survives long enough and temporarily dominates the Universe's energy density, it can induce an intermediate phase of early matter domination before the Universe becomes radiation dominated. This intermediate epoch can leave distinctive cosmological imprints, most notably through potentially observable distortions in the primordial gravitational wave spectrum, a possibility we will explore in subsequent sections. 
\begin{figure}[htb!]
\centering
\includegraphics[scale=0.7]{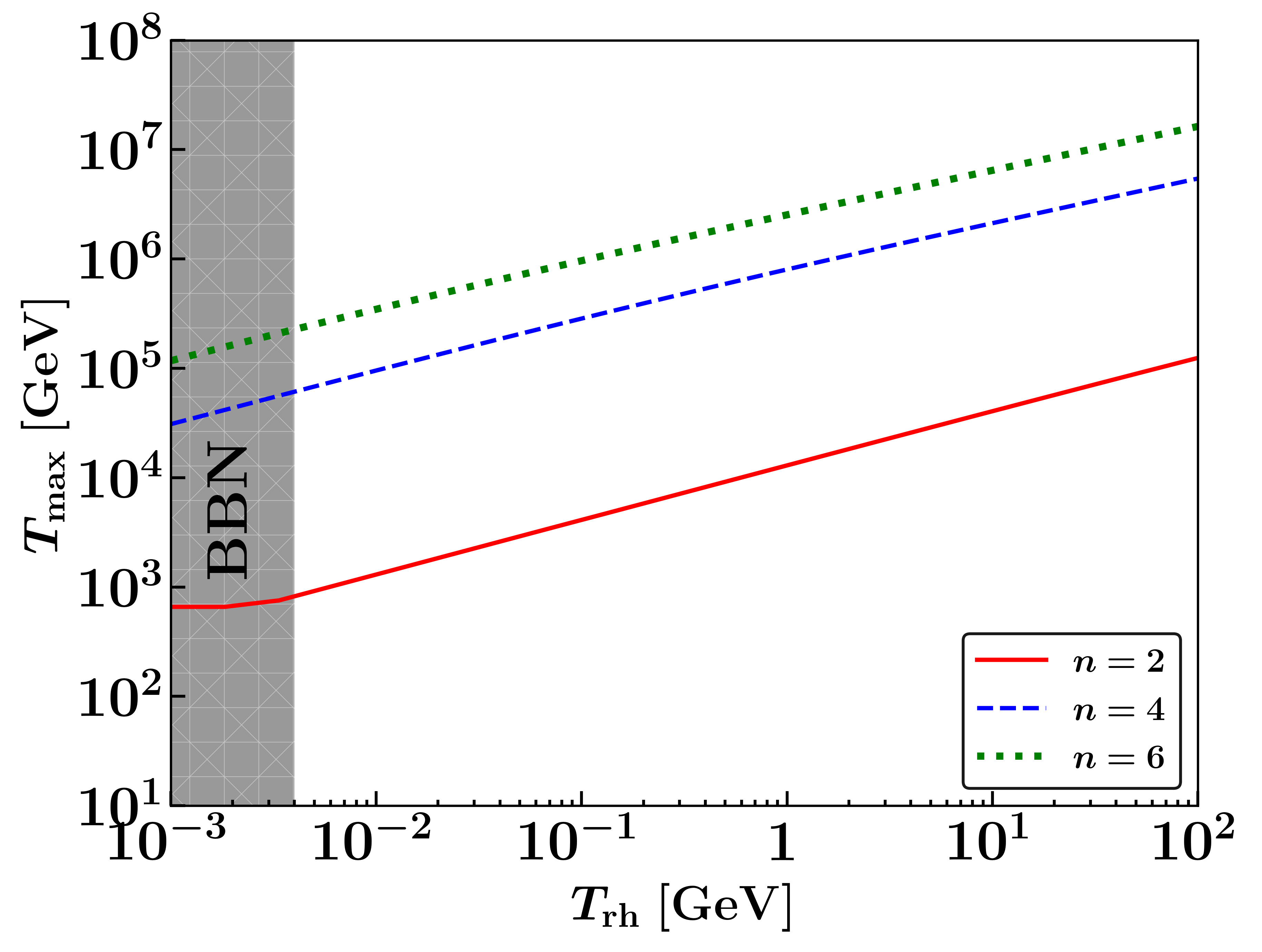}
\caption{{{\it Scenario-B}: Correlation between $\Tmax$ and $\Trh$, where different curves correspond to different choices of $n$. Here, $M_3=10^6$ GeV and $\yphn=\{10^{-6},\, 7\times 10^{-3},\,0.1\}$ for $n=\{2,\,4,\,6\}$, respectively. The gray shaded region corresponds to $\Trh<4$ MeV.}}
\label{fig:TmaxB}
\end{figure}
\begin{figure}[htb!]
\centering
\includegraphics[scale=0.47]{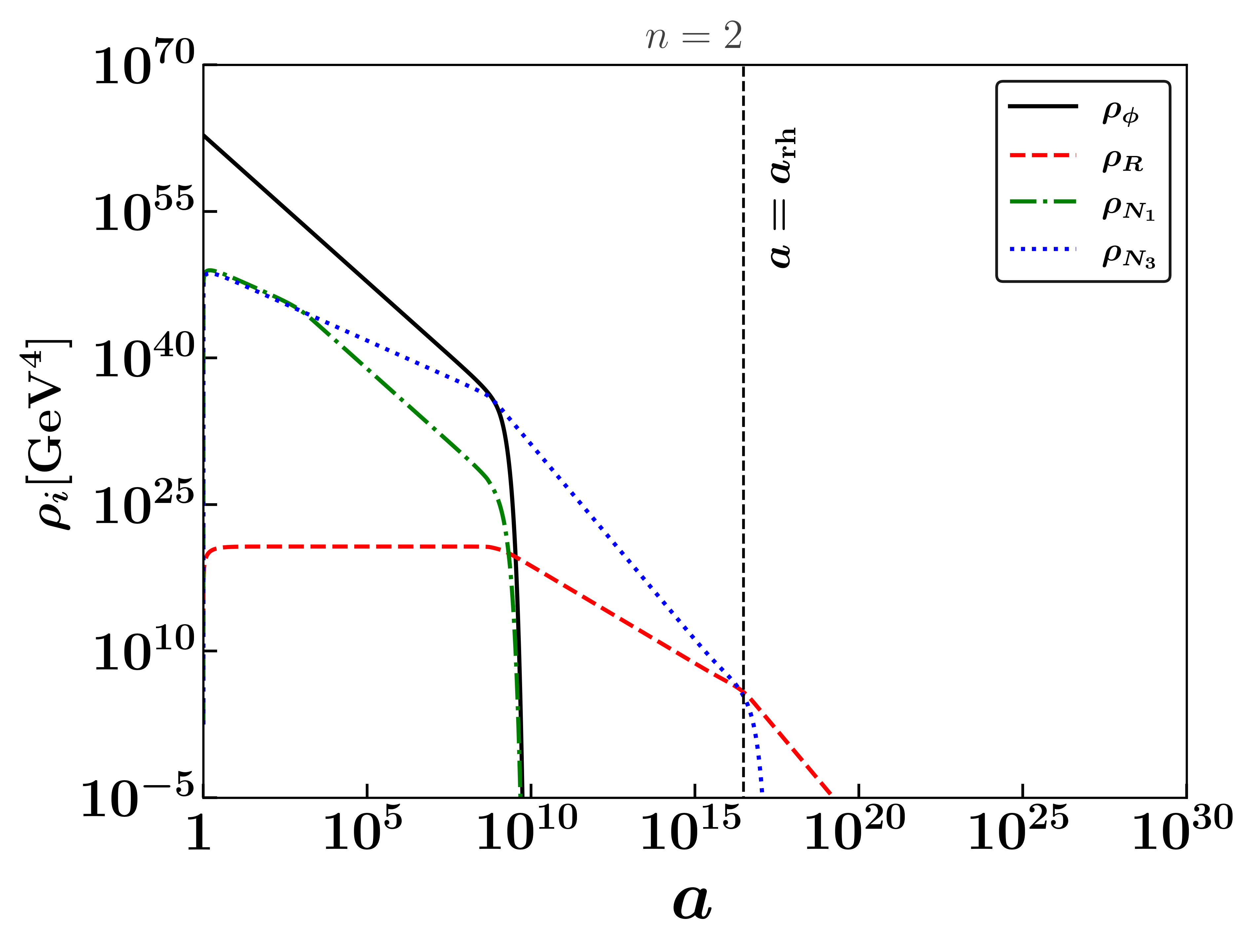}~\includegraphics[scale=0.47]{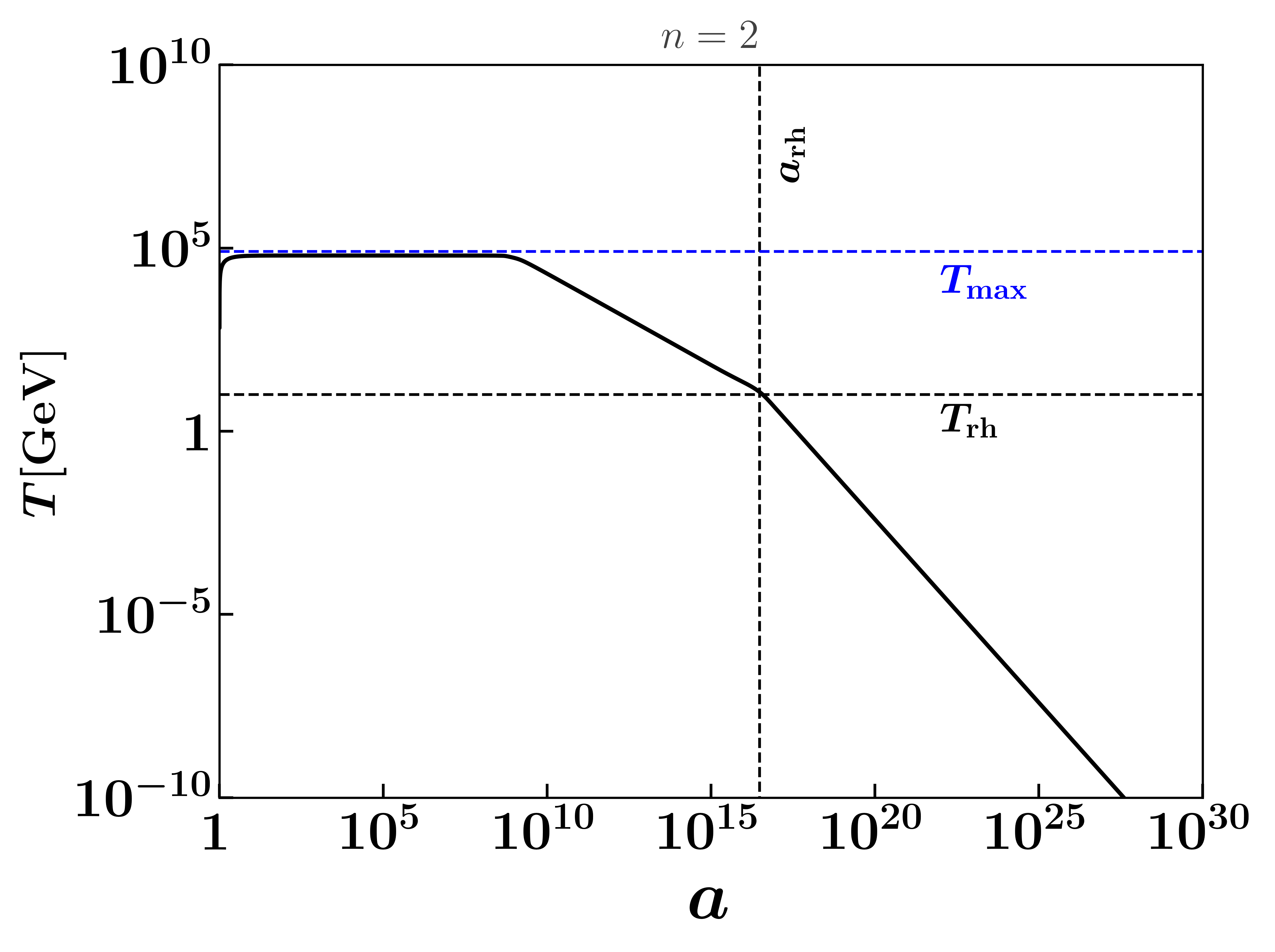}
\\[10pt]
\includegraphics[scale=0.47]{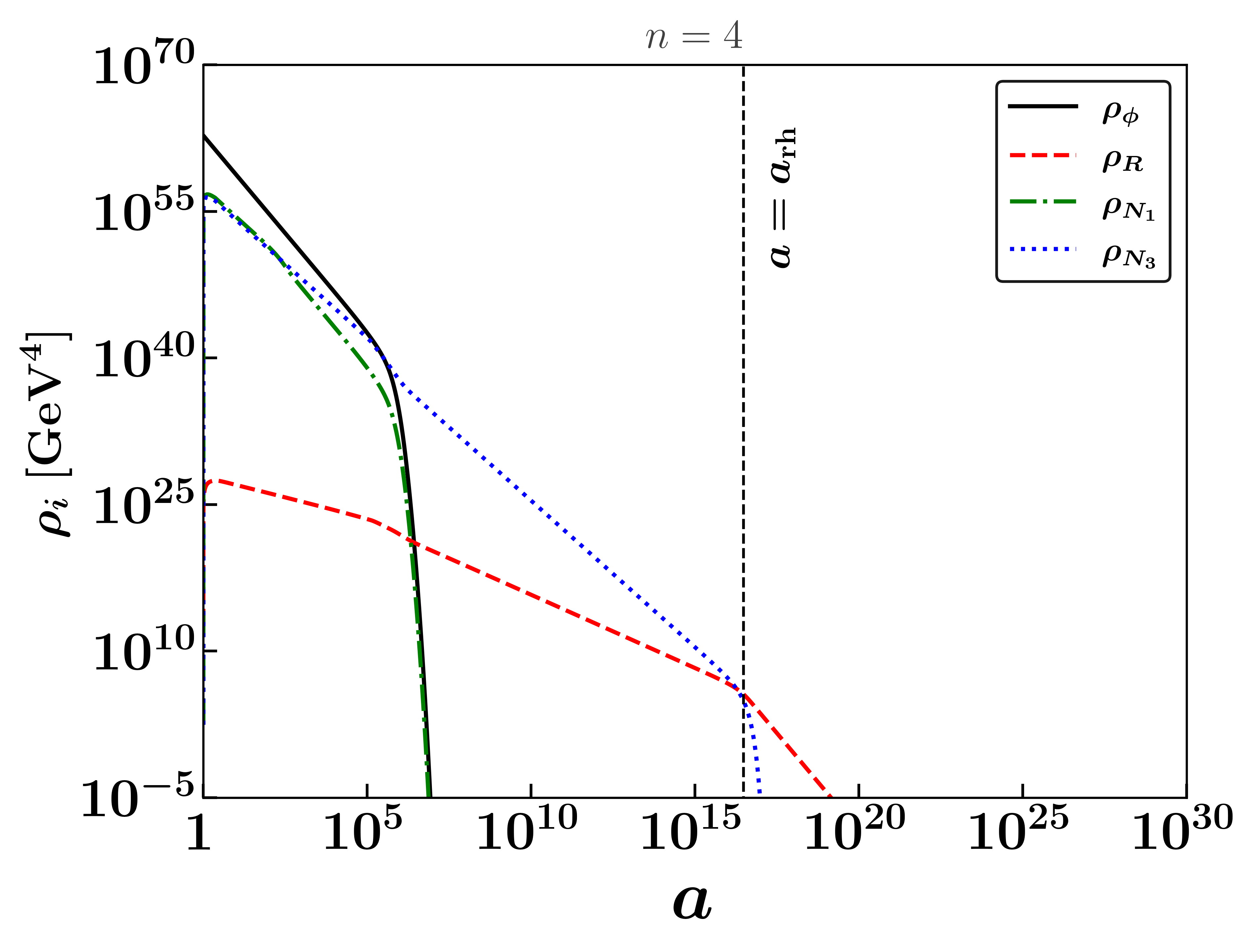}~\includegraphics[scale=0.47]{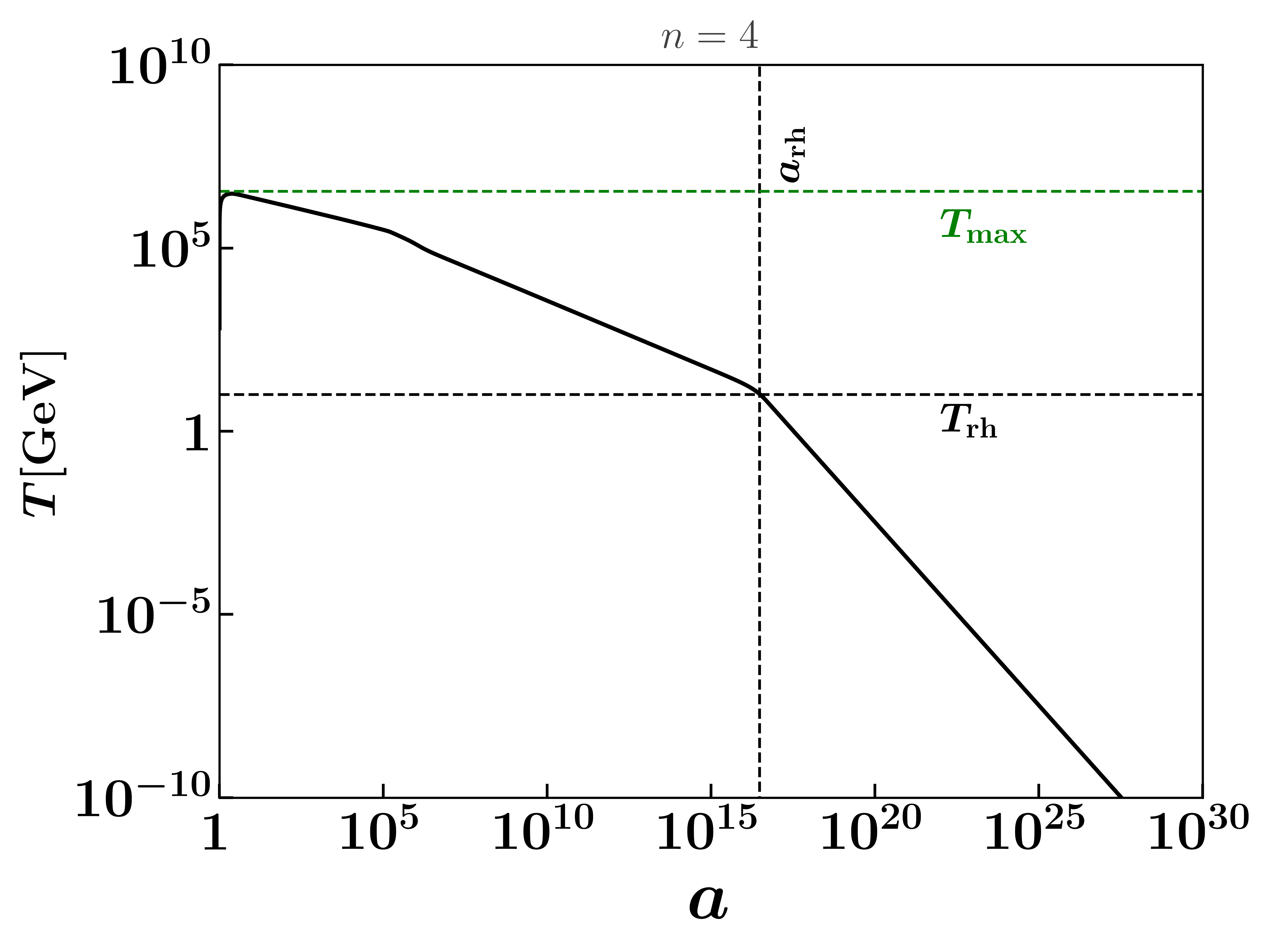}
\\[10pt]
\includegraphics[scale=0.47]{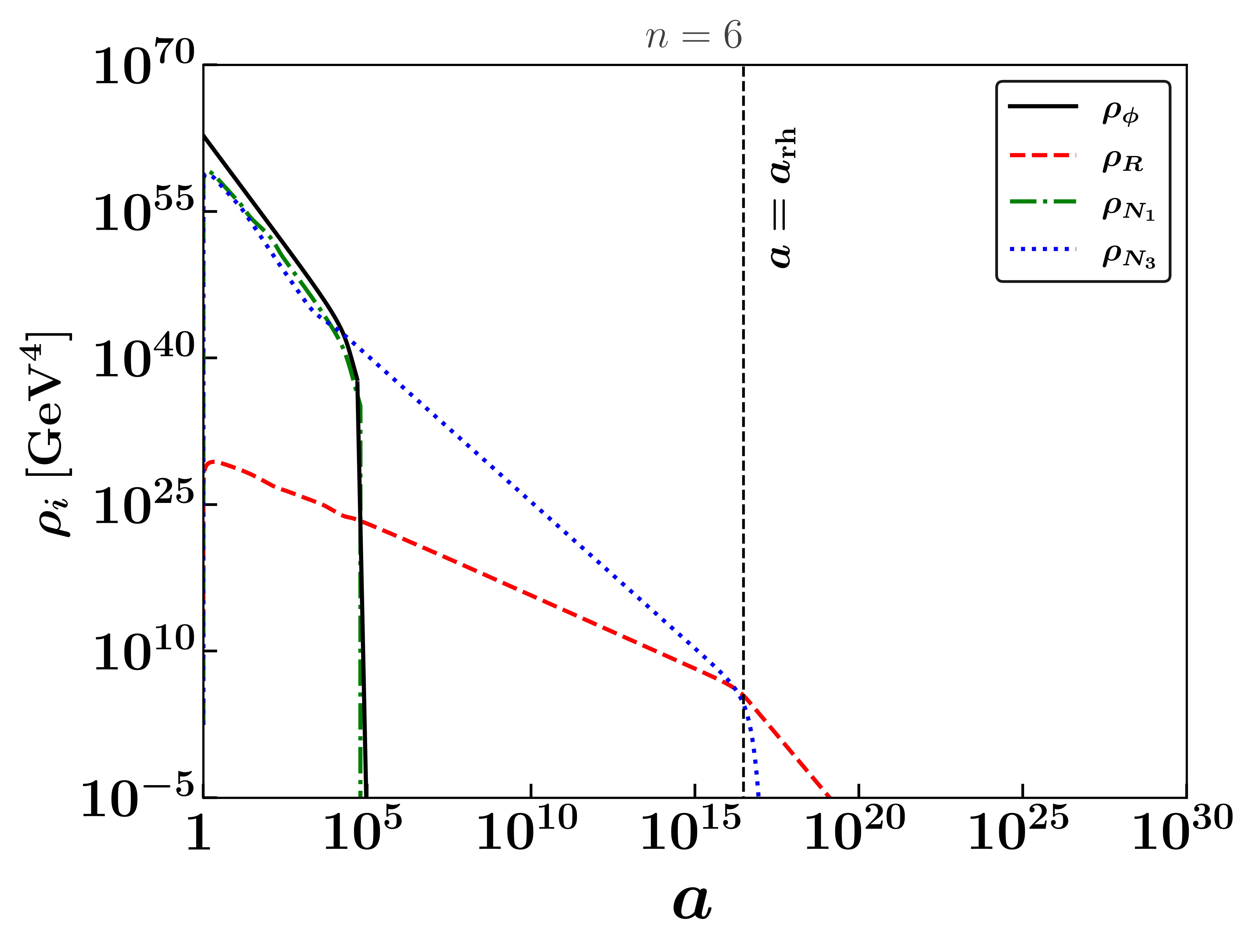}~\includegraphics[scale=0.47]{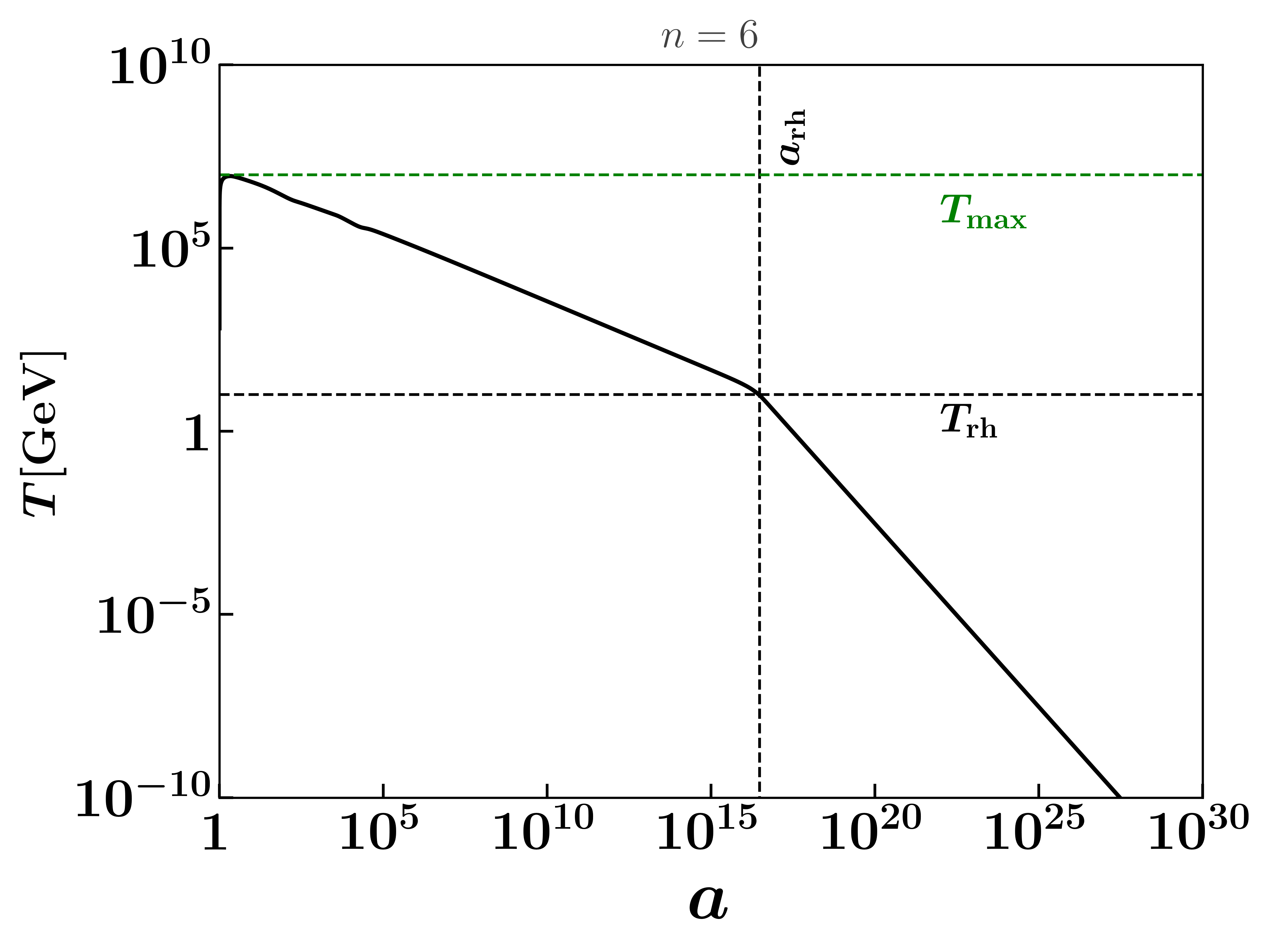}
\caption{{{\it Scenario-B:} Evolution of inflaton (black solid), radiation (red dashed) and the third RHN (blue dotted) energy densities, as a function of the scale factor for $n=\{2,\,4,\,6\}$ shown in the left column. The energy density for $N_1$ is shown via green dot-dashed curve. In the right column, we show the corresponding evolution of bath temperature as a function of the scale factor. We have taken $\yphn=\{10^{-6},\,7\times 10^{-3},\,0.1\}$ for $n=\{2,\,4,\,6\}$, respectively. In all cases we have fixed $M_1=5\times 10^{12}$ GeV, $M_3=10^6$ GeV along with $\Trh=10$ GeV.
}}
\label{fig:rho-RHN}
\end{figure}

As mentioned, since in this scenario the reheating temperature is set by the decay of $N_3$, after the Universe undergoes a brief period of $N_3$ domination, therefore
\begin{align}
\Trh\simeq\left[\left(\frac{y_{\nu 3}}{8\pi}\right)^2\,\left(\frac{45}{\pi^2\,\gs(\Trh)}\right)\,(M_3\,M_P)^2\right]^{1/4}\,,
\end{align}
where the $N_3$-SM Yukawa coupling $y_{\nu 3}$ and the mass $M_3$ are chosen such that the desired $\Trh$, is obtained, while being consistent with light neutrino data at the same time. The corresponding maximum temperature reached during reheating, $T_{\rm max}$, as a function of $\Trh$, is shown in Fig.~\ref{fig:TmaxB}. The evolution of the energy densities of the different components of the Universe for {\it Scenario-B} is shown in Fig.~\ref{fig:rho-RHN}. For all the chosen values of $n$, we observe an early matter domination during which the energy density of $N_3$ (blue dotted curve) becomes dominant. Here we fix $M_3=10^6~\GeV$ and choose $y_{\nu_3}=2\times10^{-10}$, corresponding to a reheating temperature of $\Trh=10~\GeV$. { For $n=2$, we observe a brief plateau in the evolution of the radiation energy density, and consequently in the temperature, during the interval $a_I\lesssim a\lesssim a_r$. During this epoch of inflaton domination, the RHNs remain relativistic. Therefore, $a_r$ can be determined by equating the inflaton and RHN energy densities, $\rho_\phi(a_r)=\rho_N(a_r)$, 
\begin{align}
& a_r\simeq a_I\,\left(\frac{\mathcal{H}_I}{\Gamma_{\phi\to NN}}\,\frac{n+8}{n+2}\right)^\frac{n+2}{3n}\,.
\end{align}
For $n=2$, this reduces to Eq.~(4.6) of~\cite{Bernal:2026lpr}.} Since $N_1$ has a sufficiently large coupling to the SM, it decays promptly after production, $N_3$, on the other hand, has a highly suppressed SM coupling and is therefore long-lived. Consequently, the $N_1$ population is depleted soon after the inflaton decays, with its energy density being transferred to the radiation bath. As a result, both the inflaton and $N_1$ energy densities rapidly die down together. Nevertheless, during the initial stage of reheating, the inflaton still dominates the expansion of the Universe. Consequently, we note a distinct change in the slope of the radiation energy density, which is also present in the temperature evolution, shown in the right panel. For larger values of $n$, the inflaton energy density redshifts faster [cf. Eq.~\eqref{eq:rpsol}], and therefore becomes subdominant at an earlier stage. 
\section{Sphalerons during reheating}
\label{sec:sphaleron}
In the standard radiation-dominated Universe, sphalerons remain in thermal equilibrium over a broad temperature range $10^2\,\text{GeV}\lesssim T\lesssim 10^{12}\,\text{GeV}$, and efficiently convert a pre-existing lepton asymmetry into a baryon asymmetry~\cite{Kuzmin:1985mm}. As we will demonstrate in this section, this conventional picture can be significantly modified during the reheating era. Depending on the background EoS and the specific reheating mechanism, sphaleron interactions may depart from equilibrium at temperatures much lower than $10^{12}\,\text{GeV}$.
\begin{figure}[htb!]
\centering
\includegraphics[scale=0.52]{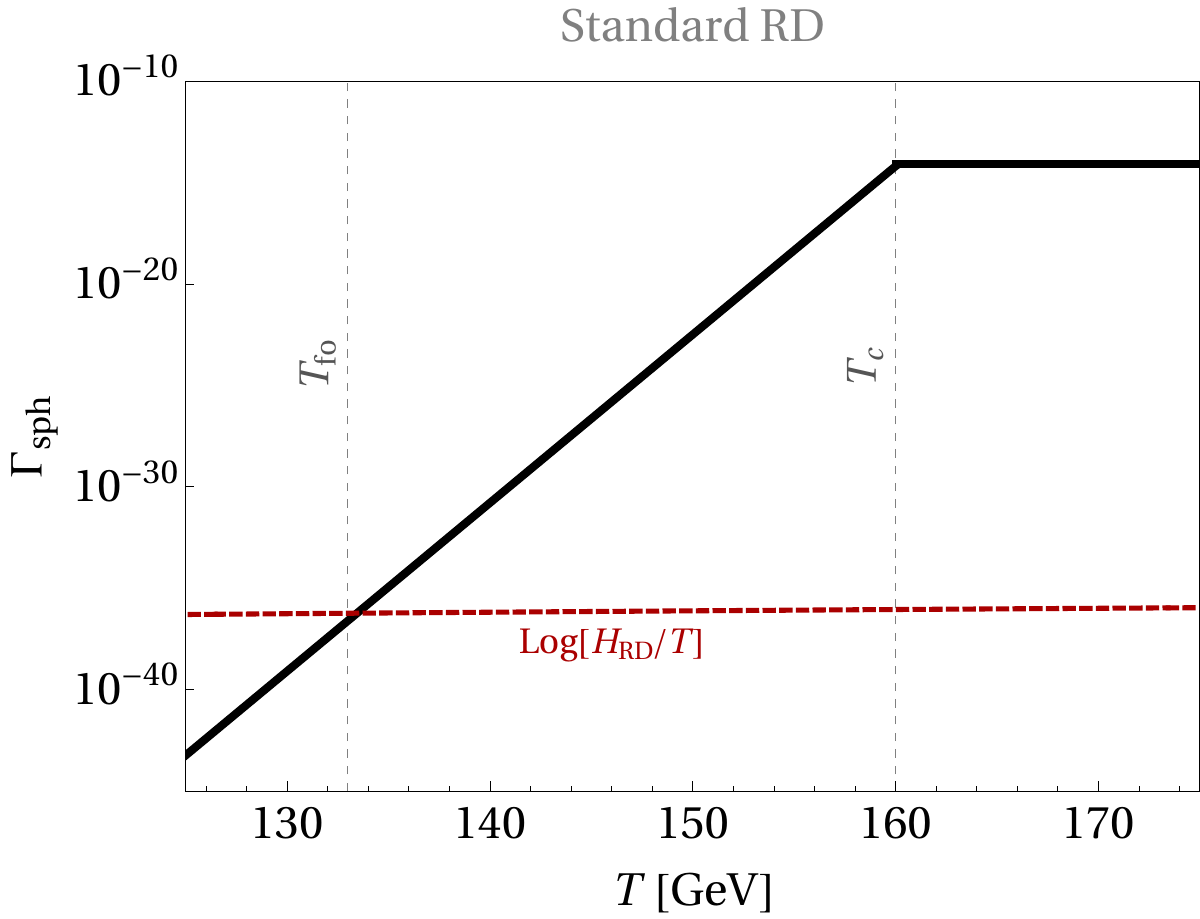}
\caption{Sphaleron rate $\Gamma_{\rm aph}$ shown in thick black (Eq.~\eqref{eq:sph-rate-den}), as a function of temperature for a standard radiation dominated Universe. The horizontal red dashed line corresponds to the Hubble rate during standard radiation domination ($H_{\rm RD}$), scaled with the temperature. The vertical dashed lines correspond to the cross-over temperature $T_c\simeq 160$ GeV and sphaleron freeze-out temperature $\Tfo\simeq 133$ GeV.}
\label{fig:sph-rd}
\end{figure}
\begin{figure}[htb!]
\centering
\includegraphics[scale=0.45]{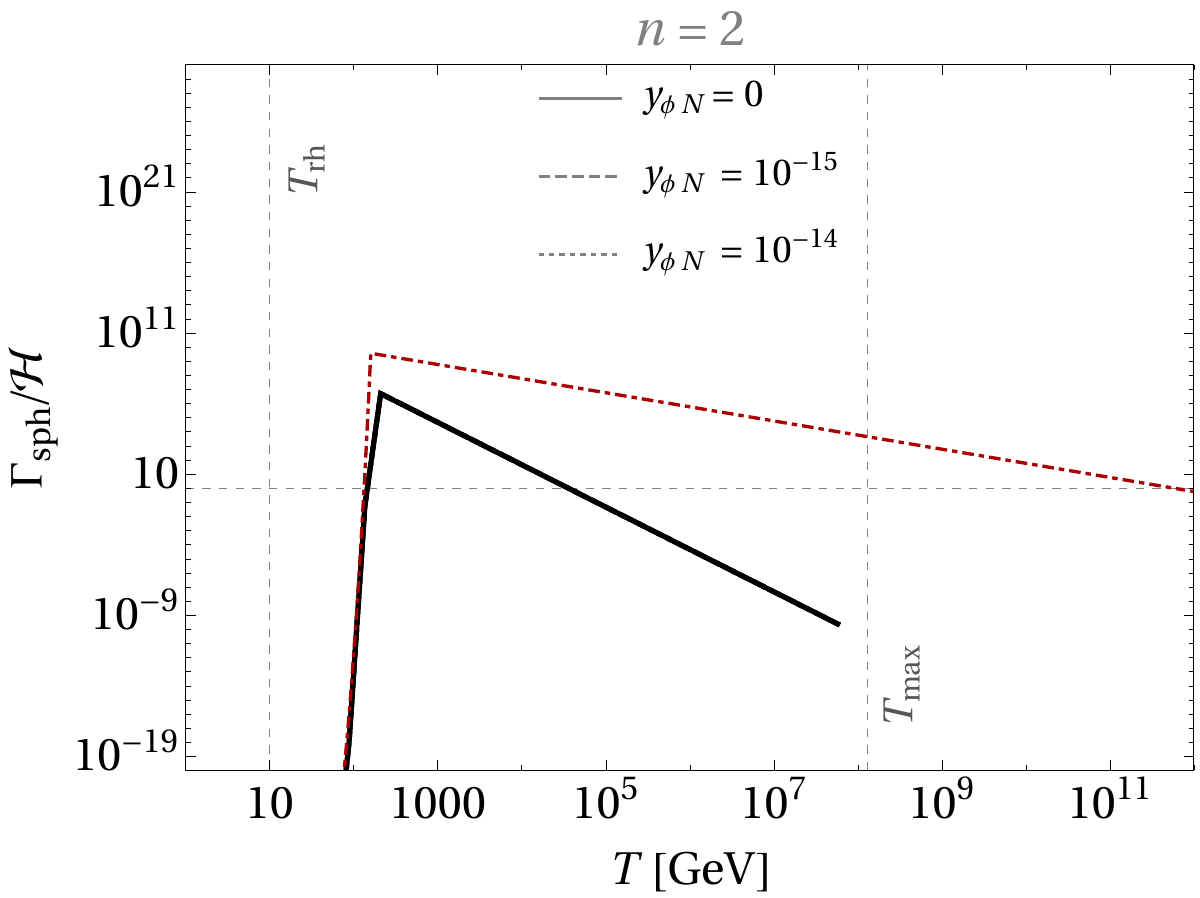}\\[10pt]
\includegraphics[scale=0.375]{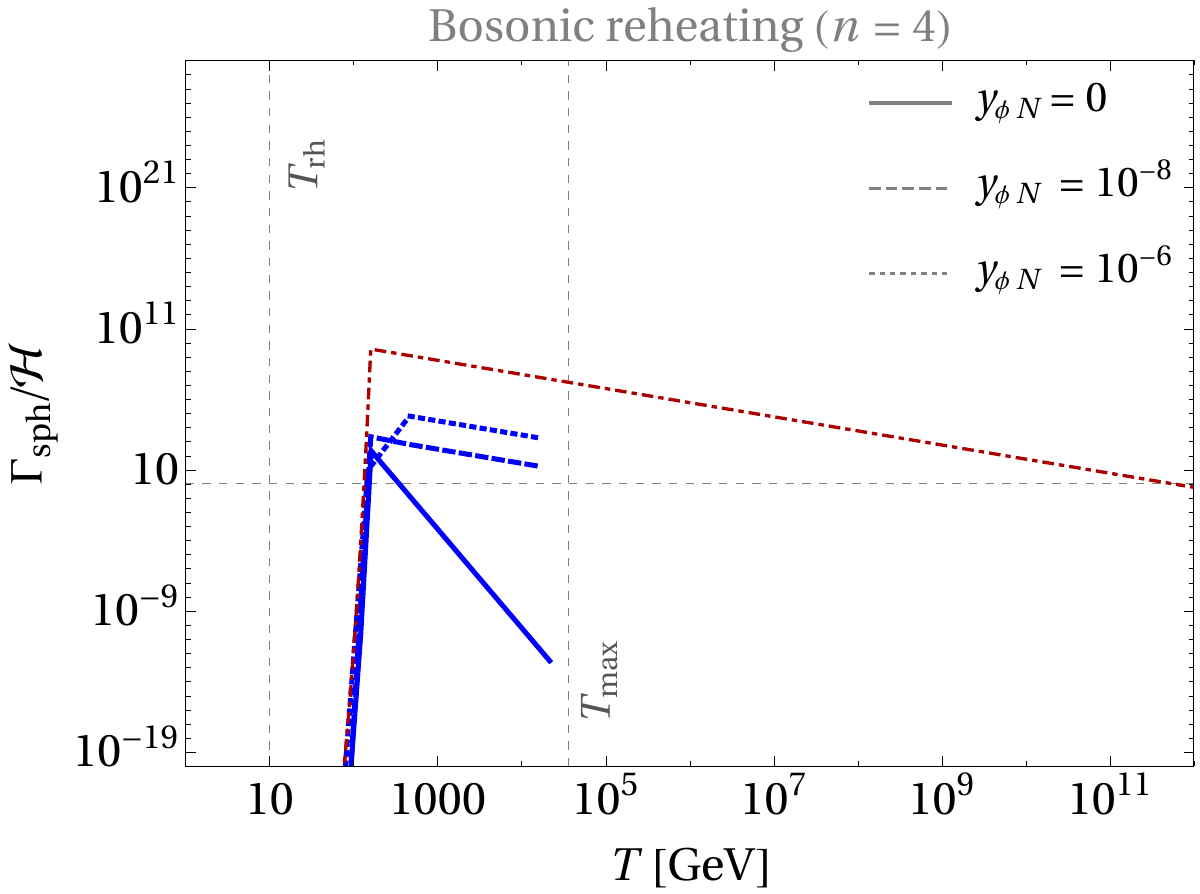}~\includegraphics[scale=0.375]{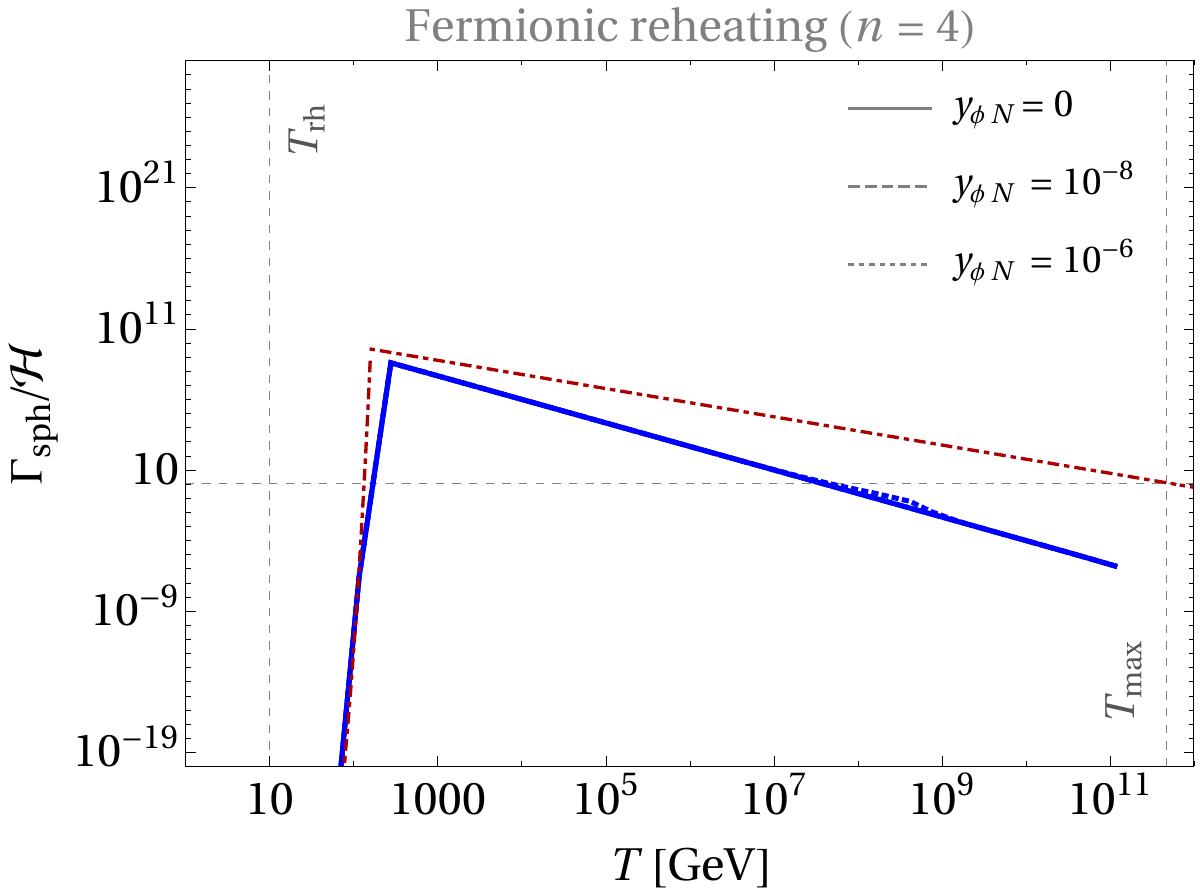}\\[10pt]
\includegraphics[scale=0.375]{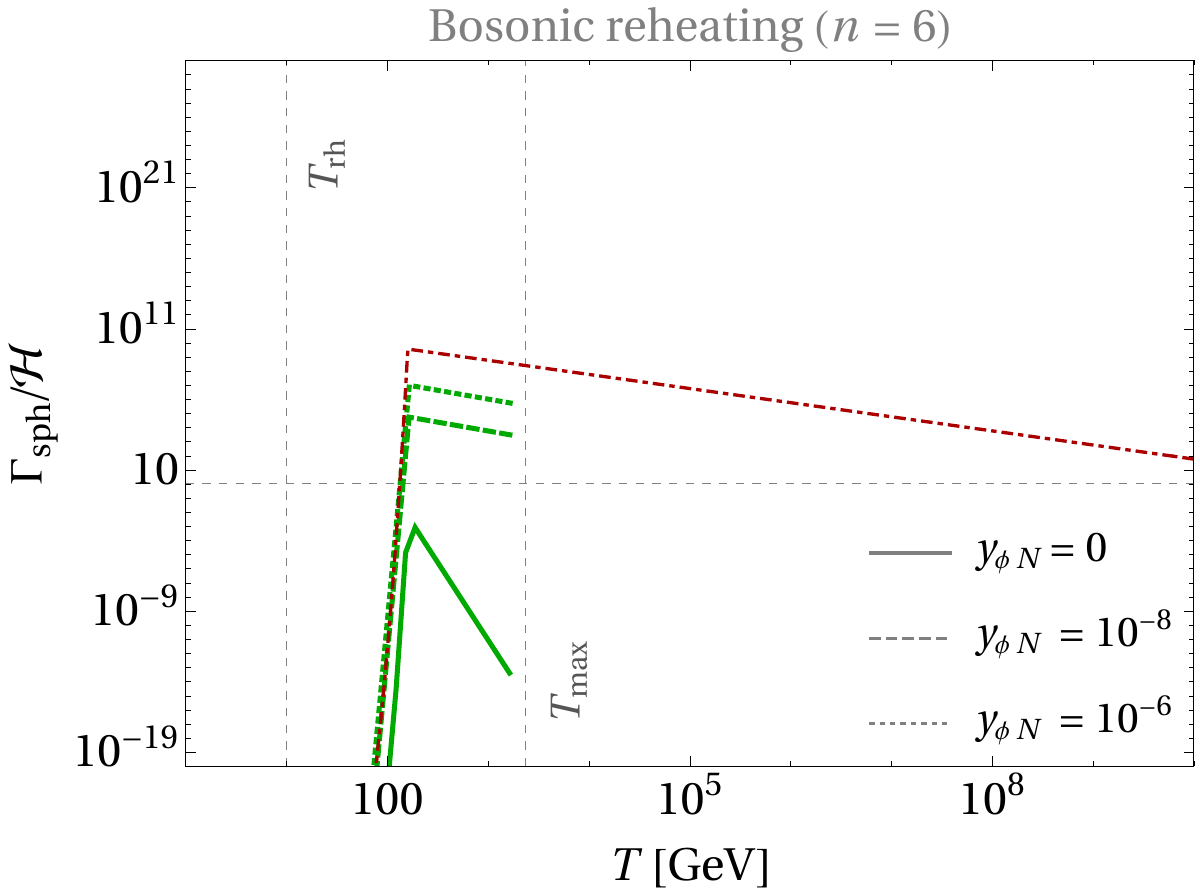}~\includegraphics[scale=0.375]{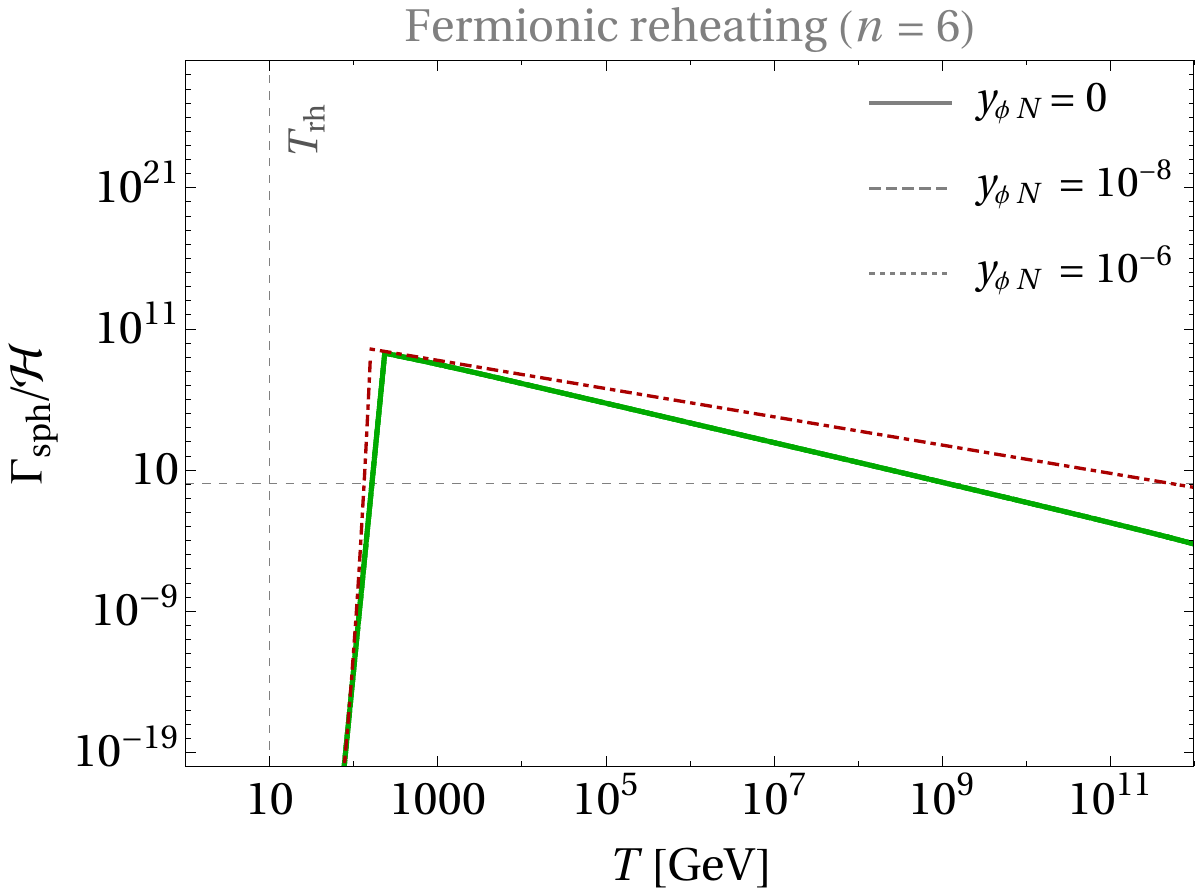}
\caption{{\it Scenario-A:} Sphaleron rate as a function of temperature for bosonic (left) fermionic (right) reheating scenarios, for different choices of $n$. The top panel shows $n=2$ case. In all cases $\Trh=10$ GeV, as shown by the vertical gray dashed line, along with corresponding $\Tmax$. The horizontal gray dashed line denotes $\Gamma_{\rm sph}/\mathcal{H}=1$. We also show the case of standard radiation domination in red dot-dashed curve.}
\label{fig:sph-rate}
\end{figure}

Following Ref.~\cite{DOnofrio:2014rug}, the sphaleron interaction rate in the minimal SM has been computed using lattice simulations of a three-dimensional effective theory. The resulting rate {\it density} $\mathcal{R}_{\rm sph}$ is given by,
\begin{align}\label{eq:sph-rate-den}
& \frac{\mathcal{R}_{\rm sph}}{T^4}\simeq
\begin{dcases}
18\,\alpha_2^5\,, & T>T_c
\\[10pt]
\exp\left(0.83\,T-147.7\right)\,, & T<T_c\,,
\end{dcases}
\end{align}
where $\alpha_2(T)\equiv g_2(T)^2/(4\pi)$, with $g_2$ being the SM gauge coupling\footnote{Considering only SM contribution, the 1-loop $\beta$-function for $g_2$ is given by, $(4\pi)^2\,\beta(g_2)=-(19/6))\,g_2^3$, where we consider $\mu=T$, $m_Z$ is the $Z$-boson pole mass in $\overline{\text{MS}}$ scheme and $g_2(m_Z)\simeq 0.653$. From the above solution we find that, $g_2$ changes by about 16\% over the temperature range $1\,\text{GeV}<T<10^{10}\,\text{GeV}$.}, and the cross-over temperature $T_c$ is taken to be 160 GeV, which is around the electroweak symmetry breaking scale. The sphaleron freeze-out temperature, on the other hand, is given by $\Tfo\sim 130$ GeV~\cite{DOnofrio:2014rug}, considering a standard radiation dominated Universe, as shown in Fig.~\ref{fig:sph-rd}. We define the sphaleron rate as,
\begin{align}
& \Gsp\equiv\frac{\mathcal{R}_{\rm sph}}{T^3}\,.     
\end{align}
For $n=2$, in case of both bosonic and fermionic reheating we find,
\begin{align}
& \frac{\Gsp}{\mathcal{H}}\simeq\sqrt{\frac{10}{\pi^2\,\gs}}\,\frac{M_P\,\Trh^2}{T^3}
\begin{dcases}
54\,\alpha_2^5\,, & T > T_c\,,
\\[10pt]
3\,\exp\left(-147.7+0.83\,T\right) & T < T_c\,.
\end{dcases}
\end{align}
For bosonic reheating scenario,
\begin{align}
&\frac{\Gsp}{\mathcal{H}}\simeq 54\,M_P\,\alpha_2^5\,\sqrt{\frac{10}{\pi^2\,\gs}}
\begin{dcases}
\Trh^6/T^7\,, & n=4\,,
\\[10pt]
\Trh^{10}/T^{11}\,, & n=6\,,
\end{dcases}
\end{align}
considering $T>T_c$. For fermionic reheating, on the other hand,
\begin{align}
&\frac{\Gsp}{\mathcal{H}}\simeq 54\,M_P\,\alpha_2^5\,\sqrt{\frac{10}{\pi^2\,\gs}}
\begin{dcases}
\Trh^{2/3}\,T^{5/3}\,, & n=4\,,
\\[10pt]
\Trh^{2/5}\,T^{7/5}\,, & n=6\,,
\end{dcases}
\end{align}
while for standard RD ($\mathcal{H}\propto T^2$), this ratio scales as $\Gsp\propto 1/T$. A common feature in all cases, as evident from the analytical estimates, is that at fixed $\Trh$, the ratio $\Gsp/\mathcal H$ decreases with increasing temperature. Consequently, at sufficiently high temperatures the sphaleron processes go out of equilibrium. 

In the standard radiation-dominated Universe, sphalerons remain in equilibrium over the temperature range $\Tfo<T\lesssim 10^{12}\,\text{GeV}$. During reheating as the Hubble rate increases with respect to standard RD scenario (at a fixed $T$), hence the sphalerons fall out of equilibrium earlier. For $n=2$, sphalerons remain in equilibrium for $\Tfo<T\lesssim\Tsph= 10^5\,\text{GeV}$, where, for a generic $n$, 
\begin{align}
&\Tsph\simeq \left(\frac{\mathcal{H}(\arh)}{18\,\alpha_2^5\,\Trh}\right)^\frac{n+2}{\alpha\,(n+2)-3n}\,, & \text{for}~~T>T_c\,,    
\end{align}
corresponds to the temperature at which sphalerons enter equilibrium during reheating. For a fixed $\Trh$, the Hubble rate for the same temperature increases with $n$. As a result, we see, for larger $n$, the sphalerons fall out of equilibrium for $\yphn=0$. In case of bosonic reheating, for $\yphn\neq0$, the RHN energy density briefly dominates over radiation (middle and bottom panels of Fig.\ref{fig:rho-bos}). During this epoch, the thermal bath is replenished gradually from RHN decays. This additional entropy injection slows down the cooling of the plasma, making the effective temperature scaling shallower than the standard inflaton-dominated reheating behavior for a finite interval. As a result, $\Gsp/\mathcal H>1$ holds for longer, allowing sphalerons to stay in equilibrium over a broader temperature interval, compared to $\yphn=0$ case. For fermionic reheating, on the other hand, the RHN energy density never dominates over radiation [cf. Fig.~\ref{fig:rho-fer}]. Moreover, the inflaton energy density redshifts approximately like radiation itself, as $T\propto a^{-3/4(-15/16)}$ for $n=4(6)$. Consequently, the thermal history remains close to the $\yphn=0$ case, and hence no significant change in the sphaleron equilibrium temperature is observed upon varying $\yphn$. It is worth noting that the abrupt ending of curves, corresponding to different $n$ and $\yphn$, is due to the fact that the temperature can not go beyond the maximum temperature [cf. Eq.~\eqref{eq:Tmax1}].

We note, even at lower reheating temperature, well below $T_c$, the sphaleron freeze-out temperature does not deviate significantly from the standard RD result, for both {\it Scenario-A} and {\it Scenario-B}. This is attributed to the fact that for $T<T_c$, the sphaleron rate anyway becomes exponentially suppressed (second line of Eq.~\eqref{eq:sph-rate-den}). Consequently, the rapid fall of $\Gamma_{\rm sph}$ dominates over the comparatively mild modification in the Hubble expansion rate during reheating, causing sphaleron decoupling to occur near the electroweak freeze-out temperature $\Tfo$ for a standard radiation dominated Universe. Therefore, the impact of the non-standard cosmological background is primarily visible at $T\gg T_c$, while the low-temperature freeze-out behavior remains largely unchanged.
\section{Leptogenesis with sub-EW-scale reheating temperature}
\label{sec:lepto}
\subsection{{\it Scenario-A:} direct reheating}
We first take up the scenario where reheating happens from direct perturbative decay of the inflaton condensate either into bosonic or into fermionic final states. We assume $2\,M_1<m_\phi(a)\ll2\,M_{2,\,3}$, such that only the lightest RHN is produced from the inflaton decay. In order to track the evolution of individual components, we solve the following set of coupled BEQs involving inflaton energy density $\rp$, radiation energy density $\rho_R$, RHN number density $n_{N_1}$, $B-L$ number density $n_{B-L}$ generated from the RHN decay~\cite{Barman:2024ujh},
\begin{align}\label{eq:BEQ-phiNN}
&\Dot{n}_{N_1}+ 3\,\mathcal{H}\,n_{N_1}= - \langle \Gamma_N\rangle\,\left(n_{N_1}-n_{N_1}^{\rm eq}\right)+\frac{\Gamma_{\phi\to N_1N_1}}{m_\phi}\,(1+w)\,\rho_\phi\,,
\nonumber\\&
\Dot{n}_{B-L}+ 3\,\mathcal{H}\,n_{B-L}=-\langle \Gamma_N\rangle\,\left[\epsilon_{\Delta L}\,(n_{N_1}-n_{N_1}^{\rm eq})+\frac{n_{N_1}^{\rm eq}}{2\,n_\ell^{\rm eq}}\,n_{B-L}\right]\,,
\nonumber\\&
3\,\mathcal{H}^2\,M_P^2= \rho_\phi+\rho_R+ n_{N_1}\,E_N\,,
\nonumber\\&
E_N^2=M_N^2 + p_N^2\,,
\end{align}
along with Eq.~\eqref{eq:drhodt} and \eqref{eq:rR}. { Here, the RHN 3-momenta $p_N$ is given by~\cite{Bandyopadhyay:2026gci}, \begin{align}
& p_N=
\begin{dcases}
m_\phi(a)/2\,, & a<a_D\,,
\\[10pt]
\left(m_\phi(a_D)/2\right)\,(a_D/a)\,, & a>a_D\,,
\end{dcases}
\end{align}
where $a_D$ corresponds to the scale factor at which $\Gamma_{\phi\to N_1\,N_1}\simeq\mathcal{H}$. For $a<a_D$, the RHNs remain relativistic. For $a>a_D$, we approximate their production as occurring around $a_D$, after which their momentum redshifts as $p_N(a)\propto a^{-1}$. We have verified that this approximation agrees reasonably well with the numerical results of~\cite{Bernal:2026lpr}.} The thermally averaged $N_1\to\ell\,H$ decay width is defined as,
\begin{align}
\langle \Gamma_N\rangle= \frac{K_1(M_1/T)}{K_2(M_1/T)}\frac{M_1}{8\pi} \left(y_\nu^\dagger y_\nu\right)_{11}\,,  
\end{align}
with $K_i$'s being modified Bessel functions of $i$-th kind. The equilibrium number density of any species $j$ is given by
\begin{align}
    n_j^{\rm eq}= \frac{g_j\,T^3}{2 \pi^2} \left(\frac{M_j}{T}\right)^2 K_2(M_j/T),
\end{align}
with $g_j$ representing the degrees of freedom of the $j$-species. Depending on the background equation of state, sphaleron interactions are in equilibrium as shown in Fig.~\ref{fig:sph-rate}. In that case they convert a fraction of a non-zero $B-L$ asymmetry into a baryon asymmetry via
\begin{align}\label{eq:YB0}
Y_B\simeq (28/79)\times Y_{B-L}=\frac{8\,N_F+4\,N_H}{22\,N_F+13\,N_H}\,Y_{B-L}\,,    
\end{align}
where $N_F$ is the number of fermion generations and $N_H$ is the number of Higgs doublets, which in our case: $N_F = 3,\,N_H = 1$. In leptogenesis, where purely a lepton asymmetry is generated, $B-L=-L$, is then converted into the baryon asymmetry via sphaleron transition~\cite{Buchmuller:2004nz}. Finally, the observed baryon asymmetry of the Universe is given by $Y_B^0 \simeq 8.75\times 10^{-11}$. 
\begin{figure}[htb!]
\centering
\includegraphics[scale=0.375]{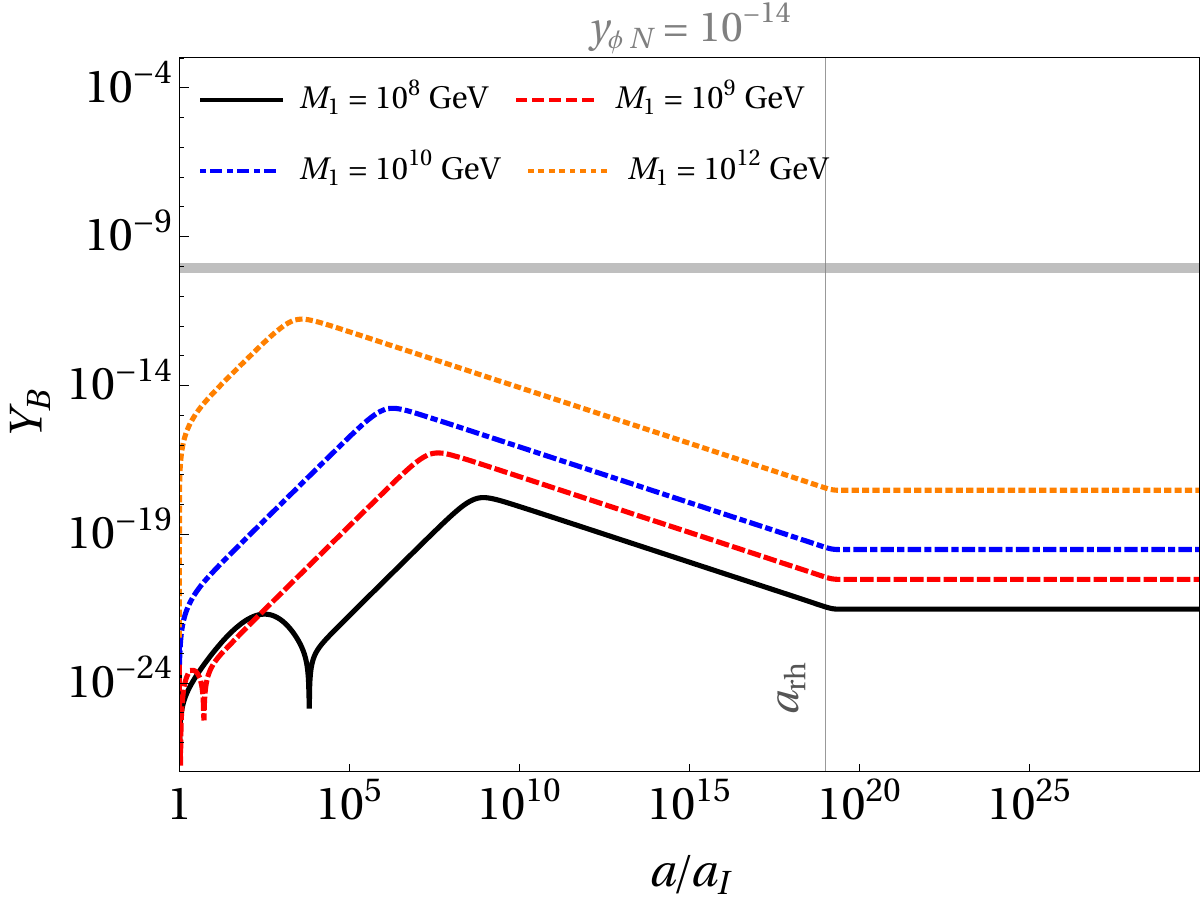}~\includegraphics[scale=0.375]{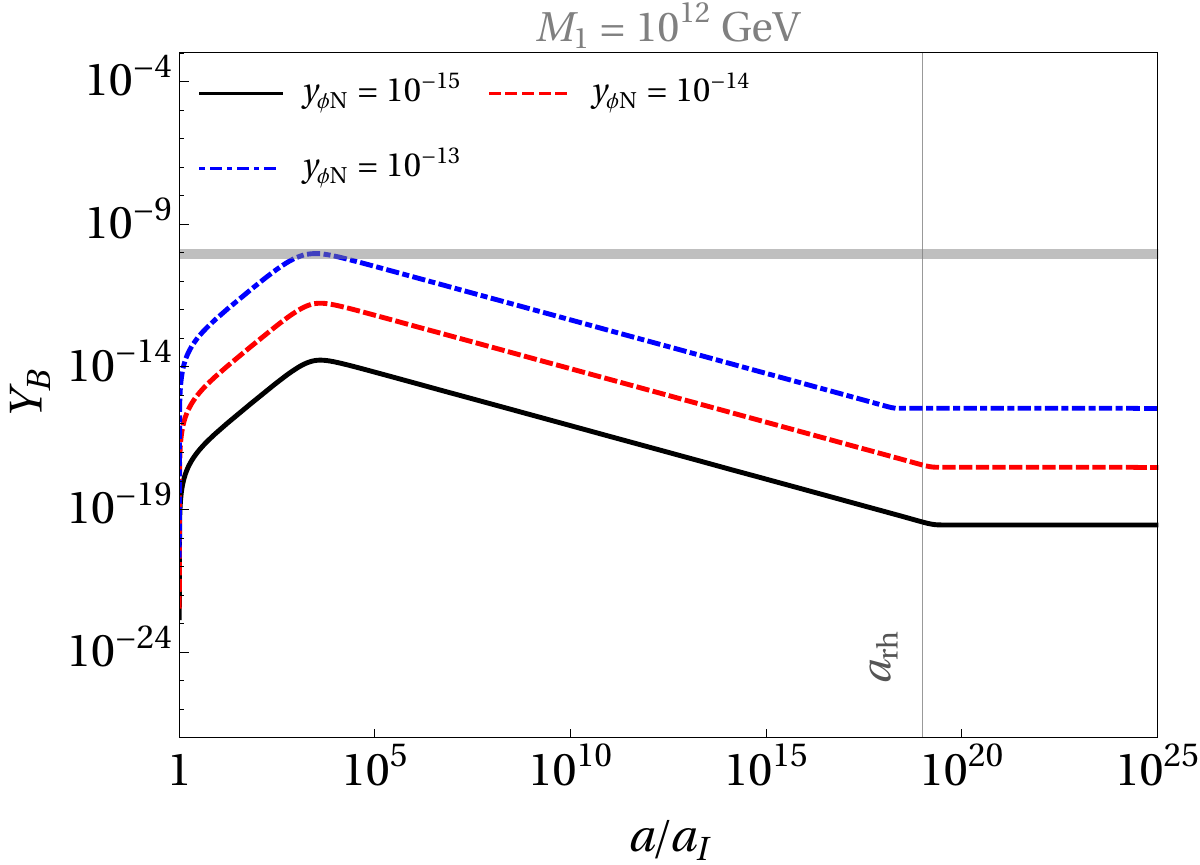}
\caption{{\it Scenario-A}: Evolution of baryon asymmetry as a function of the scale factor for $n=2$, for a fixed $\yphn=10^{-14}$ (left panel) and for a fixed RHN mass of $M_1=10^{12}$ GeV (right panel). Here we have fixed $\Trh=10$ GeV. The horizontal gray line corresponds to the observed baryon asymmetry.}
\label{fig:bosYBn2}
\end{figure}
\begin{figure}[htb!]
\centering
\includegraphics[scale=0.375]{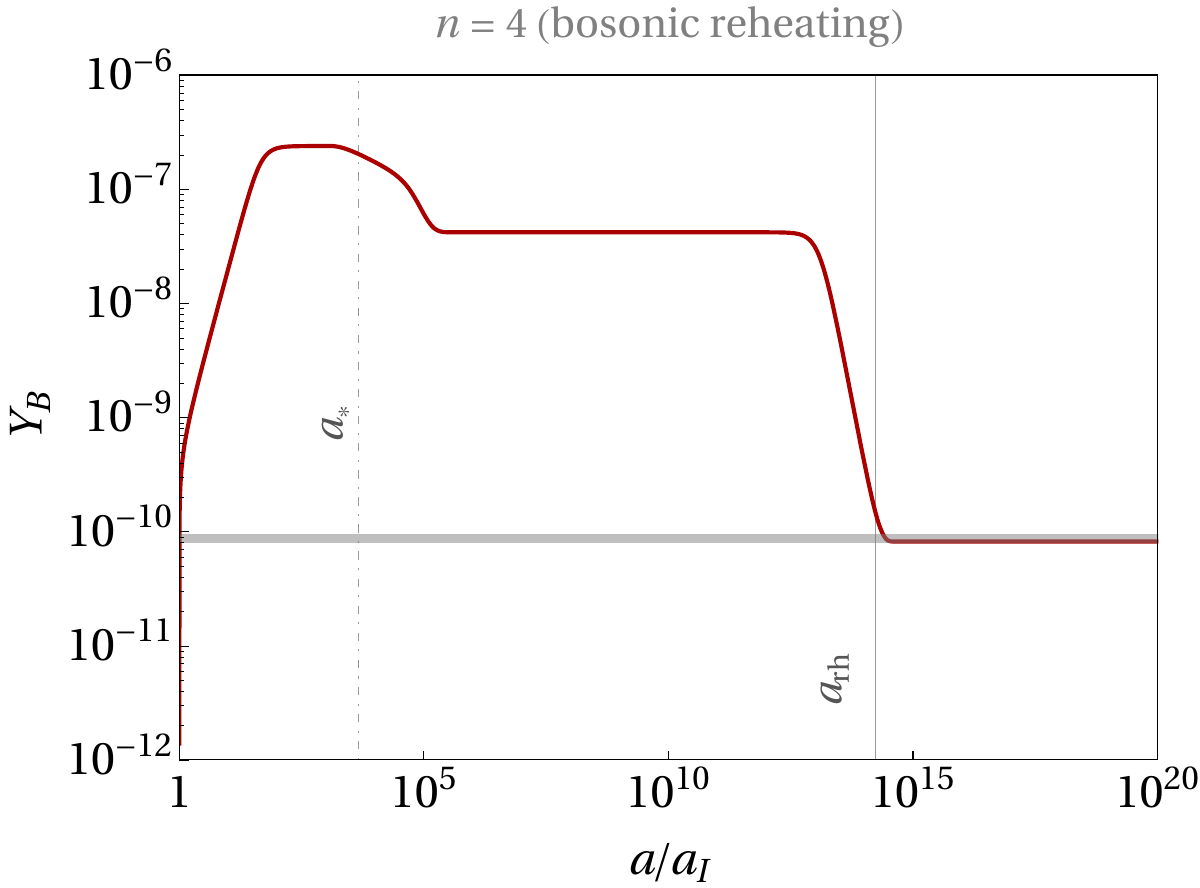}~\includegraphics[scale=0.375]{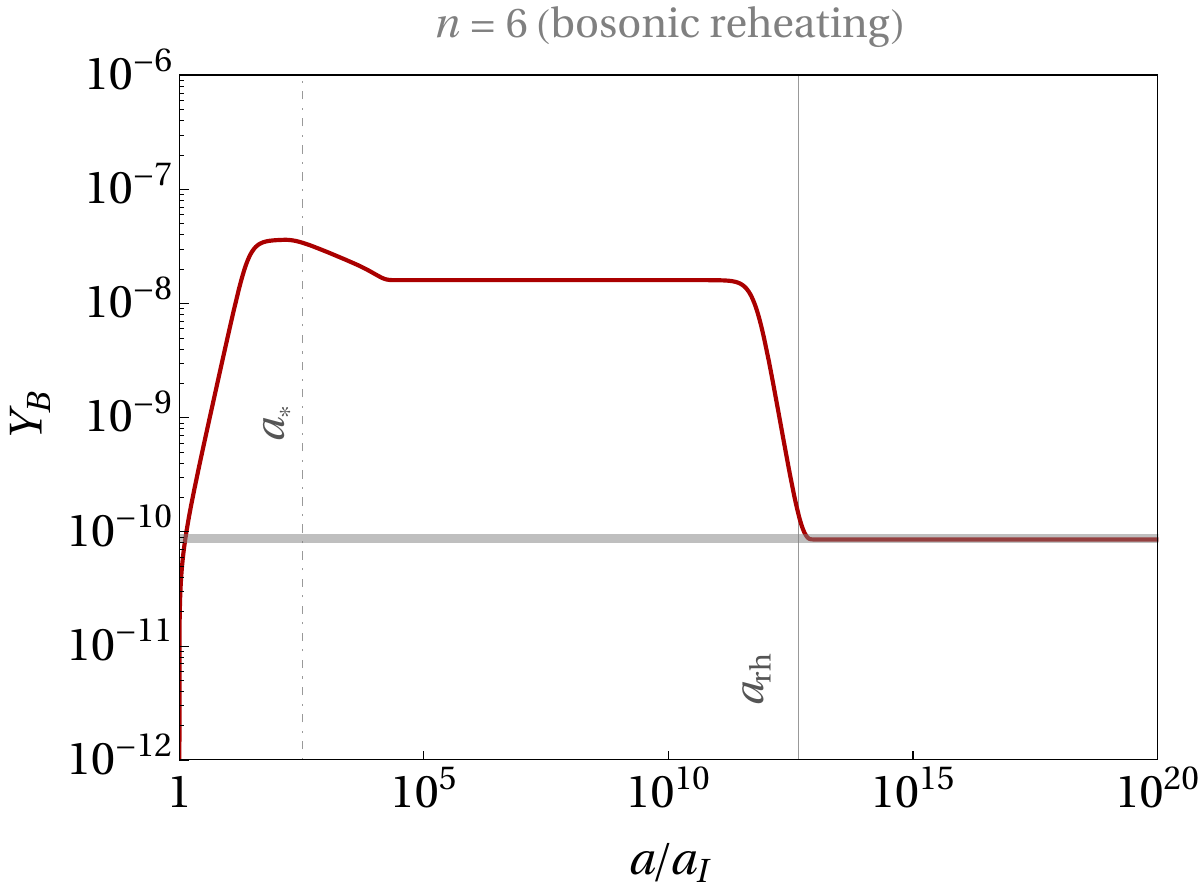}
\\[10pt]
\includegraphics[scale=0.375]{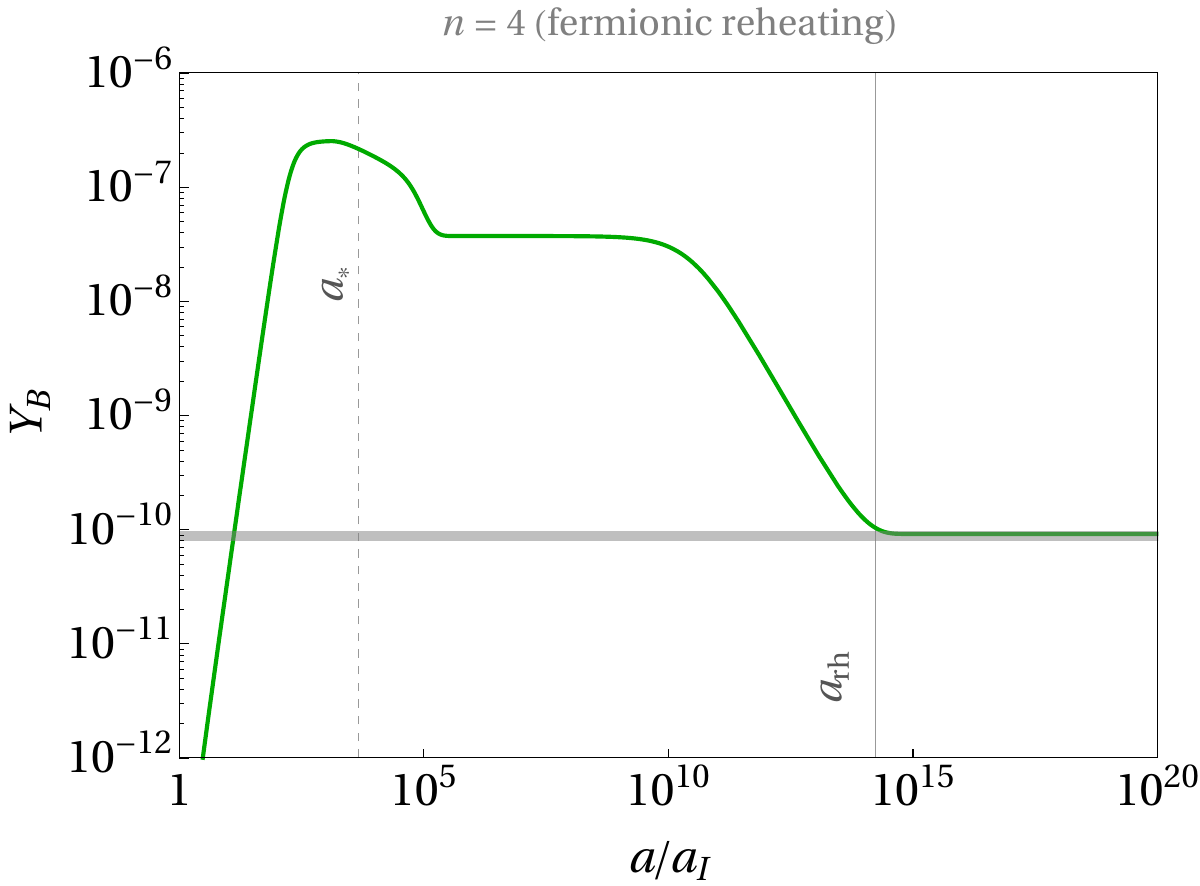}~\includegraphics[scale=0.375]{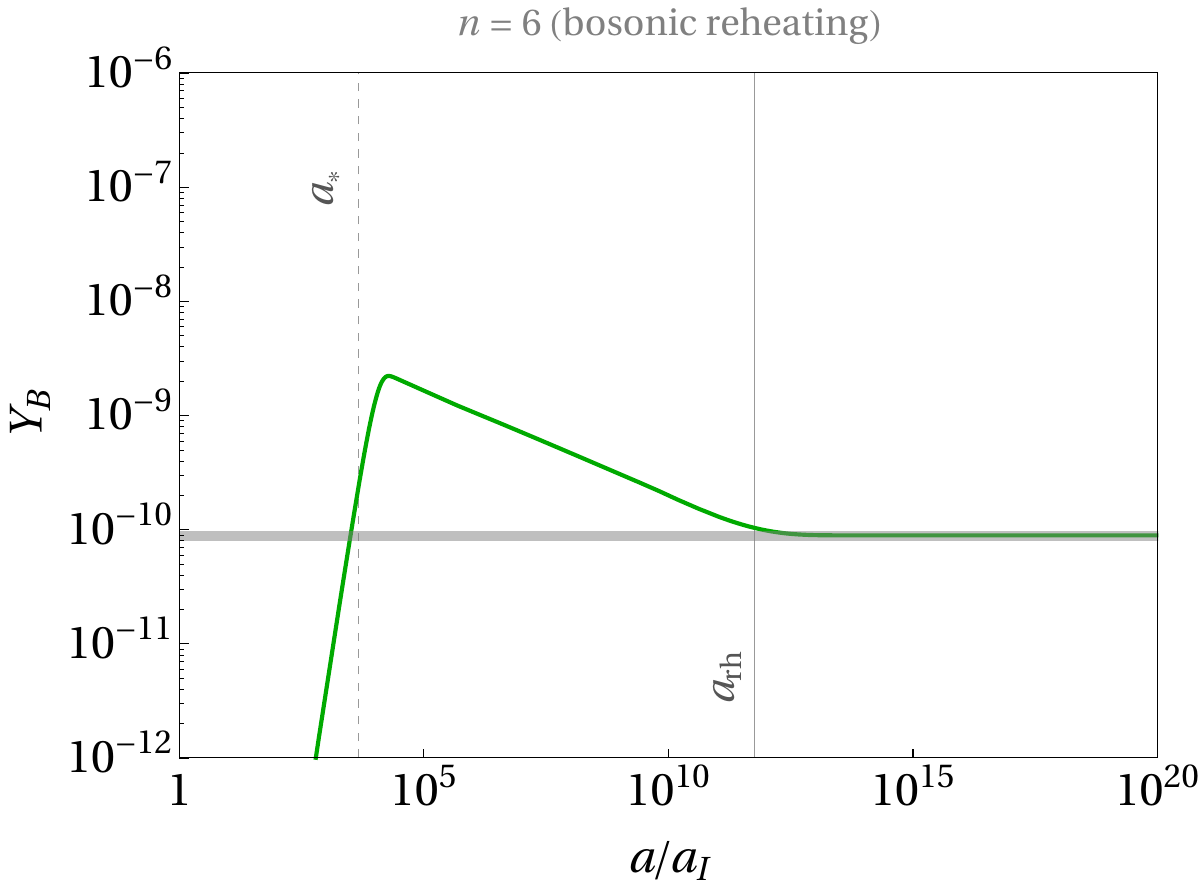}
\caption{{\it Scenario-A}: Evolution of baryon asymmetry as a function of the scale factor for bosonic (top panel)  fermionic reheating (bottom panel) cases. Here we have fixed $\Trh=10$ GeV and $M_1=10^{10}$ GeV. The horizontal gray line corresponds to the observed baryon asymmetry. The coupling $\yphn$ has been adjusted accordingly to satisfy the final baryon asymmetry (see text for details).}
\label{fig:bosYB}
\end{figure}
\begin{figure}[htb!]
\centering
\includegraphics[scale=0.375]{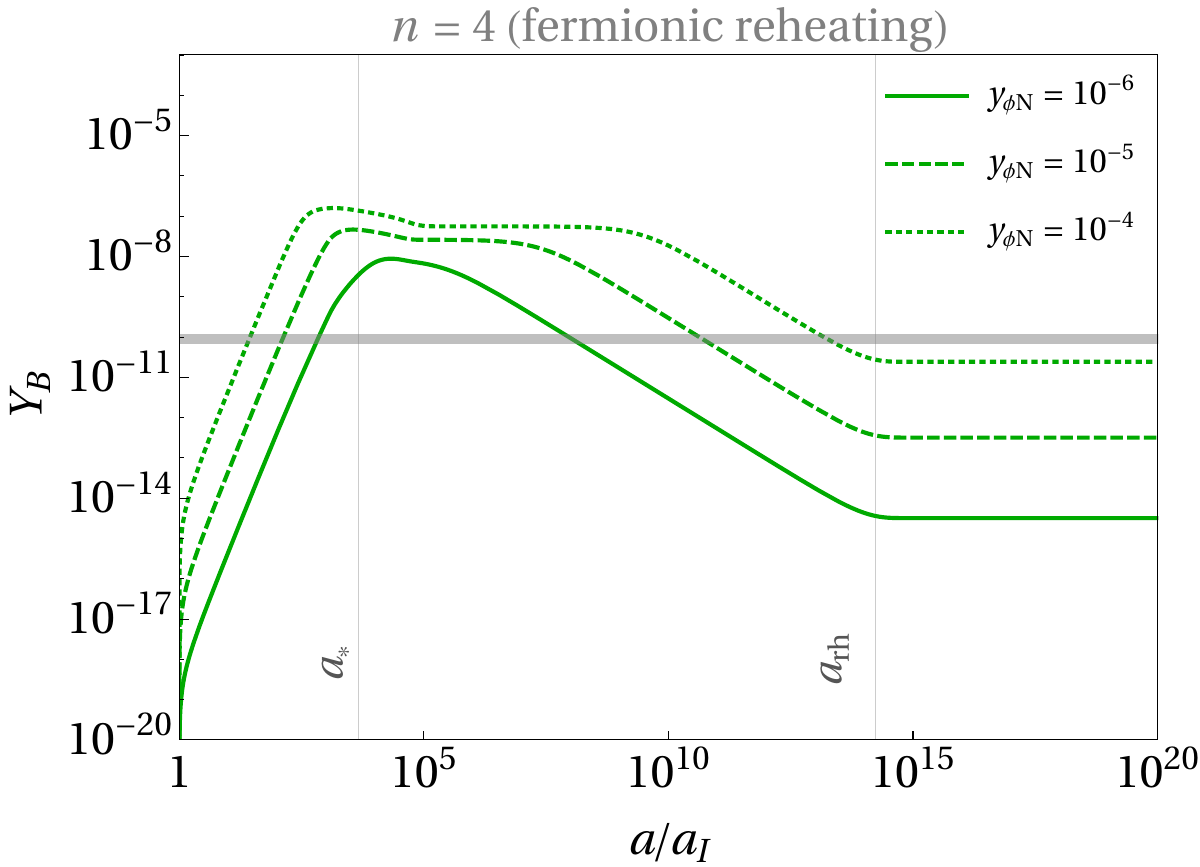}~\includegraphics[scale=0.375]{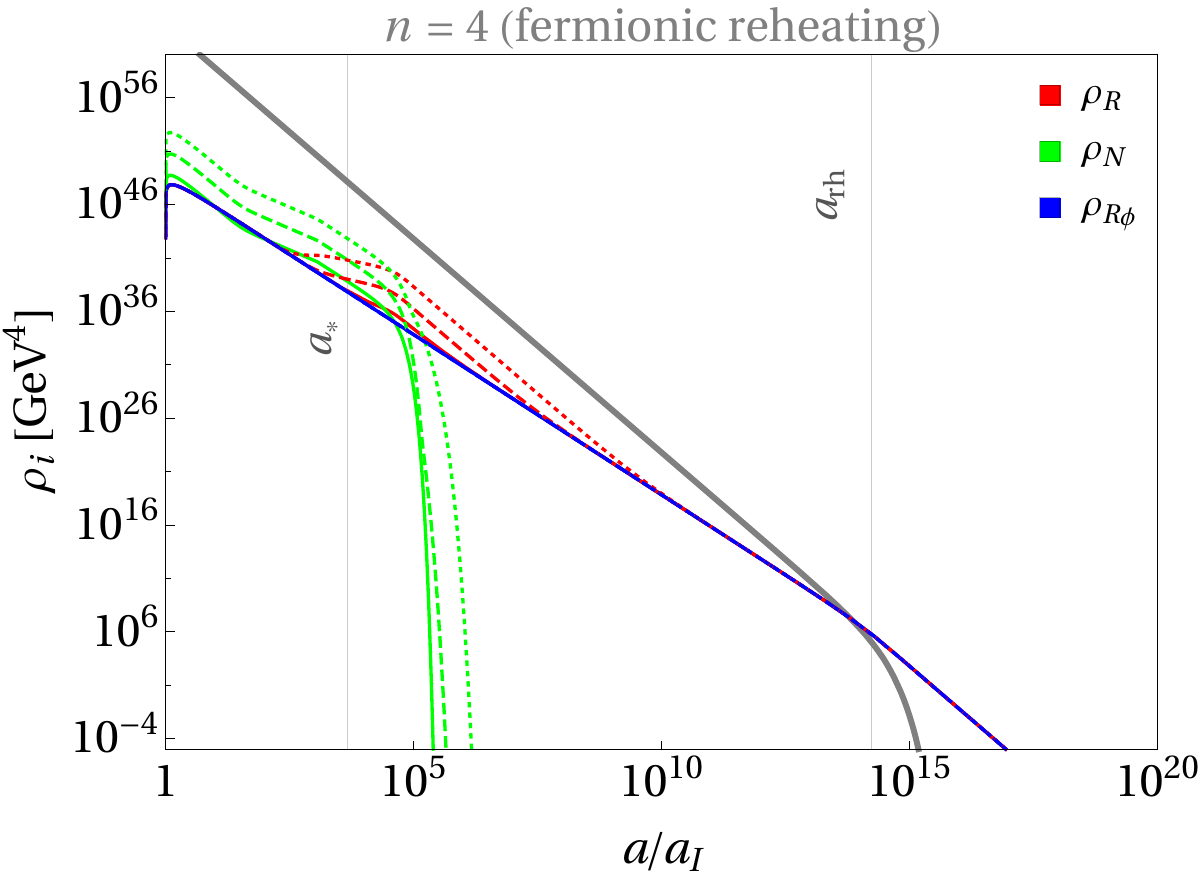}\\[10pt]
\includegraphics[scale=0.375]{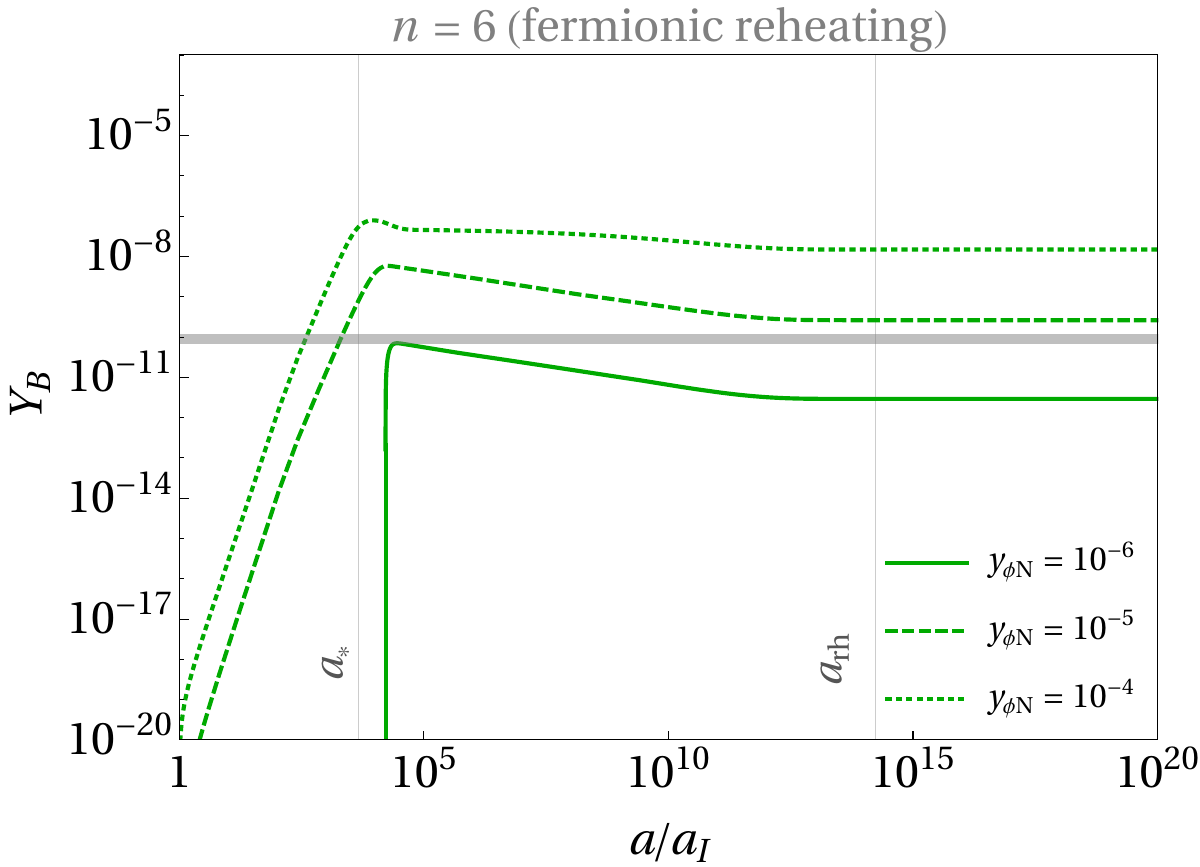}~\includegraphics[scale=0.375]{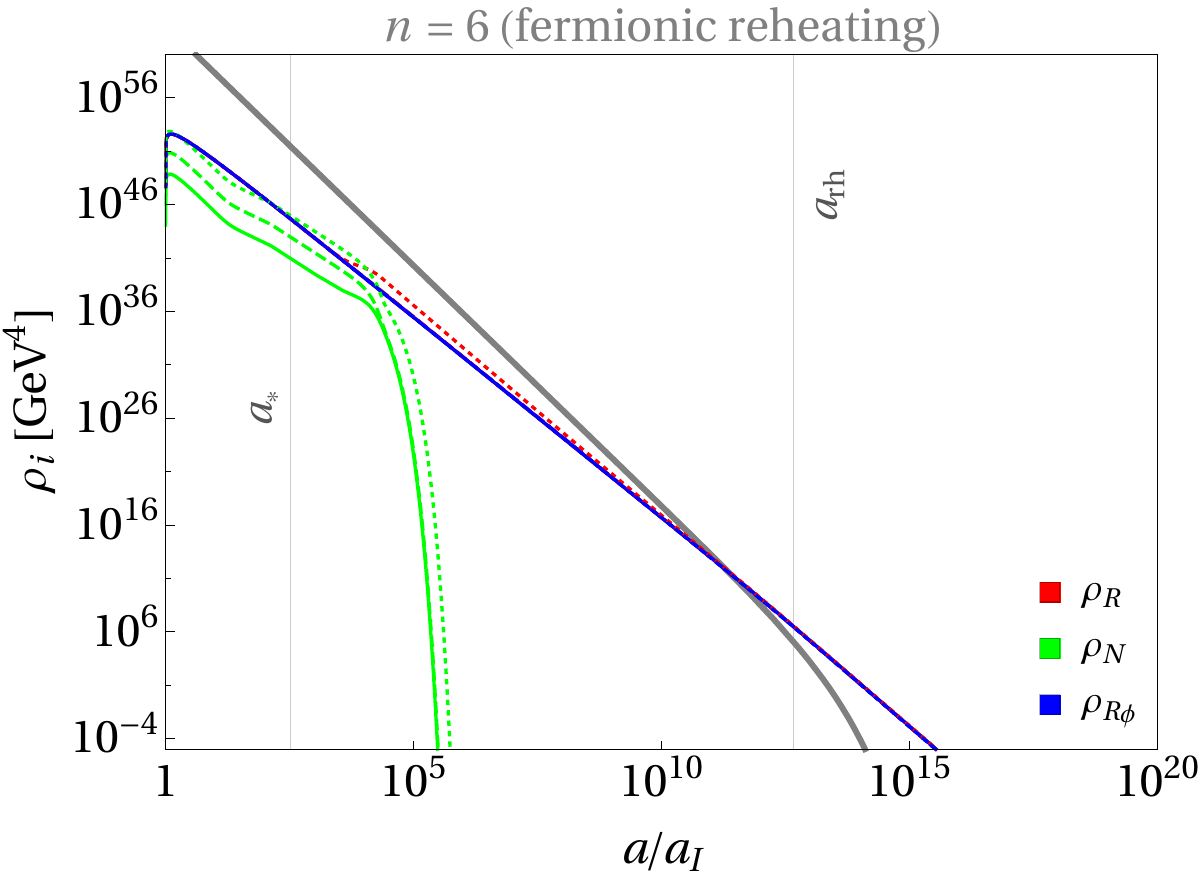}
\caption{{\it Scenario-A}: Evolution of $Y_B$, as a function of the scale factor, corresponding to $n=4$ (top left panel) and $n=6$ (bottom left panel), for different choices of $\yphn$. In the right panel we show corresponding evolution of energy densities for different components, for the same choice of couplings. In all cases we have fixed $M_1=10^{10}$ GeV and $\Trh=10$ GeV. We have considered fermionic reheating in all cases for illustration.}
\label{fig:compare}
\end{figure}
\begin{figure}[htb!]
\centering
\includegraphics[scale=0.375]{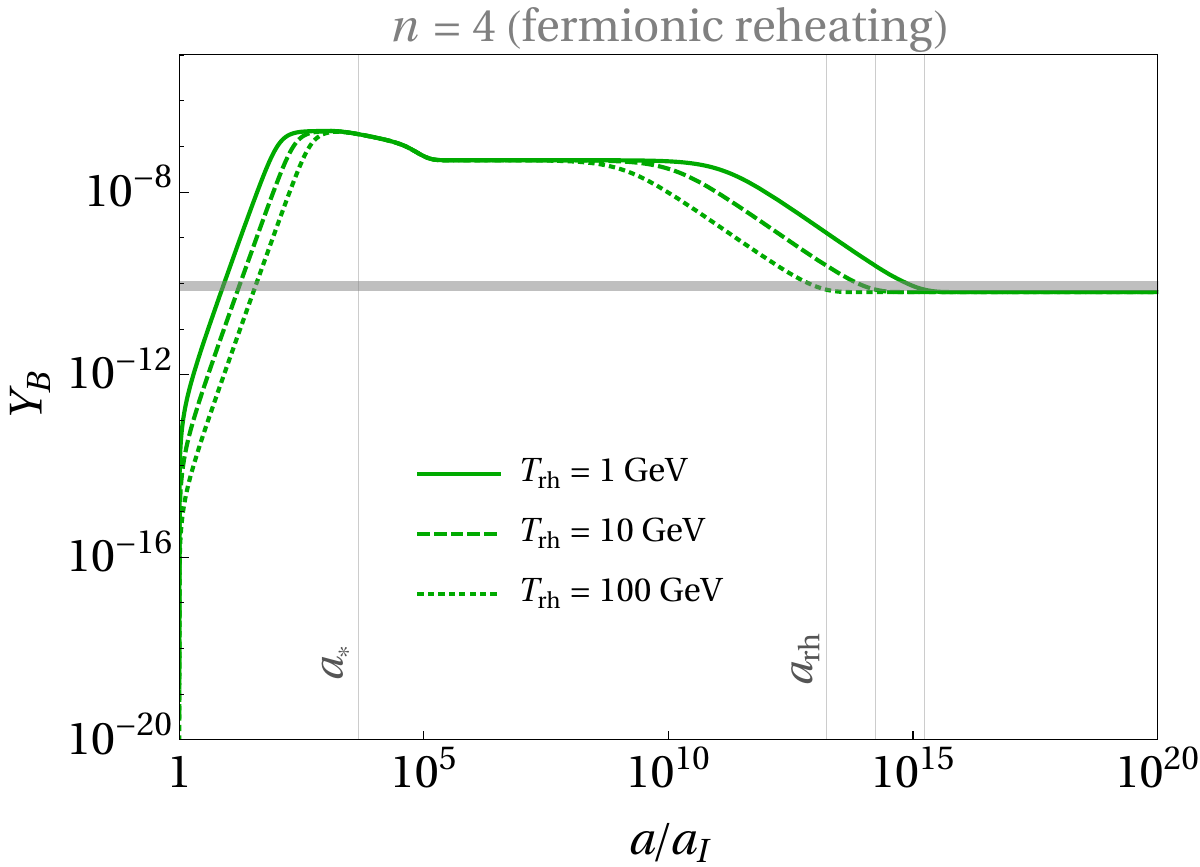}~\includegraphics[scale=0.375]{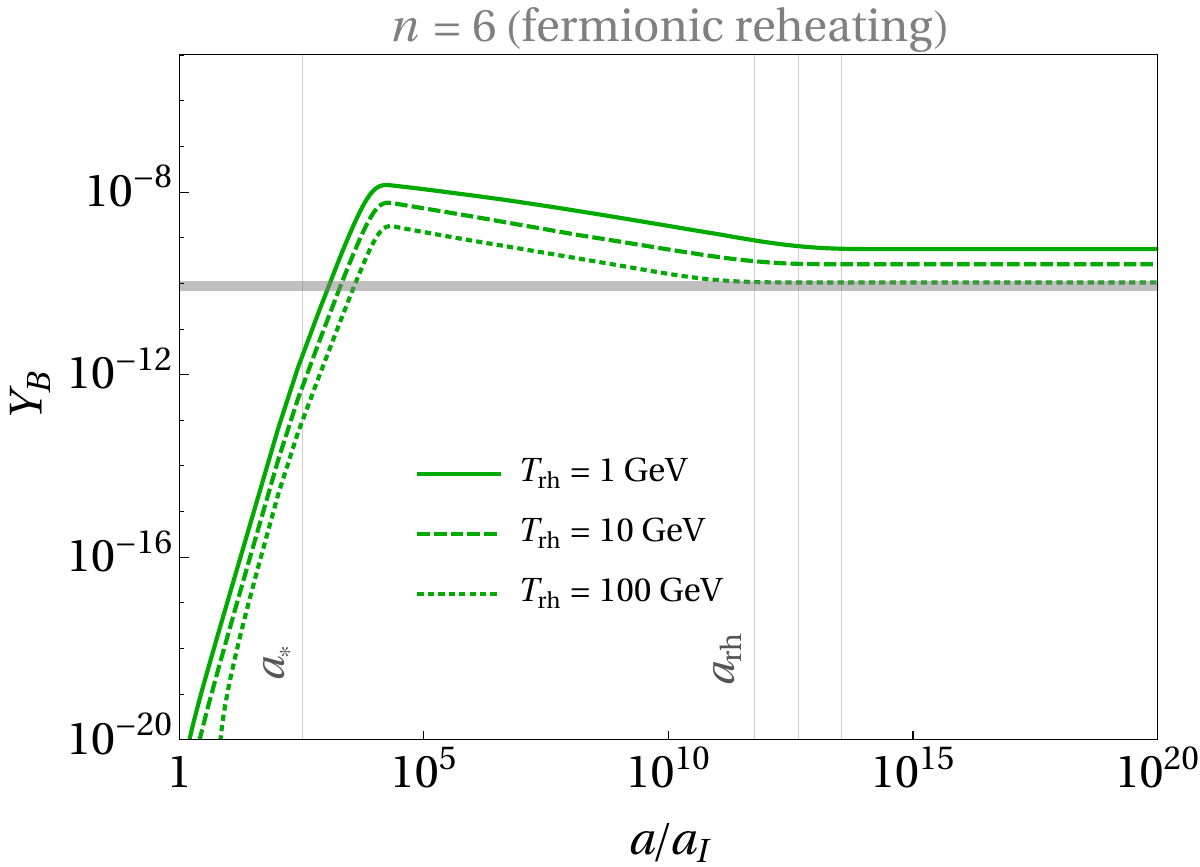}\
\caption{{\it Scenario-A}: Evolution of $Y_B$, as a function of the scale factor, corresponding to $n=4$ (left panel) and $n=6$ (right panel), for different choices of $\Trh$, and $\yphn=\{2\times 10^{-4},\,10^{-5}\}$. In all cases we have fixed $M_1=10^{10}$ GeV. We have considered fermionic reheating in all cases for illustration.}
\label{fig:compare2}
\end{figure}

If RHNs are produced solely through perturbative inflaton decay during reheating, an approximate analytical expression for their number density can be derived. Below we consider two cases, corresponding to quadratic potential ($n=2$), and potentials steeper than quadratic.    
\subsubsection{Case of quadratic potential ($n=2$)}
We first consider the case of a quadratic inflaton potential ($n=2$). In this case, RHN production from inflaton decay remains kinematically allowed as long as $m_\phi(a)>2\,M_1$. The lepton asymmetry generated from RHN decay is converted into baryon asymmetry while electroweak sphalerons remain in thermal equilibrium, i.e., until the sphaleron freeze-out temperature $T=\Tfo$. If reheating is still ongoing beyond this point, the continued production of radiation increases the entropy, thereby diluting the already-generated baryon asymmetry. The RHN number density at sphaleron freeze-out is approximately
\begin{align}\label{eq:nNrh} 
n_{N_1}(\afo)\equiv\frac{N_{N_1}(\afo)}{\afo^3}\simeq\frac{n\,\yphn^2}{8\,\pi}\,M_P^2\,\mathcal{H}_I\,\left(\frac{\Tfo}{\Tmax}\right)^4\,\left[1-\left(\frac{\Tfo}{\Tmax}\right)^4\right] 
\end{align}
where $N_{N_1}=n_{N_1}a^3$ denotes the comoving RHN number, and we have assumed $N_{N_1}(a_I)=0$. Here, $\mathcal{H}_I\equiv \mathcal{H}(a_I)$ is the Hubble scale at the end of inflation (see Appendix~\ref{sec:inflation} for the CMB constraints on $\mathcal{H}_I$). The baryon asymmetry generated at $T=\Tfo$ is subsequently diluted if reheating ends at a lower temperature, $\Trh<\Tfo$. The final baryon asymmetry at the end of reheating is therefore
\begin{align} 
& Y_B\equiv\frac{n_B}{s}\Big|_{\Trh}\simeq\frac{28}{79}\,\epsilon_{\Delta L}\,\frac{n_{N_1}}{s}\Big|_{\Tfo}\times\frac{S(\Tfo)}{S(\Trh)}\simeq\frac{315\,\yphn^2\,\epsilon_{\Delta L}}{158\,\pi^3\,\gss(\Trh)}\,\frac{\mathcal{H}_I\,M_P^2}{\Tfo^3}\,\left(\frac{\Tfo}{\Tmax}\right)^4\,\left(\frac{\Trh}{\Tfo}\right)^5\,, 
\end{align} 
since $\Trh\ll\Tmax$. Using the above expression we obtain, 
\begin{align} 
& \Trh\simeq 7.5\times 10^4\,\text{GeV}\left(\frac{10^{-5}}{\epsilon_{\Delta L}}\right)^{1/3}\,\left(\frac{10^{-14}}{\yphn}\right)^{1/3}\,, 
\end{align}
requiring the observed baryon asymmetry. This result implies that, for a quadratic inflaton potential, the observed baryon asymmetry cannot be reproduced if $\Trh<\Tfo$, assuming reheating proceeds through inflaton decay into SM final states. The main obstacle is the large entropy dilution, $S(\Tfo)/S(\Trh)\sim
\left(\Trh/\Tfo\right)^5\simeq 2.7\times10^{-6}$, for $\Trh=10$ GeV. The dilution becomes even stronger for lower $\Trh$. As a result, although RHNs with masses as large as $10^{12}$ GeV can still be produced during reheating, the entropy generated after sphaleron freeze-out suppresses the final baryon asymmetry well below the observed value. This can be seen from Fig.~\ref{fig:bosYBn2}, where we numerically obtain the evolution of BAU as a function of the scale factor corresponding to $n=2$, for different choices of $\yphn$ and RHN masses. We see, in all cases, the final asymmetry is several orders of magnitude below the observed value for $\Trh<\Tfo$, due to huge entropy dilution. In all cases, we ensure that the reheating happens entirely through inflaton decay into the SM final states.
\subsubsection{Case of potential beyond quadratic ($n>2$)}
For $n>2$, the production ceases at $a=\ast<\afo$ [cf. Eq.~\eqref{eq:ast}], with corresponding number density,
\begin{align}
& n_{N_1}(\ast)=\frac{a_I^3\,\yphn^2}{24\pi}\,\frac{\rho_I}{\mathcal{H}_I}\,\left[\left(\frac{\ast}{a_I}\right)^\frac{6}{n+2}-1\right]\,,    
\end{align}
where $\ast$ is given by Eq.~\eqref{eq:ast}. Following Eq.~\eqref{eq:Tevol}, one can further write $\ast=a_I\,\left(\Tmax/\Tst\right)^{1/\alpha}$, leading to $T_\star=\Trh\,\left(\Tmax/\Trh\right)^{1/\alpha}\,\left(2\,m_I/M_1\right)^\frac{n+2}{3\,(2-n)}$, as the temperature of the thermal bath at $a=\ast$. Once the RHN production from inflaton decay is over, the RHN number density simply gets diluted as $a^{-3}$. Therefore, the final baryon asymmetry at the end of reheating becomes,
\begin{align}\label{eq:YBn46}
& Y_B\equiv\frac{n_B}{s}\Big|_{\Trh}\simeq\frac{28}{79}\,\epsilon_{\Delta L}\,\frac{n_{N_1}(\ast)}{s(\arh)}\,\left(\frac{\ast}{\arh}\right)^3=\frac{28}{79}\,\epsilon_{\Delta L}\,\frac{n_{N_1}(\ast)}{s(\arh)}\,\left(\frac{\ast}{a_I}\right)^3\,\left(\frac{\rp(\arh)}{\rho_I}\right)^{\frac{n+2}{2n}}
\nonumber\\&
=\frac{7\,\sqrt{135}}{79\pi\times (60/\pi^2)^{1/n}\,\gss(\Trh)}\,\frac{n_{N_1}(\ast)}{\Trh^3}\,\epsilon_{\Delta L}\,\left(\frac{\ast}{a_I}\right)^3\,\left(\frac{\gs(\Trh)\,\Trh^4}{\rho_I}\right)^\frac{n+2}{2n}
\nonumber\\&\propto
\begin{dcases}
\Trh^0\,, & \text{for}~n=4\,,
\\[10pt]
\Trh^{-1/3}\,, & \text{for}~n=6\,,
\end{dcases}
\end{align}
which shows $\Trh$-independence for $n=4$, while for $n=6$, it is a decreasing with $\Trh$. Note that, $S(\ast)/S(\arh)=S(\ast)/S(\afo)\times S(\afo)/S(\arh)$, implying the entropy dilution from $\afo$ to $\arh$ is already contained within the total dilution factor. Consequently, unlike the $n=2$ case, where RHN production can happen all throughout the history of reheating, even after $\afo$, there is no need to apply a second dilution factor for $n>2$, starting at sphaleron freeze-out. It is important to note that the result obtained in Eq.~\eqref{eq:YBn46} is independent of the underlying reheating mechanism, once $\Trh$ is fixed. The final asymmetry is completely decided by $\yphn$ and $M_1$, for a given $\Trh$. However, $Y_B\sim n_B/T(a)^3$, and the bath temperature evolve differently under bosonic and fermionic reheating scenarios, thereby affecting the evolution of $Y_B$ with the scale factor (temperature). This can clearly be seen from Fig.~\ref{fig:bosYB}, where we have fixed $\Trh=10$ GeV and chosen $\yphn$ accordingly such that the final asymmetry satisfies the observed value. The dependence of the baryon asymmetry on $\yphn$ considering fermionic reheating, for a given $\Trh=10$ GeV, is shown in Fig.~\ref{fig:compare}. As illustrated in the left panels, the final baryon asymmetry increases with increasing $\yphn$ for both $n=4$ and $n=6$, in agreement with Eq.~\eqref{eq:YBn46}. A notable feature is the appearance of an intermediate plateau in the evolution of the baryon asymmetry for larger values of $\yphn$, particularly for $n=4$. This occurs because the radiation produced from RHN decays temporarily dominates over the radiation generated directly from inflaton decay, as shown by the red curves in the right panels. During this temporary RHN-decay-dominated radiation production, both the entropy injection and the asymmetry are sourced by the same RHN decay process, i.e., $\dot{n}_{B-L}\propto \Gamma_{N_1}\,n_{N_1},\,\dot{\rho}_R^{(N_1)}\propto \Gamma_{N_1}\,n_{N_1}$, causing the baryon yield to evolve only weakly. This feature is more pronounced for $n=4$ than for $n=6$, since in the latter case the radiation energy density from inflaton decay redshifts nearly as free radiation, whereas for $n=4$ it decreases more rapidly with the scale factor, allowing the RHN-generated radiation to dominate for a longer duration. The dependence of the final baryon asymmetry on $\Trh$ is shown in Fig.~\ref{fig:compare2} for a fixed value of $\yphn$. Once again, we see, for $n=4$, the final asymmetry is independent of $\Trh$, whereas for $n=6$, increasing $\Trh$ suppresses the baryon asymmetry, consistent with the scaling $Y_B\propto\Trh^{-1/3}$ predicted by Eq.~\eqref{eq:YBn46}. The bosonic reheating scenario exhibits the same qualitative behavior, so we do not show the corresponding results separately. 
\subsection{{\it Scenario-B:} reheating via RHN decay}
\begin{figure}[htb!]    \centering\includegraphics[width=0.5\linewidth]{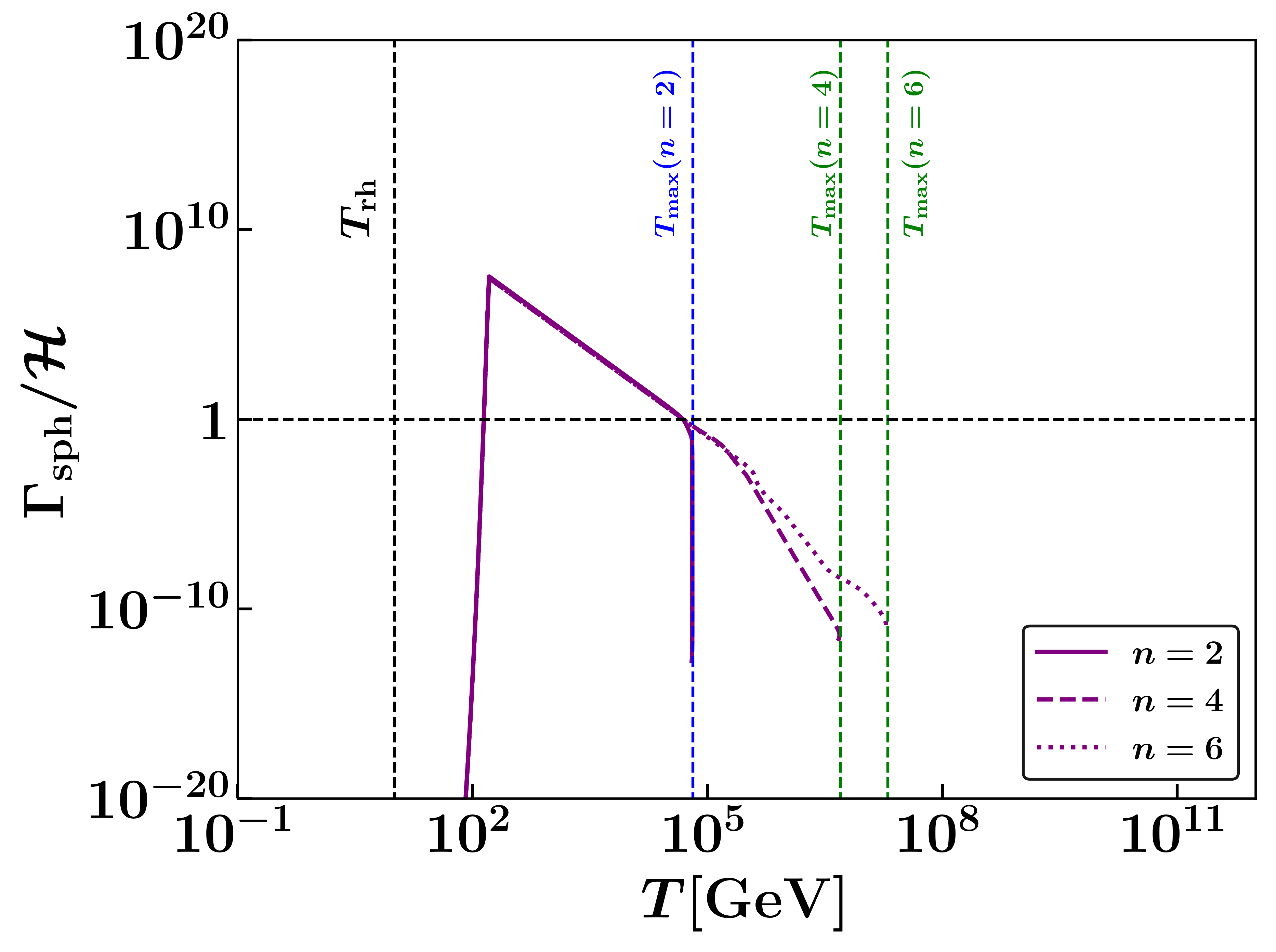}~\includegraphics[width=0.5\linewidth]{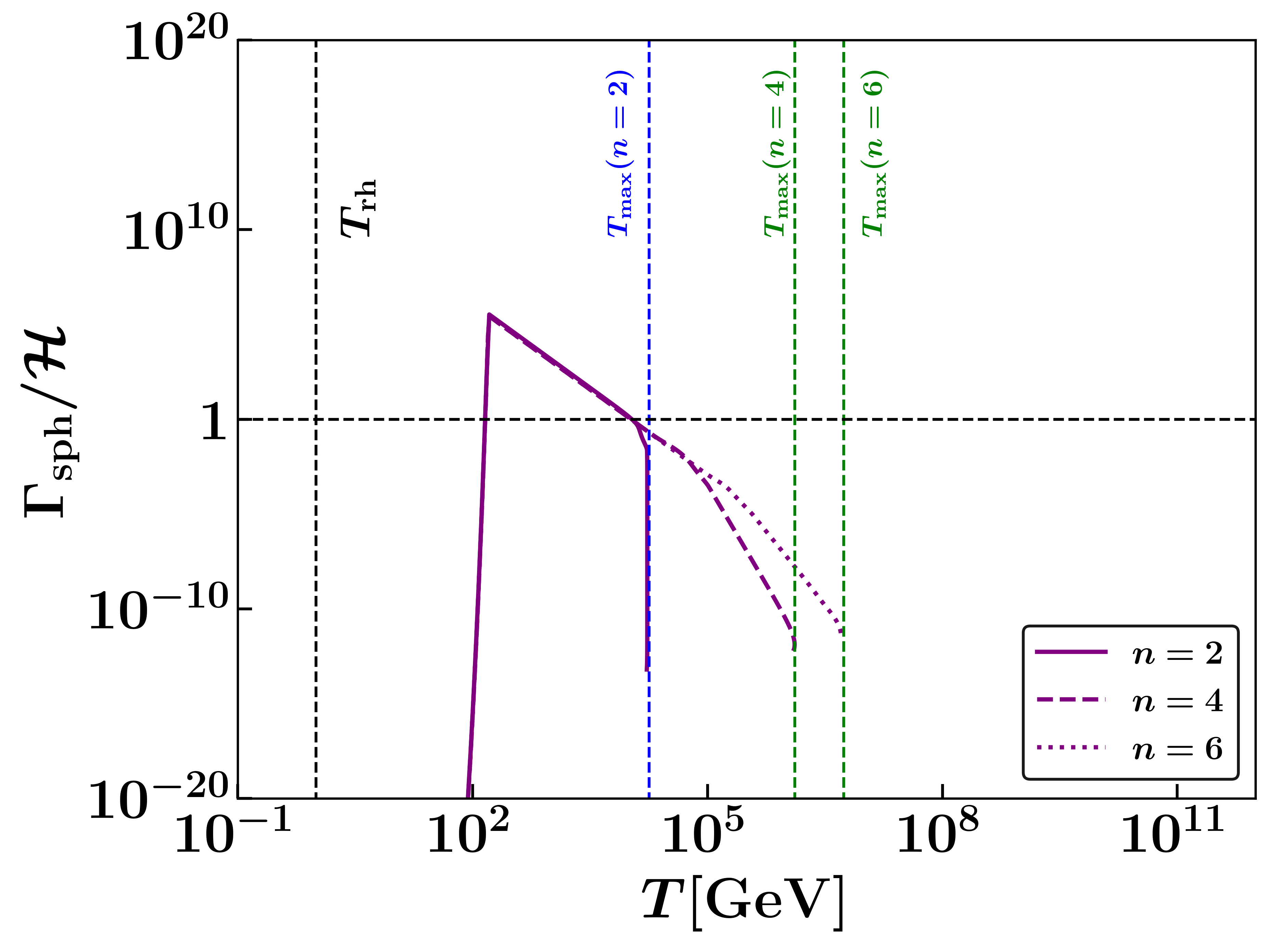}
\caption{{{\it Scenario-B:} Sphaleron rate as a function of temperature. The red dot-dashed line corresponds to standard radiation domination. Reheating temperature is fixed at $T_{\rm rh}=\{10,\,1\}$ GeV in the left and in the right panel, respectively. The vertical dashed lines correspond to $T=\Tmax$, for each $n$. We fix $y_{\nu_3}=\{2\times 10^{-10},\,2\times10^{-11}\}$ in the left and in the right panel, respectively. For all cases we choose $M_3=10^6$ GeV.}}
\label{fig:sph-rate-RHN}
\end{figure}

As discussed in subsection~\ref{sec:caseB}, the RHN $N_3$ plays the role of the long-lived species whose decay into SM final states eventually completes the reheating process, while the other two RHNs participate in leptogenesis. In this case, therefore, we consider $m_\phi(a)>2\,M_{1,\,3}$ and the lepton asymmetry is primarily generated by the decay of $N_1$. We solve the same set of coupled BEQs given in Eq.~\eqref{eq:BEQ-phiNN}, with the key modification that the inflaton condensate decays exclusively into RHN final states, while reheating of the Universe is driven entirely by the decay of long-lived $N_3$. It is important to emphasize that $N_3$ does not significantly contribute to the generation of the baryon asymmetry, since its coupling to the SM is extremely suppressed ($y_{\nu_3}\lesssim\mathcal{O}(10^{-10})$) in order to realize a low reheating temperature. Such a tiny coupling is inadequate to produce the observed asymmetry. Consequently, the asymmetry is generated predominantly through the decay of $N_1$, whereas the late decay of $N_3$ mainly dilutes the previously produced asymmetry until reheating is completed. Once again in Fig.~\ref{fig:sph-rate-RHN}, we see, the temperature over which sphaleron rate remains in equilibrium during reheating is modified compared to standard cosmology.   
\begin{figure}[htb!]
    \centering    \includegraphics[scale=0.55]{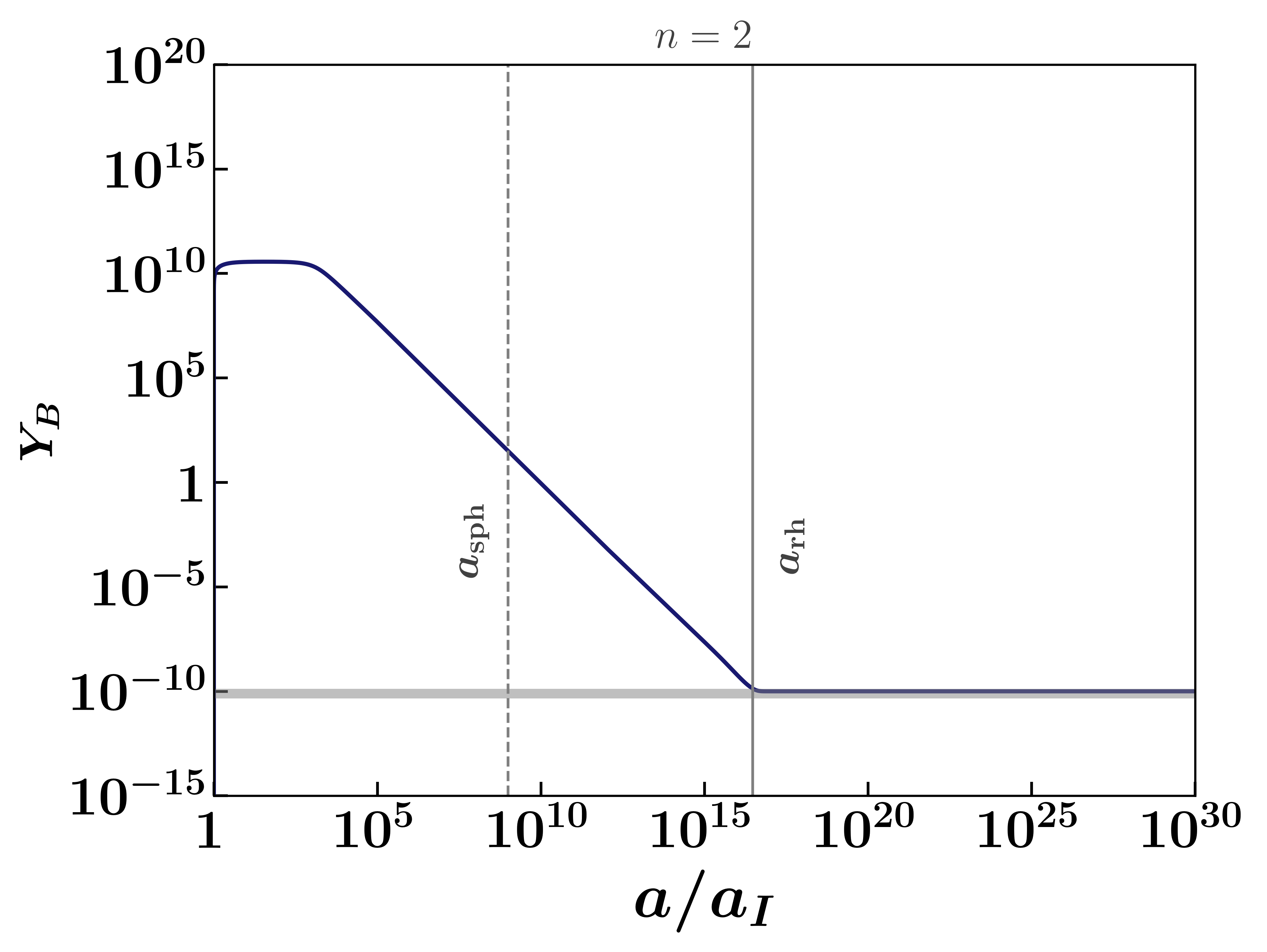}\\[10pt]    \includegraphics[scale=0.47]{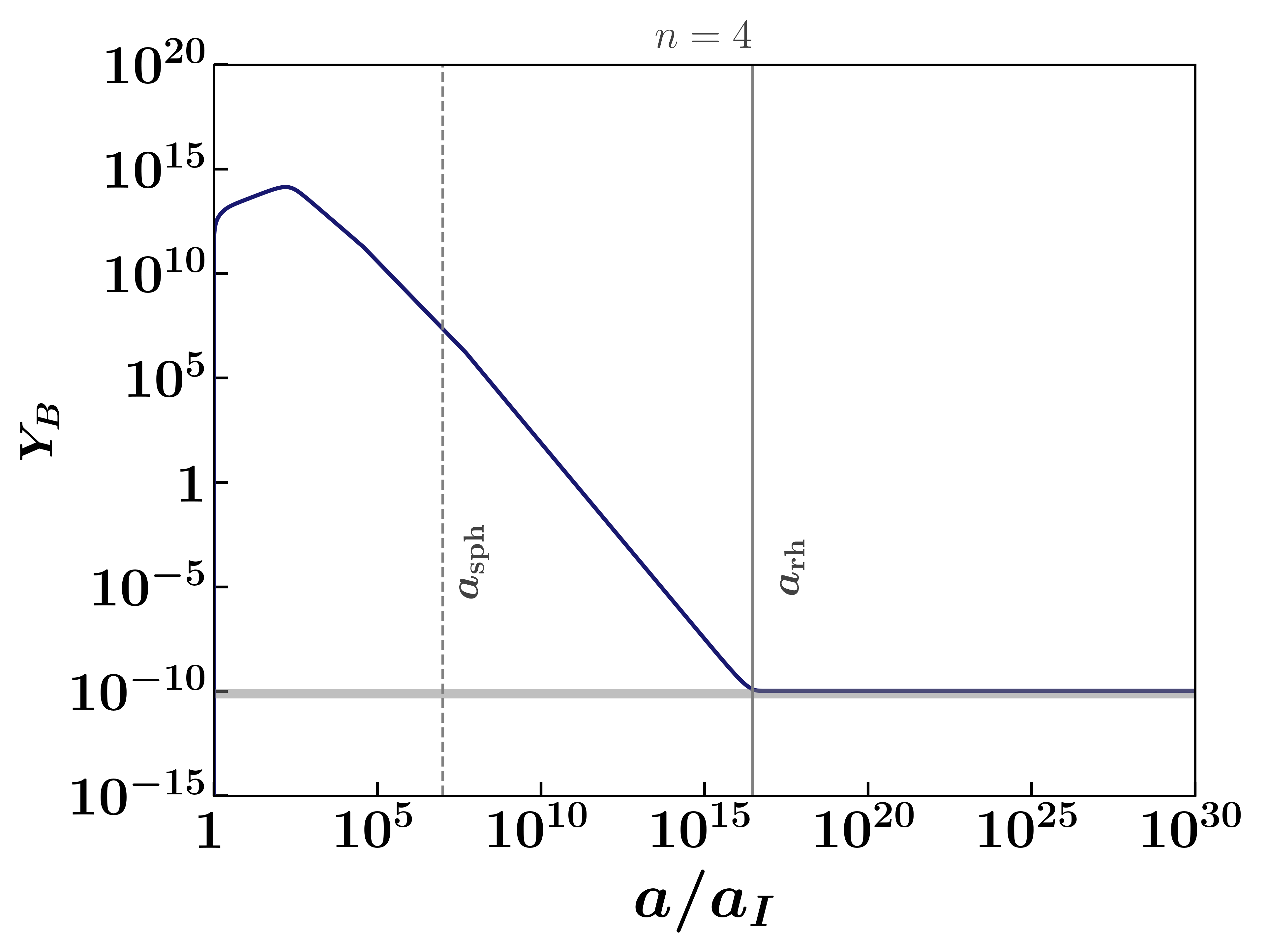}~\includegraphics[scale=0.47]{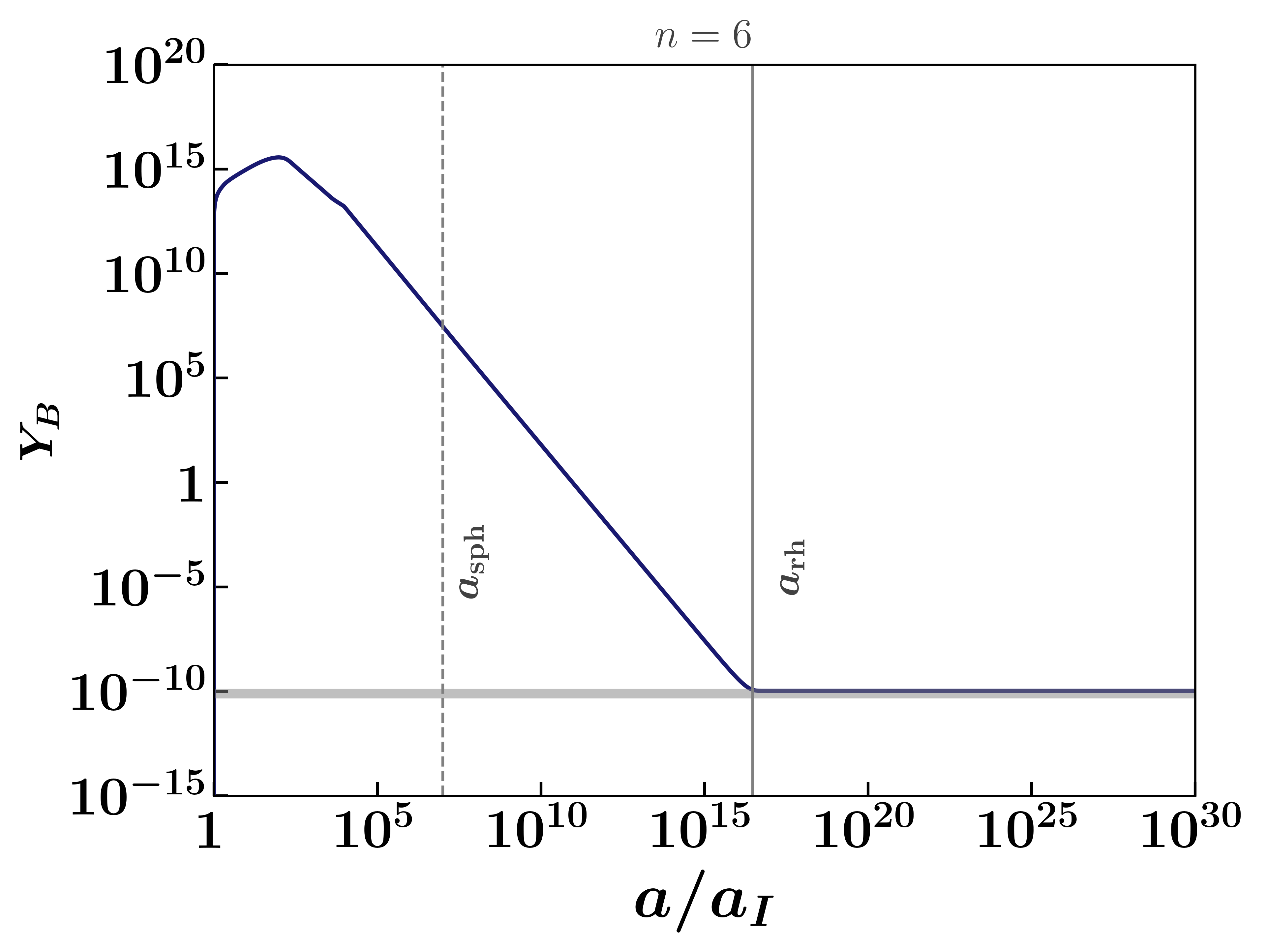}
    \caption{{{\it Scenario-B:} Evolution of baryon asymmetry as a function of scale factor. Here we fix $M_1=5\times 10^{12}$ GeV, $M_3 =10^6$ GeV with $y_{\nu_3}=2\times 10^{-10}$, such that $\Trh=10$ GeV. We consider $\yphn=\{10^{-6},\,7\times 10^{-3},\,0.1\}$ for $n=\{2,\,4,\,6\}$, respectively.
    }}
    \label{fig:scBYB}
\end{figure}
\begin{figure}[htb!]
    \centering    \includegraphics[scale=0.52]{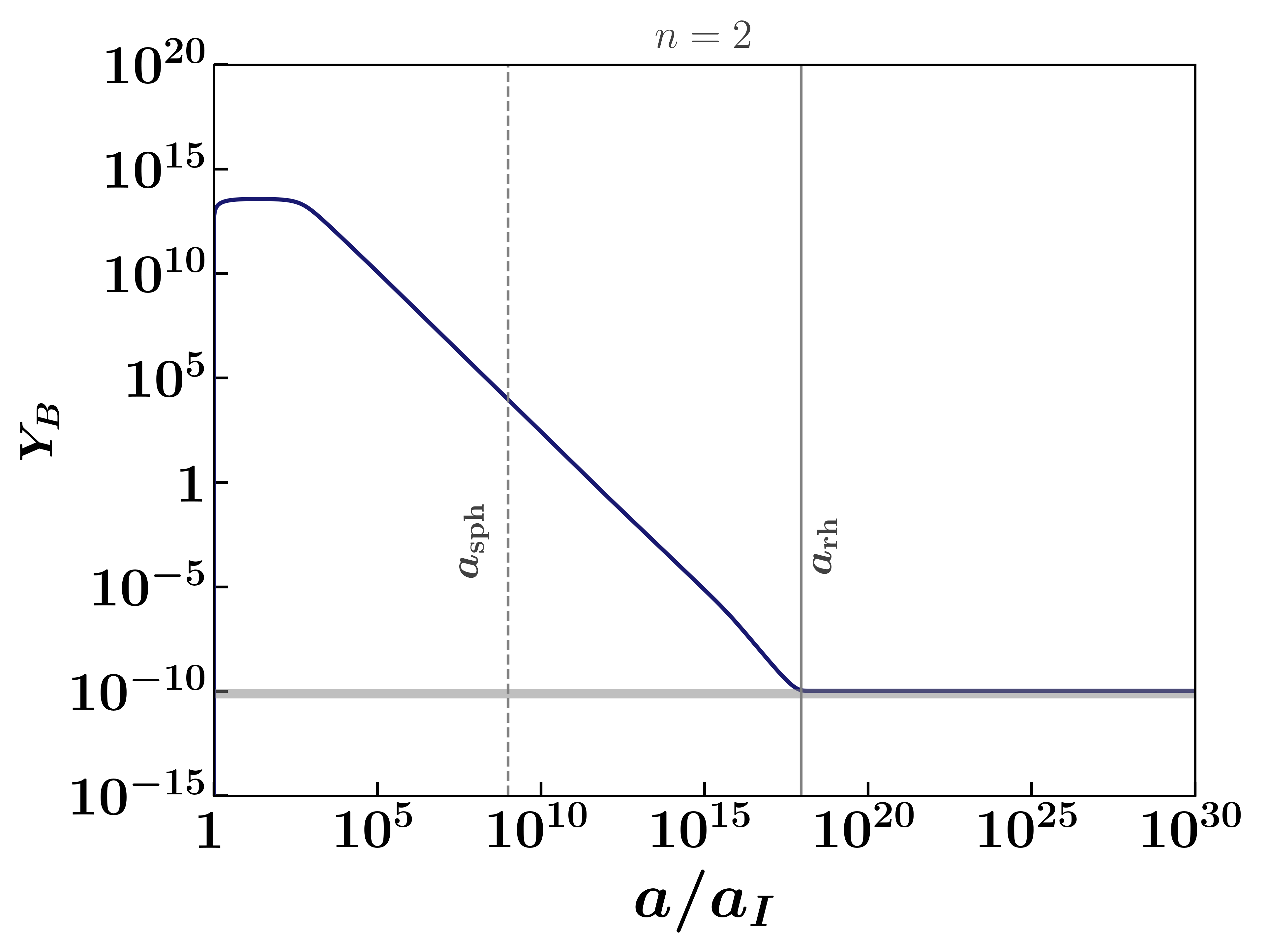}\\[10pt]    \includegraphics[scale=0.45]{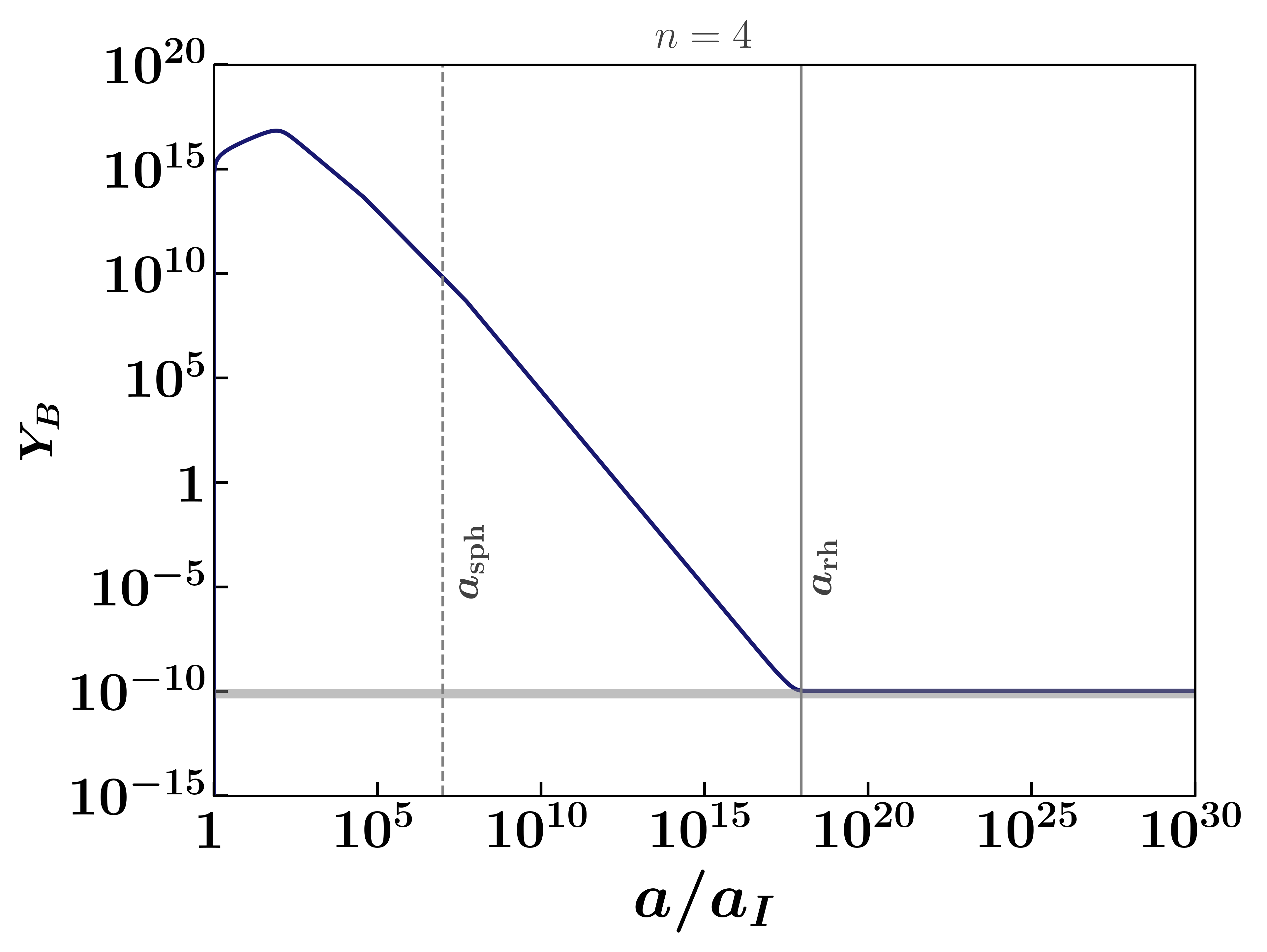}~\includegraphics[scale=0.45]{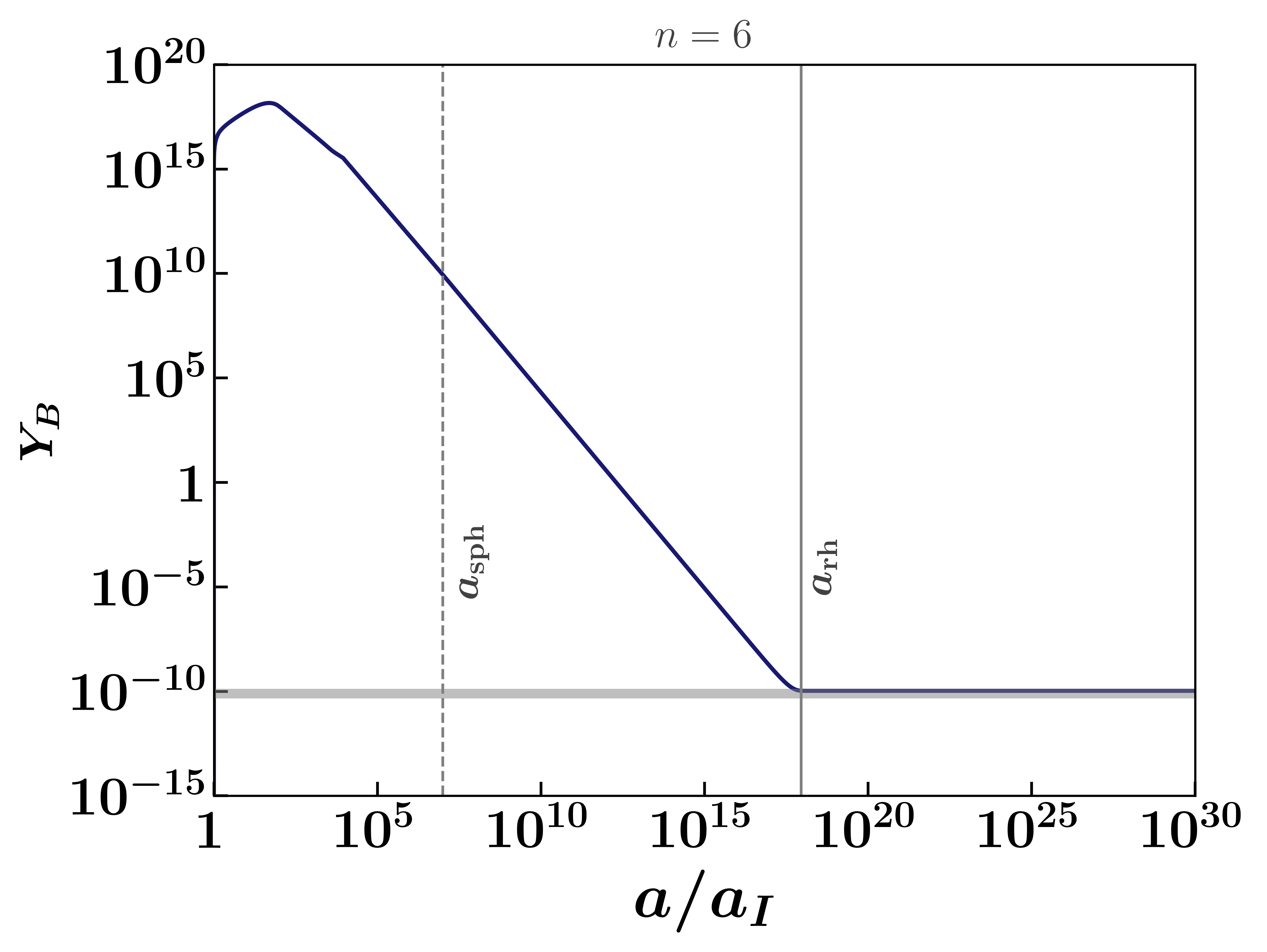}
    \caption{{{\it Scenario-B:} Same as Fig.~\ref{fig:scBYB}, but with $\Trh=1$ GeV. Here, we take $y_{\phi N}=\{10^{-6},\,7\times 10^{-3},\,0.1\}$ corresponding to $n=\{2,\,4,\,6\}$, respectively, while fixing $M_{1}=10^{13}$ GeV, $M_3 = 10^6$ GeV and $y_{\nu_{3}}=2\times 10^{-11}$, with corresponding $\Trh=1$ GeV in all cases.
    }}
    \label{fig:scBYB2}
\end{figure}
In this case, as already shown in Fig.~\ref{fig:rho-RHN}, there exists a scale factor $a=\amd$ at which $\rho_\phi(\amd)=\rho_{N_3}(\amd)$. Beyond this point, the Universe enters an $N_3$-dominated phase. To derive an approximate analytical expression for the final baryon asymmetry, we assume that all $N_1$ particles decay during inflaton domination. The generated asymmetry is then only diluted until $\arh$. Throughout the interval $\amd< a\leq\arh$, the Universe remains under $N_3$ (matter) domination. The final baryon asymmetry can therefore be written as,
\begin{align}
& Y_B\equiv\frac{n_B}{s}\Big|_{\Trh}\simeq\frac{28}{79}\,\epsilon_{\Delta L}\,\frac{n_{N_1}(\amd)}{s(\arh)}\,\left(\frac{\amd}{\arh}\right)^3=\frac{28}{79}\,\epsilon_{\Delta L}\,\frac{n_{N_1}(\amd)}{s(\arh)}\,\left(\frac{\rho_{N_3}(\arh)}{\rho_{N_3}(\amd)}\right)
\nonumber\\&
=\frac{28}{79}\,\epsilon_{\Delta L}\,\frac{n_{N_1}(\amd)}{s(\arh)}\,\left(\frac{\rR(\arh)}{\rho_{N_3}(\amd)}\right)
\end{align}
where
\begin{align}
& n_{N_1}(\amd)=\frac{a_I^3\,\yphn^2}{24\pi}\,\frac{\rho_I}{\mathcal{H}_I}\,\left[\left(\frac{\amd}{a_I}\right)^\frac{6}{n+2}-1\right]\,,
\end{align}
which is obtained by considering $N_1$ production during the inflaton-dominated era, $a_I<a\leq \amd$. To determine the $N_3$ energy density at $a=\amd$, we solve
\begin{align}
& \dot{\rho}_{N_3}+3\mathcal{H}\,\rho_{N_3}\simeq\Gamma_{\phi\to N_3}\,\rho_{N_3}\implies a^3\,\rho_{N_3}=\int_{a_I}^a\,da'\,a'^2\,\frac{\Gamma_{\phi\to N_3\,N_3}(a')}{\mathcal{H}(a')}\,\rp(a')\,da'\,,
\end{align}
where the $N_3$ decay term has been neglected since $\Gamma_{N_3}\ll\Gamma_\phi$. This is required for $N_3$ to be sufficiently long-lived to dominate the energy density and subsequently reheat the Universe. Performing the integration gives,
\begin{align}
& \rho_{N_3}(\amd)\simeq
\frac{\yphn^2}{8\pi}\,\frac{m_I\,\rho_I}{\mathcal{H}_I}
\begin{dcases}
\frac{4}{3}\,\left(\frac{a_I}{\amd}\right)^{3/2}\,
\left[1-\left(\frac{a_I}{\amd}\right)^{3/2}\right]\,,& n=2\,,
\\[12pt]
\frac{n+2}{3\,(n-4)}\,\left[\left(\frac{a_I}{\amd}\right)^3-\left(\frac{a_I}{\amd}\right)^\frac{6\,(n-1)}{n+2}\right]\,, & n>4\,,
\\[12pt]
\left(\frac{a_I}{\amd}\right)^3\,\ln\left(\frac{\amd}{a_I}\right)\,, & n=4\,.
\end{dcases}     
\end{align}
We thus arrive at the expression for final baryon asymmetry as,
\begin{align}\label{eq:YB-RHN}
& Y_B\simeq \frac{28}{79}\,\epsilon_{\Delta L}\,\frac{\mathcal{H}_I^2\,M_P^2\,a_I^3}{m_I\,\rho_I}\,\Trh\times
\nonumber\\&
\begin{dcases}
\frac{9}{8}\,\left(\frac{a_I}{\amd}\right)^{-3/2}\,
\left[1-\left(\frac{a_I}{\amd}\right)^{3/2}\right]^{-1}\,\left[\left(\frac{\amd}{a_I}\right)^{3/2}-1\right]\,, & n=2\,,
\\[12pt]
\frac{9}{4}\,\frac{n\,(n-4)}{n+2}\,\left[\left(\frac{a_I}{\amd}\right)^3-\left(\frac{a_I}{\amd}\right)^\frac{6\,(n-1)}{n+2}\right]^{-1}\,\left[\left(\frac{\amd}{a_I}\right)^\frac{6}{n+2}-1\right]\,, & n>4\,,
\\[12pt]
\left[\frac{1}{3}\,\left(\frac{a_I}{\amd}\right)^3\,\ln\left(\frac{\amd}{a_I}\right)\right]^{-1}\,\left[\left(\frac{\amd}{a_I}\right)-1\right]\,, & n=4\,.
\end{dcases}
\end{align}
Interestingly, the final baryon asymmetry is largely insensitive to $\yphn$ once $\Trh$ and RHN masses are fixed. The dependence on $\yphn$ enters only through $\amd$, which is approximately given by
\begin{align}
\amd/a_I\simeq
\begin{dcases}
\left[\frac{(C_I/\rho_I)\,\yphn^2}{3/4+(C_I/\rho_I)\,\yphn^2}\right]^{2/3}\,, & n=2\,,
\\[12pt]
\left[\yphn^2\,(C_I/\rho_I)\,\frac{n+2}{3\,(n-4)}\right]^\frac{n+2}{3\,(n-2)}\,, & n>4\,,
\\[12pt]
\left[\frac{7\,(C_I/\rho_I)\,\yphn^2}{W_{-1}\left(-7\,(C_I/\rho_I)\,\yphn^2\right)}\right]^{1/7}\,, & n=4\,,
\end{dcases}
\end{align}
where, $C_I/\rho_I=m_I/(8\pi\,\mathcal{H}_I)\simeq\mathcal{O}(10^{-2})$, and $W_{-1}[...]$ is the $-1$ branch of the Lambert-$W$ function. In the second line, we have assumed $(a_I/\amd)^\frac{6\,(n-1)}{n+2}\ll(a_I/a)^3$, which is a good approximation for $a_I\lesssim \amd$ (for $a_I\ll \amd$, both terms contribute with equal weight). With this approximation it is possible to derive a closed form of $\amd/a_I$ for $n>4$, otherwise we encounter a transcendental equation that can only be solved numerically. We also note that, as $n$ increases, $a_{\rm md}$ shifts to smaller values, consistent with the behaviour already observed in Fig.~\ref{fig:rho-RHN}. As a result, the entropy dilution persists for a longer period. Therefore, a larger value of $\yphn$ is required for increasing $n$ so that more RHNs are produced per inflaton decay, ultimately yielding the observed BAU. 

The evolution of the BAU with the scale factor is shown in Fig.~\ref{fig:scBYB} and Fig.~\ref{fig:scBYB2} for $\Trh=10$ GeV and $\Trh=1$ GeV, respectively. We find, a larger $n$ requires a stronger inflaton--RHN coupling to generate the observed final BAU, consistent with the discussion above. Unlike {\it Scenario-A}, here we do not observe a plateau-like behaviour in the BAU evolution. This is because there is no competition between radiation produced by RHN decay and that produced by inflaton decay, as the inflaton decays entirely into RHNs. Instead, the generated asymmetry initially increases during the RHN production phase until $a\simeq\amd$, after which it is continuously diluted by the substantial entropy injection from the decay of the long-lived RHN. Following Fig.~\ref{fig:sph-rate-RHN}, we also indicate by $\asph$, the scale factor at which the sphaleron processes enter equilibrium during RHN-reheating.
\section{Detection prospects for low-reheating leptogenesis}
\label{sec:detection}
The RHN mass scale required for successful baryogenesis in the previous section lies far beyond the sensitivity of current collider experiments searching for RHN signatures. Consequently, we turn to cosmological observables as potential probes of the present framework. One such observable is the primordial gravitational wave (PGW) background, which can be directly connected to the scale of leptogenesis in scenarios with an inflationary origin~\cite{Berbig:2023yyy,Barman:2024slw,Barman:2024ujh,Borboruah:2024eha,Borboruah:2025hai}. PGW signals, associated with leptogeneis, may also arise from, for example, Bremsstrahlung processes~\cite{Ghoshal:2022kqp,Datta:2024tne}, cosmic strings~\cite{Dror:2019syi,Blasi:2020wpy,Samanta:2020cdk,Chianese:2024gee,Datta:2025vyu}, or colliding domain walls~\cite{Barman:2022yos,Borah:2025bfa}. In what follows, however, we focus on PGWs of inflationary origin, which naturally link our scenario to gravitational wave observations without getting into further model building. As we will illustrate, the PGW spectral shape gives rise to characteristic red and blue-tilt, depending on the reheating scenarios discussed in subsections.~\ref{sec:caseA} and \ref{sec:caseB}.
\subsection{Inflationary gravitational wave spectrum}
The PGW spectrum with an inflationary origin have been extensively studied, for example, in Refs.~\cite{Giovannini:1998bp,Giovannini:1999bh,Riazuelo:2000fc,Seto:2003kc,Boyle:2007zx,Stewart:2007fu,Li:2021htg,Artymowski:2017pua,Caprini:2018mtu,Bettoni:2018pbl,Figueroa:2019paj,Opferkuch:2019zbd,Bernal:2020ywq,Ghoshal:2022ruy,Caldwell:2022qsj,Gouttenoire:2021jhk,Haque:2021dha,Maity:2024cpq}. The present day GW background is conveniently described by its energy density per logarithmic interval in comoving wavenumber,
\begin{equation}
\Omega_{\rm GW}(a,k)\equiv \frac{1}{\rho_c}\frac{d\rho_{\rm GW}}{d\ln k}\,,
\qquad 
\rho_c=3\mathcal{H}^2M_P^2\,.
\end{equation}
This quantity measures the fraction of the critical energy density stored in gravitational waves at a given scale $k$. For primordial tensor fluctuations, it can be expressed as
\begin{equation}
\Omega_{\rm GW}(a,k)=\frac{1}{12}\left(\frac{k}{a\mathcal{H}}\right)^2 
\mathcal{P}_{T,\rm prim}(k)\,\mathcal{T}(a,k)\,,
\end{equation}
where \(\mathcal{P}_{T,\rm prim}\) is the primordial tensor power spectrum generated in the early universe, and \(\mathcal{T}(a,k)\) encodes the subsequent cosmological evolution after horizon reentry. The primordial tensor spectrum is parameterized as,
\begin{equation}
\mathcal{P}_{T,\rm prim}(k)=
r\,\mathcal{P}_\zeta(k_\star)
\left(\frac{k}{k_\star}\right)^{n_T}\,,
\end{equation}
where the pivot scale is \(k_\star=0.05\,{\rm Mpc}^{-1}\), the scalar amplitude is \(\mathcal{P}_\zeta(k_\star)\simeq 2.1\times10^{-9}\), and \(r\) is the tensor-to-scalar ratio. Since the observational bound \(r<0.036\) implies a negligible tensor tilt (\(n_T\simeq-r/8\)), we take \(n_T=0\), corresponding to an approximately scale-invariant primordial GW spectrum. The transfer function is defined as,
\begin{equation}
\mathcal{T}(a,k)=\frac12\left(\frac{a_{\rm hc}}{a}\right)^2\,,
\end{equation}
where $a_{\rm hc}$ is the scale factor at horizon crossing, determined by
\begin{equation}
a_{\rm hc}\mathcal{H}(a_{\rm hc})=k\,.
\end{equation}
Using this, the present GW abundance becomes
\begin{equation}
\Omega_{\rm GW}(k)=
\frac{1}{24}
\left(\frac{k}{a_0 \mathcal{H}_0}\right)^2
\mathcal{P}_{T,\rm prim}(k)
\left(\frac{a_{\rm hc}}{a_0}\right)^2\,.
\end{equation}
\subsubsection{{\it Scenario-A:} direct reheating}
The nature of the final GW spectrum depends on when each mode reenters the horizon. Modes reentering during the standard radiation-dominated era preserve the usual flat primordial shape, while modes entering during the non-standard pre-BBN epoch characterized by the parameter $n$, experience an additional enhancement. Including this effect, for {\it Scenario-A}, the present GW spectral energy density reads,
\begin{align}
\Omega_{\rm GW}(a_{\rm hc}) \simeq &
\ \Omega_\gamma^{(0)}
\frac{\mathcal{P}_{T,\rm prim}}{24}
\frac{g_*(T_{\rm hc})}{2}
\left(\frac{g_{*s}(T_0)}{g_{*s}(T_{\rm hc})}\right)^{4/3}
\nonumber\\
&\times
\begin{dcases}
\frac{g_*(T_{\rm rh})}{g_*(T_{\rm hc})}
\left(\frac{g_{*s}(T_{\rm hc})}{g_{*s}(T_{\rm rh})}\right)^{4/3}
\left(\frac{a_{\rm rh}}{a_{\rm hc}}\right)^{\frac{2(n-4)}{n+2}},
& a_I<a_{\rm hc}\le a_{\rm rh}\,,\\[10pt]
1,
& a_{\rm rh}\le a_{\rm hc}\le a_{\rm eq}\,,
\end{dcases}
\end{align}
where $\Omega_\gamma^{(0)}=2.47\times10^{-5}h^{-2}$ is the present photon density fraction. The observed GW frequency today corresponding to a mode that reentered at $a_{\rm hc}$ is given by,
\begin{align}
f(a_{\rm hc})=\frac{k}{2\pi a_0}=\mathcal{F}(g_*,g_{*s})\,T_0\,\sqrt{\frac{\mathcal{H}_I}{M_P}}\,\sqrt{h_r}
\times
\begin{dcases}
\left(\frac{a_{\rm rh}}{a_{\rm hc}}\right)^{\frac{2(n-1)}{n+2}},
& a_I<a_{\rm hc}\le a_{\rm rh}\,,\\[10pt]
\frac{a_{\rm rh}}{a_{\rm hc}},
& a_{\rm rh}\le a_{\rm hc}\le a_{\rm eq}\,,
\end{dcases}
\end{align}
with
\begin{align}
& h_r\equiv \frac{\mathcal{H}(a_{\rm rh})}{\mathcal{H}_I}\,, & \mathcal{F}(\gs,\,\gss)=\frac16 \sqrt{\frac{\gs(\Trh)}{10}} \left(\frac{\gss(T_0)}{\gss(\Trh)}\right)^\frac13\,\left(\frac{90}{\pi^2\,\gs(\Trh)}\right)^{1/4}\,.
\end{align}
Expressed in frequency space, the GW spectrum becomes
\begin{align}\label{eq:pgw1}
\Omega_{\rm GW}(f)\simeq &
\ \Omega_\gamma^{(0)}
\frac{\mathcal{P}_{T,\rm prim}}{24}
\frac{g_*(T_{\rm hc})}{2}
\left(\frac{g_{*s}(T_0)}{g_{*s}(T_{\rm hc})}\right)^{4/3}
\nonumber\\
&\times
\begin{dcases}
\frac{g_*(T_{\rm rh})}{g_*(T_{\rm hc})}
\left(\frac{g_{*s}(T_{\rm hc})}{g_{*s}(T_{\rm rh})}\right)^{4/3}
\left(\frac{f}{f_{\rm rh}}\right)^{\frac{n-4}{n-1}},
& f_{\rm rh}\le f<f_{\rm max}\,,\\[10pt]
1,
& f_{\rm eq}\le f\le f_{\rm rh}\,,
\end{dcases}
\end{align}
where 
\begin{equation}\label{eq:frh}
f_{\rm rh}=
\mathcal{F}(g_*,g_{*s})
T_0\,\sqrt{\frac{\mathcal{H}_I}{M_P}}\,\sqrt{h_r}\,,
\end{equation}
and the highest observable primordial frequency is
\begin{equation}
f_{\rm max}=f(a_I)=
\frac{\mathcal{H}_I}{2\pi}\frac{a_I}{a_0}\,.
\end{equation}
In summary, frequencies below $f_{\rm rh}$ follow the standard nearly scale-invariant primordial spectrum, whereas modes with $f_{\rm rh}<f<f_{\rm max}
$, re-enter during the early nonstandard epoch and acquire a blue tilt, $\Omega_{\rm GW}(f)\propto f^{\frac{n-4}{n-1}}$. Thus, larger $n$ leads to a stronger enhancement of the GW amplitude at high frequencies, significantly improving the prospects for detection in future GW observatories.
\subsubsection{{\it Scenario-B:} reheating via RHN decay}
\begin{figure}[htb!]
\centering
\includegraphics[scale=0.74]{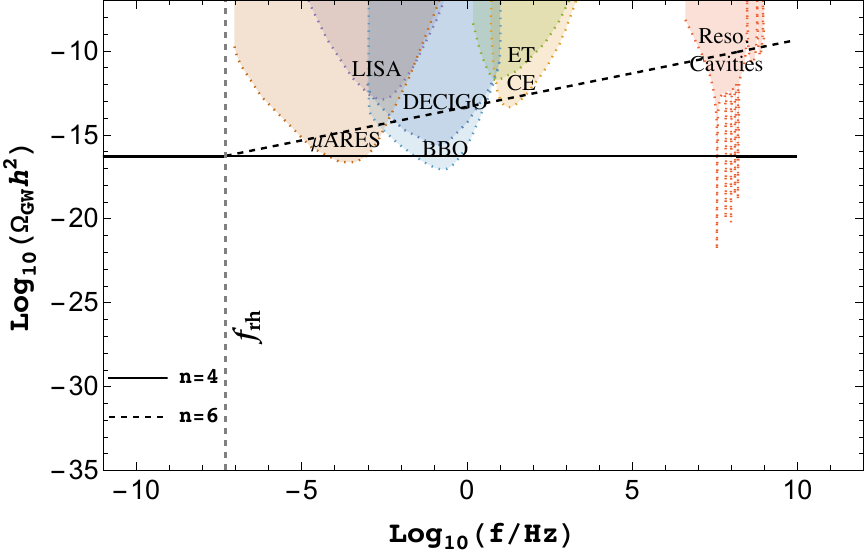}\\[10pt]
\includegraphics[scale=0.74]{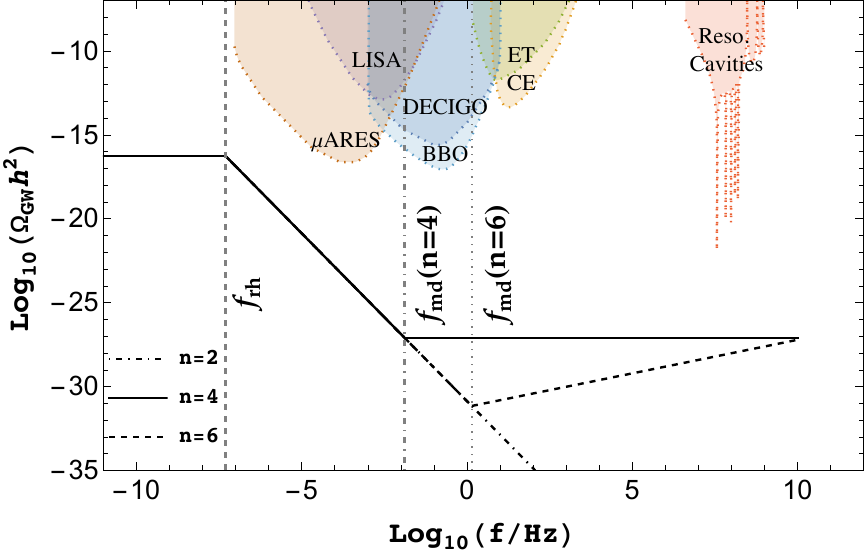}
\caption{Top panel: The blue-tilted primordial GW spectrum corresponding to {\it Scenario-A}, due to tensor perturbation for $n=\{4,\,6\}$, shown via black solid and black dashed lines, respectively. We project sensitivity curves from several future experiments, for example, the Big Bang Observer (BBO)~\cite{Crowder:2005nr, Corbin:2005ny}, ultimate DECIGO (uDECIGO)~\cite{Seto:2001qf, Kudoh:2005as}, LISA~\cite{LISA:2017pwj}, $\mu$Ares~\cite{Sesana:2019vho}, the cosmic explorer (CE)~\cite{Reitze:2019iox}, the Einstein Telescope (ET)~\cite{Hild:2010id, Punturo:2010zz, Sathyaprakash:2012jk, Maggiore:2019uih}, and from resonant cavities~\cite{Herman:2022fau}. Bottom panel: Same as top, but for {\it Scenario-B}. In both cases we have fixed $\Trh=10$ GeV, the vertical broken lines represent GW frequency corresponding to the onset of radiation domination ($\frh$) [cf. Eq.~\eqref{eq:frh}] and matter domination ($f_{\rm md}$) [cf. Eq.~\eqref{eq:fmd}].} 
\label{fig:pgw}
\end{figure}

Due to the presence of the intermediate RHN-dominated epoch, after the end of inflaton domination, the GW spectral energy density in this case reads,
\begin{align}
\Omega_{\rm GW}(a_{\rm hc}) \simeq &
\ \Omega_\gamma^{(0)}
\frac{\mathcal{P}_{T,\rm prim}}{24}
\frac{g_*(T_{\rm hc})}{2}
\left(\frac{g_{*s}(T_0)}{g_{*s}(T_{\rm hc})}\right)^{4/3}
\nonumber\\
&\times
\begin{dcases}
\frac{g_*(T_{\rm rh})}{g_*(T_{\rm hc})}
\left(\frac{g_{*s}(T_{\rm hc})}{g_{*s}(T_{\rm rh})}\right)^{4/3}
\left(\frac{\arh}{a_{\rm md}}\right)^{-1}\,\left(\frac{a_{\rm md}}{a_{\rm hc}}\right)^{\frac{2(n-4)}{n+2}},
& a_I<a_{\rm hc}\le a_{\rm md}\,,\\[10pt]
\frac{\gs(\Trh)}{\gs(T_{\rm hc})}\,\left(\frac{g_{*s}(T_{\rm hc})}{g_{*s}(T_{\rm rh})}\right)^{4/3}\left(\frac{\arh}{a_{\rm hc}}\right)^{-1}\,,& a_{\rm md}\leq a_{\rm hc}\le \arh\,,\\[10pt]
1,
& a_{\rm rh}\le a_{\rm hc}\le a_{\rm eq}\,,
\end{dcases}
\end{align}
where $a_{\rm md}$ is the onset of RHN-domination. Once again, in terms of frequency, this takes the form,
\begin{align}\label{eq:pgw2}
\Omega_{\rm GW}(f)\simeq &
\ \Omega_\gamma^{(0)}
\frac{\mathcal{P}_{T,\rm prim}}{24}
\frac{g_*(T_{\rm hc})}{2}
\left(\frac{g_{*s}(T_0)}{g_{*s}(T_{\rm hc})}\right)^{4/3}
\nonumber\\
&\times
\begin{dcases}
\frac{g_*(T_{\rm rh})}{g_*(T_{\rm hc})}
\left(\frac{g_{*s}(T_{\rm hc})}{g_{*s}(T_{\rm rh})}\right)^{4/3}\left(\frac{f_{\rm md}}{f_{\rm rh}}\right)^{-2}
\left(\frac{f}{f_{\rm md}}\right)^{\frac{n-4}{n-1}},
& f_{\rm md}\le f<f_{\rm max}\,,\\[10pt]
\frac{g_*(T_{\rm rh})}{g_*(T_{\rm hc})}
\left(\frac{g_{*s}(T_{\rm hc})}{g_{*s}(T_{\rm rh})}\right)^{4/3}\left(\frac{f}{f_{\rm rh}}\right)^{-2}\,,& f_{\rm rh}\leq f < f_{\rm md}\,,\\[10pt]
1,
& f_{\rm eq}\le f\le f_{\rm rh}\,,
\end{dcases}
\end{align}
where 
\begin{equation}\label{eq:fmd}
    f_{\rm md} = \left(\frac{a_{\rm rh}}{a_{\rm md}}\right)^{1/2}\,f_{\rm rh}\,,
\end{equation}
represents the observed GW frequency today corresponding to the mode that crosses the horizon at $a_{\rm md}$. Due to intermediate matter dominated background cosmology, the spectrum in {\it Scenario-B} exhibits a characteristic red tilt with $\Omega_{\rm GW} \propto f^{-2}$, following the second line of Eq.~\eqref{eq:pgw2}, as one can see from the right panel of Fig.~\ref{fig:pgw}. The red-tilted spectrum continues up to $f_{\rm md}$, where it undergoes another spectral break, resulting in a red tilt, flat spectrum, or blue tilt depending on $n = \{2,\,4,\,6\}$, respectively, following the first line of Eq.~\eqref{eq:pgw2}. To assess the detectability of the predicted GW signal, we compute the signal-to-noise ratio (SNR), $\mathcal{S}$, by integrating over the detector's observation time $t_{\rm obs}$ and frequency range $[f_{\rm min},\,f_{\rm max}]$~\cite{Maggiore:1999vm,Allen:1996vm,Allen:1997ad}:
\begin{align}
\mathcal{S}=\sqrt{n_{\rm det}\,t_{\rm obs}\,\int_{f_{\rm min}}^{f_{\rm max}} df \left[\frac{\Omega_{\rm GW}(f)\,h^2}{\Omega_{\rm expt}(f)\,h^2}\right]^2}\,.
\end{align}
Here, $n_{\rm det}=1$ for auto-correlation searches and $n_{\rm det}=2$ for cross-correlation searches for a stochastic GW background. We consider a GW signal to be detectable if it satisfies $\mathcal{S}>10$. As we will demonstrate, the region of parameter space consistent with the observed baryon asymmetry can also produce detectable GW signals, with $\mathcal{S}\gtrsim10$, at several future GW observatories. The prospects for detection depend on the values of $n$ and $\Trh$.
\begin{figure}[htb!]
\centering
\includegraphics[scale=0.375]{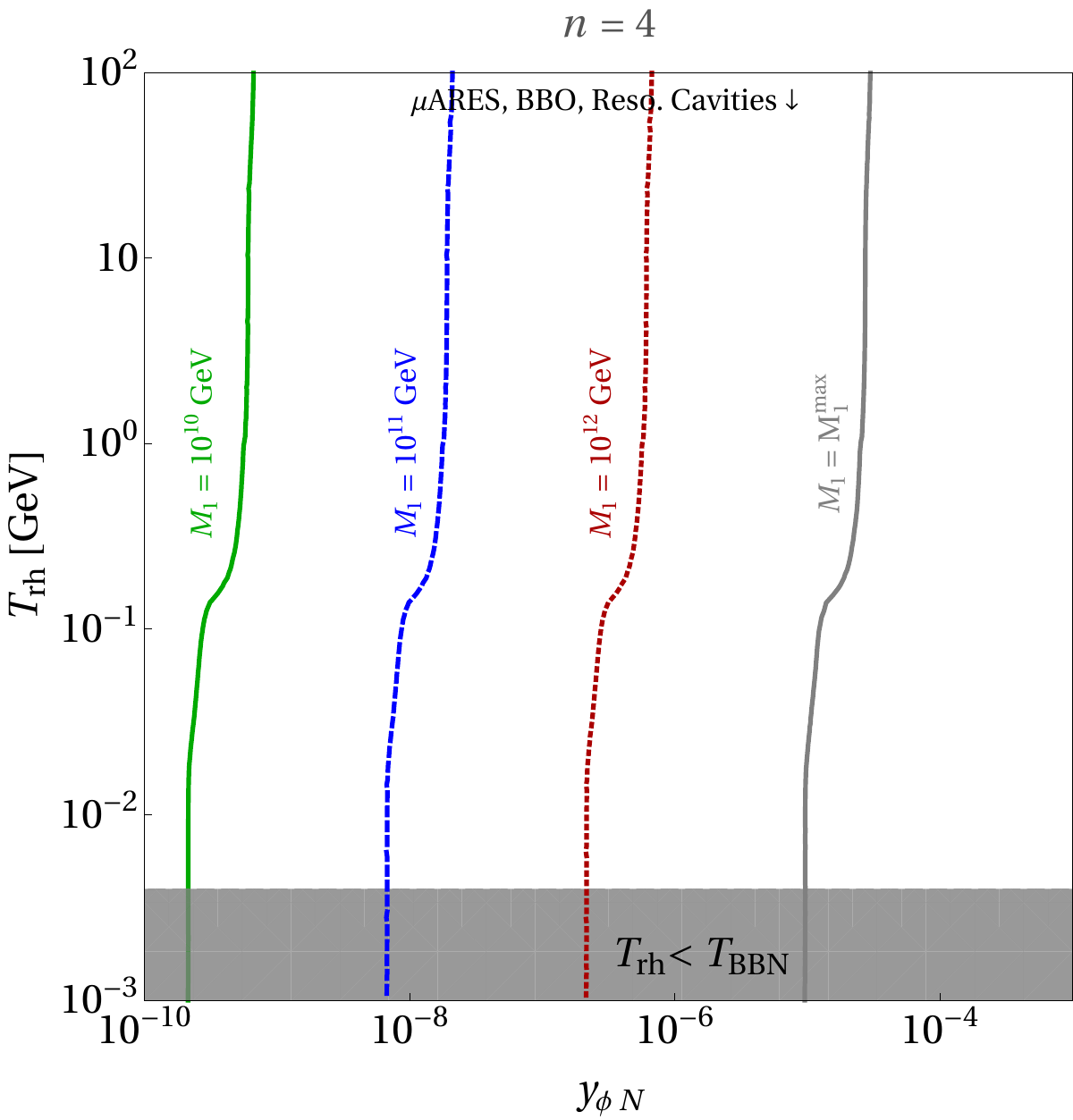}~\includegraphics[scale=0.375]{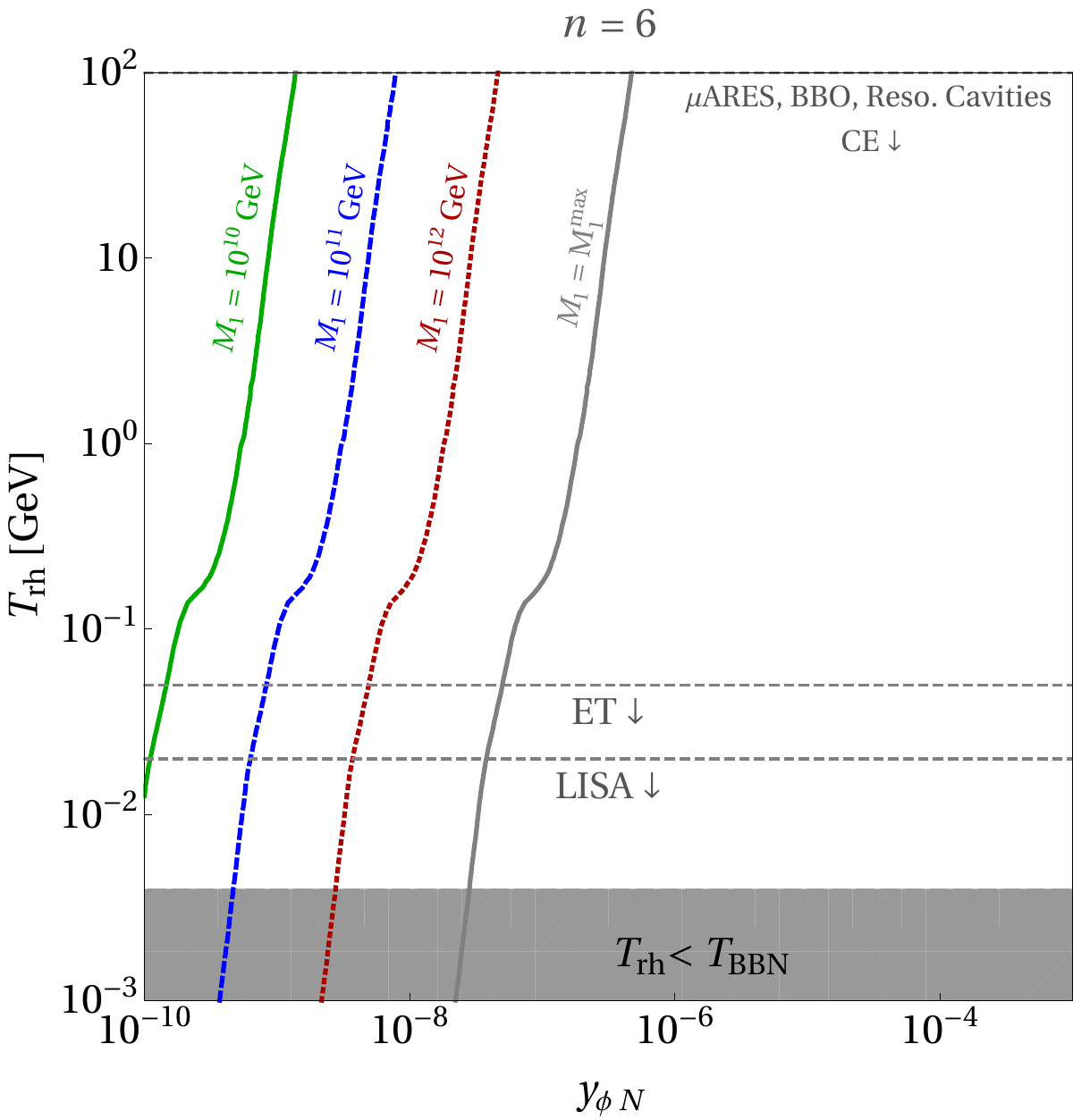}
\caption{{\it Scenario-A:} Contours for the observed BAU for $n=\{4,\,6\}$ in the left and in the right panel, respectively. Different contours correspond to different RHN masses, as indicated in the plots.  The gray shaded region correspond to the BBN bound $\Trh=4$ MeV. We also show the sensitivity reach of different future GW detectors (for $\mathcal{S}\geq10$) in probing $\Trh$ via primordial GW spectrum corresponding to different $n$'s.
}
\label{fig:SumPlt-caseA}
\end{figure}
\begin{figure}[htb!]
    \centering
    \includegraphics[scale=0.5]{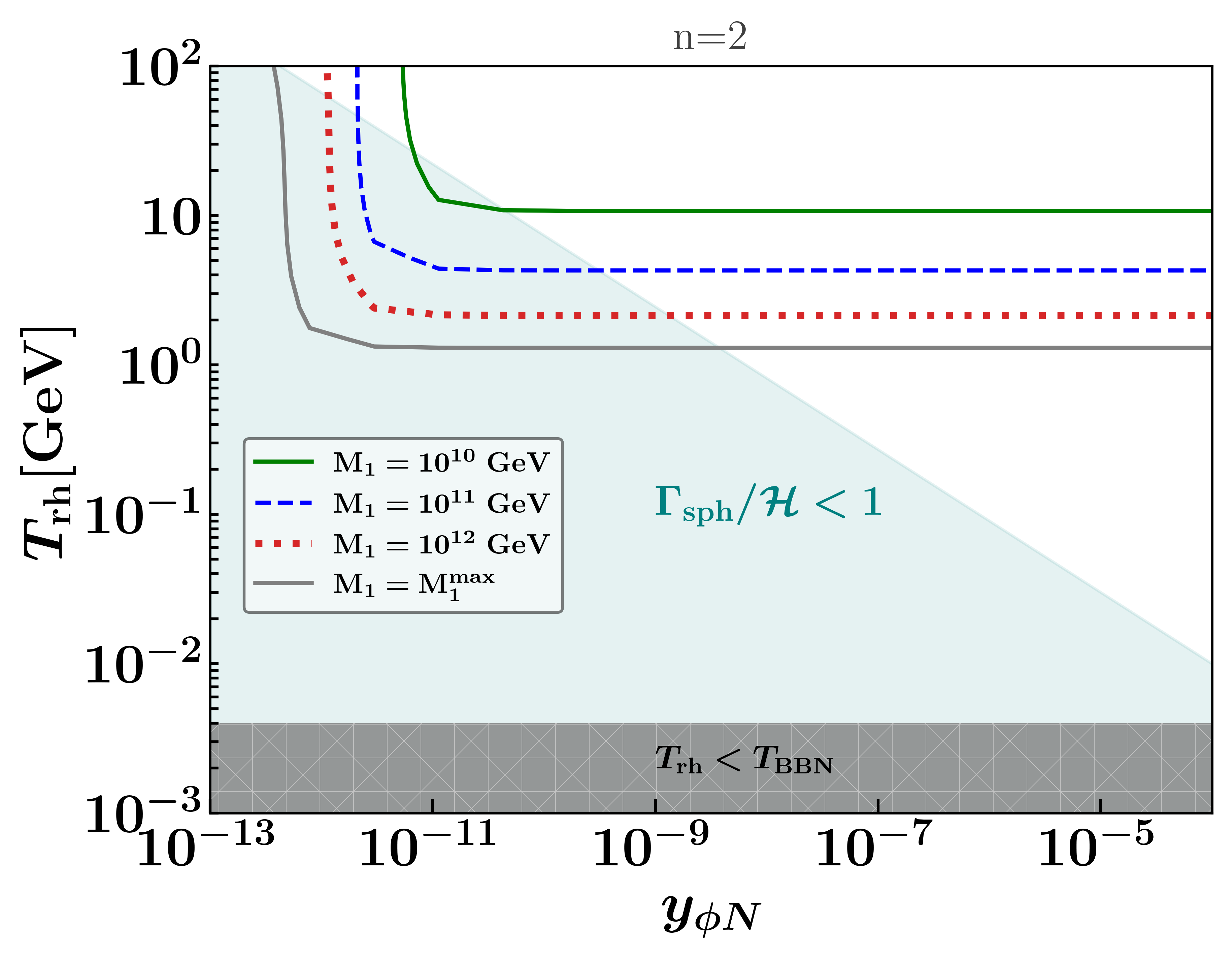}\\[10pt]
    \includegraphics[scale=0.45]{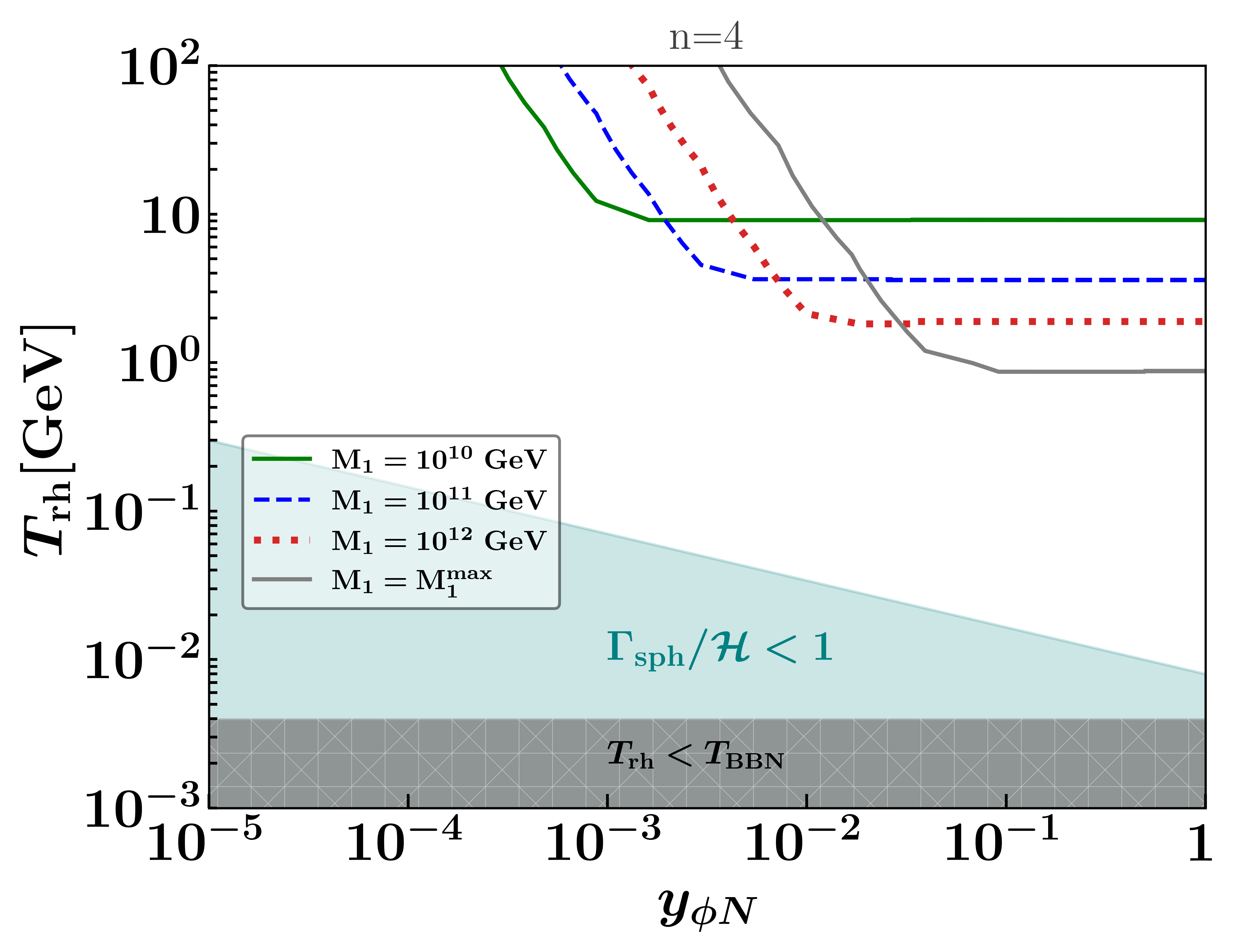}~\includegraphics[scale=0.45]{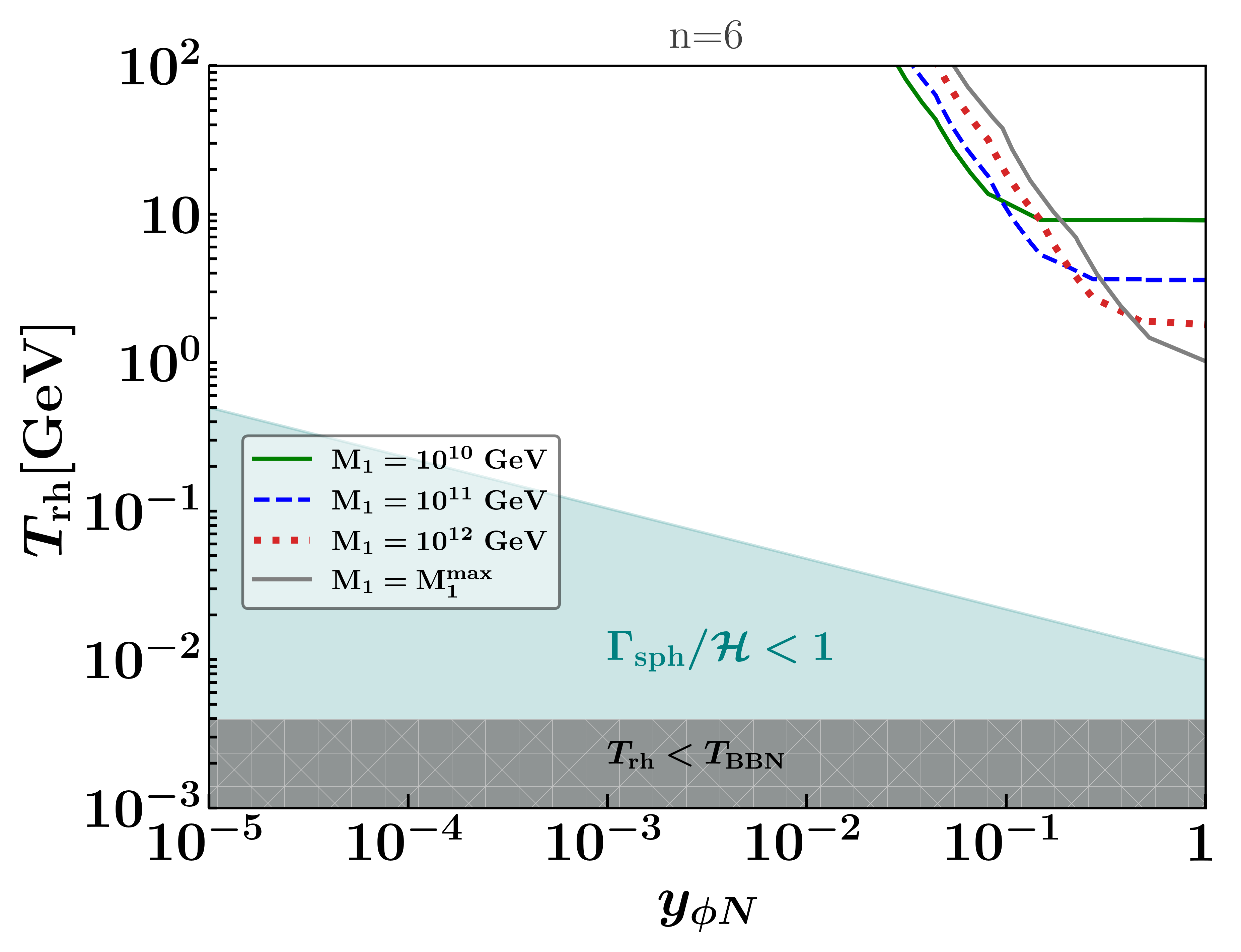}
    \caption{Same as Fig.~\ref{fig:SumPlt-caseA}, but for {\it Scenario-B}.}
    \label{fig:SumPlt-caseB}
\end{figure}

From the top panel of Fig.~\ref{fig:pgw}, one can see that {\it Scenario-A} has promising discovery prospects at several future GW detectors due to the blue-tilted nature of the spectrum for $n>4$. As already shown in Fig.~\ref{fig:bosYB}, an RHN with mass $10^{10}$ GeV can successfully generate the observed baryon asymmetry at $\Trh=10$ GeV for $n=6$, in both bosonic and fermionic reheating scenarios. It is important to note, however, that such a blue-tilted GW spectrum is not unique to low-reheating leptogenesis. Rather, it is a generic feature of a pre-BBN epoch dominated by a background equation of state stiffer than radiation ($w>1/3$). In contrast, the detection prospects for {\it Scenario-B} are much weaker, as one can notice from the bottom panel. The intermediate matter-dominated epoch introduces a strong red tilt in the resulting PGW spectrum, significantly suppressing the signal. As a result, detecting such spectra appears extremely challenging, even for future GW observatories. It is worth emphasizing that this large red-tilted part of the spectrum arises from our choice of $\Trh<T_{\rm sph}$. If this condition is relaxed, the corresponding GW spectrum can become detectable at several proposed GW detectors. Therefore, within {\it Scenario-B}, simultaneously realizing low-reheating leptogenesis and an observable inflationary GW signal is highly difficult. It is important to emphasize, however, that such GW signatures are not unique probes of low-reheating leptogenesis; rather, they probe the pre-BBN cosmological history (more generally, a stiffer than radiation background EoS). Nevertheless, these signatures can still serve as indirect indicators of a low-reheating leptogenesis scenario.   
\subsection{A summary of constraints and probes}
A scan over the two-dimensional parameter space in the $\left[\Trh-\yphn\right]$ plane, for different choices of the RHN mass, is shown in Fig.~\ref{fig:SumPlt-caseA} and Fig.~\ref{fig:SumPlt-caseB} for {\it Scenario-A} and {\it Scenario-B}, respectively. As discussed earlier, for $n=2$ the observed baryon asymmetry cannot be reproduced for $\Trh\lesssim\mathcal{O}(100)$ GeV in {\it Scenario-A}. We therefore focus on {\it Scenario-A} with $n=4$ and $n=6$. As shown in the left panel, the contours corresponding to the observed BAU are nearly independent of $\Trh$. The change in their slopes around $\Trh\sim0.1\,\text{GeV}$ arises from the sudden increase in the effective number of entropy degrees of freedom, $\gss(\Trh)$, which changes from $\gss\simeq10$ to $\gss\simeq100$. Along each contour, the light neutrino data are simultaneously satisfied. For a given contour, both the RHN mass and the CI parameters are fixed, implying that the CP asymmetry ($\epsilon_{\Delta L}$) is also fixed. For a fixed value of $n$, increasing the RHN mass causes RHN production from inflaton decay to terminate earlier [cf.~Eq.~\eqref{eq:ast}]. As a result, the generated asymmetry is diluted over a longer period. To compensate for this additional dilution and reproduce the observed BAU, more RHNs must be produced per inflaton decay, requiring a larger value of $\yphn$. RHN production, however, becomes kinematically forbidden once $M_1\gtrsim m_I/2\simeq\left[1.15,\,1.53\right]\times 10^{13}$ GeV, for $n=\{4,\,6\}$. This upper limit, $M_1^{\rm max}\equiv m_I/2$, is indicated by the gray contour in both panels. Finally, for $n=6$, the contours exhibit the expected scaling $\yphn\propto\Trh^{1/6}$, in agreement with Eq.~\eqref{eq:YBn46}. We would like to emphasize once again that the final asymmetry in this case is independent of the underlying reheating mechanism, and is decided entirely by $\yphn,\,\Trh,\,M_1$ and the CI parameters. Therefore, the summary plot in Fig.~\ref{fig:SumPlt-caseA} applies to both bosonic and fermionc reheating dynamics.

For {\it Scenario-B}, $\Trh$ is entirely determined by the coupling between the long-lived RHN and the SM, which has been varied over a range $y_{\nu_3}\in\left[2.6 \times 10^{-15}-2.6\times 10^{-8}\right]$. The contours of right baryon asymmetry is entirely fixed by $\Trh$, being largely independent of $\yphn$, a nature that agrees with Eq.~\eqref{eq:YB-RHN}. For $n>2$, the allowed region shifts towards larger values of $\yphn$. This behavior is consistent with the results shown in Figs.~\ref{fig:scBYB} and \ref{fig:scBYB2}. All RHNs have a universal coupling $\yphn$ to the inflaton. This results in a lower bound on $\yphn$ below which sufficient $\Trh$ cannot be achieved, making it impossible to generate the desired BAU. Consequently, for all values of $n$, the curves become approximately independent of $\Trh$ below a certain value of $\yphn$. The green shaded region in each plot shows the region of parameter space where the sphaleron falls out of equilibrium during reheating [cf. Eq.~\eqref{fig:sph-rate-RHN}], and therefore ruled out. 

The entire parameter space consistent with the observed BAU lies within the projected sensitivity of several future GW detectors in {\it Scenario-A}, depending on the values of $n$ and $\Trh$. For $n=6$, the blue-tilted primordial inflationary GW spectrum provides a promising opportunity to probe reheating temperatures $\Trh \lesssim 10^6$ GeV with DECIGO, CE, ET, LISA, BBO, $\mu$ARES, and resonant cavity detectors [cf. Fig.~\ref{fig:pgw}]. In contrast, for $n=4$, where the primordial GW spectrum is nearly scale-invariant that could be tested by BBO, $\mu$ARES, and resonant cavity detectors, covering a wide range of frequencies. In all cases we ensure that the SNR $\mathcal{S}\geq 10$ is satisfied, corresponding to each experiment. On the other hand, in {\it Scenario-B}, as discussed earlier, the intermediate matter-like RHN-dominated era suppresses the GW spectrum over the frequency range accessible to current and planned GW experiments. Consequently, this scenario is not expected to be probed by any existing or proposed GW detector. 
\section{Conclusions}
\label{sec:concl}
The baryon asymmetry of the Universe may have been dynamically generated during the reheating epoch following inflation in a single-field inflationary paradigm. In this work, we have investigated this possibility within a low-reheating-temperature scenario, where the reheating temperature lies well below the temperature of electroweak phase transition. We consider leptogenesis as the mechanism for asymmetry generation, with heavy RHNs produced either thermally from the radiation bath or non-thermally through the perturbative decay of the inflaton condensate during reheating. We show that the final baryon asymmetry is sensitive to both the equation of state of the Universe during reheating and the inflaton--RHN coupling. These factors determine how long the electroweak sphaleron processes remain in equilibrium during reheating, thereby affecting the efficiency of converting the generated lepton asymmetry into the observed baryon asymmetry. Our study reveals several interesting features:
\begin{itemize}
\item For {\it Scenario-A}, where the Universe is reheated after inflation through inflaton decays into the SM final states at the bottom of a monomial potential $V(\phi)\sim\phi^n$, we find that the final baryon asymmetry (i) is always underproduced for $n=2$ when $\Trh\lesssim\mathcal{O}(100)$ GeV, (ii) is nearly independent of the reheating temperature for $n=4$, and (iii) decreases as the reheating temperature increases for $n=6$.
\item For {\it Scenario-B}, where reheating occurs through RHN decay following a brief period of RHN domination, the final baryon asymmetry is found to be largely independent of the inflaton-RHN coupling.
\end{itemize}
In all cases, we find that the production of RHNs from the radiation bath remains subdominant. As a result, leptogenesis is predominantly non-thermal. We identify the regions of parameter space consistent with the observed baryon asymmetry and demonstrate that a significant portion of this parameter space can be probed by future GW detectors through the characteristic blue-tilted inflationary GW spectrum. Our results therefore provide a testable framework for leptogenesis with sub-electroweak-scale reheating, establishing a compelling connection between particle physics and the cosmology of the early Universe.  

Finally, let us highlight a few limitations of our analysis. In the RHN-induced reheating scenario, we have considered a relatively large inflaton--RHN Yukawa coupling, reaching values as high as $0.4$ (for $n=6$), in order to generate the observed baryon asymmetry. However, such large couplings inevitably trigger non-perturbative particle production, leading to a {\it preheating} phase\footnote{Production of heavy RHNs during preheating leading to leptogensis has been studied, for example, in~\cite{Giudice:1999fb,Peloso:2000hy,Kanemura:2025rct}.}. It is well known that, for inflaton potentials with $n\gtrsim 3$, the EoS parameter during reheating asymptotically approaches $w \to 1/3$ due to inflaton fragmentation and parametric resonance effects~\cite{Amin:2010dc,Lozanov:2019jxc,Garcia:2023eol,Garcia:2023dyf}. This indicates that preheating can play a significant role in rapidly driving the Universe toward radiation domination. Nevertheless, the complete transfer of the inflaton energy density still requires perturbative decays through trilinear interactions between the inflaton and its decay products, which are expected to dominate the final stage of reheating~\cite{Kofman:1985aw,Kofman:1997yn,Greene:1998nh}. Moreover, studies performed for a quadratic inflaton potential~\cite{Drewes:2017fmn,Drewes:2019rxn} have shown that perturbative reheating remains valid only for inflaton--fermion Yukawa couplings $\lesssim 10^{-5}$, consistent with the values adopted for $n=2$ in {\it Scenario-A}. { On the other hand, perturbative reheating through the inflaton–Higgs trilinear interaction, $\mu\,\phi\,|H|^2$, requires $\mu/m_\phi\lesssim 10^{-5}$~\cite{Drewes:2017fmn,Drewes:2019rxn}. In our scenario, for $n=2$ and $\Trh<100$ GeV, we find $\mu/m_\phi\lesssim\mathcal{O}(10^{-10})$, which is well below this bound. In contrast, for $n>2$, non-perturbative effects become unavoidable. Note that, this trilinear interaction is gauge-invariant which, in general, cannot be forbidden without introducing additional symmetries or assumptions. Nevertheless, the coupling strength can be chosen such that the decay of the inflaton remains within the perturbative regime. However, parametric resonance can be efficient when the inflaton field amplitude is sufficiently large just after inflation. As the amplitude of the inflaton oscillations decreases (with expansion of the Universe), the resonance becomes inefficient and eventually shuts off. The subsequent stage of reheating can then be described by perturbative inflaton decay. In fact, the reheating temperature is typically determined during these final stages, when perturbative decay dominates.} A proper treatment of these effects requires a dedicated analysis, which we plan to perform in a future work. We also do not include the so-called ``flavour effects''~\cite{Nardi:2005hs,DeSimone:2006nrs,Nardi:2006fx,Blanchet:2006be,Blanchet:2006ch,Davidson:2008bu,Blanchet:2011xq,Moffat:2018wke,Granelli:2021fyc} in our analysis of leptogenesis, which arise when the individual charged lepton Yukawa interactions come into equilibrium at different temperature depending on their interaction rates. Instead, we restrict ourselves to the standard vanilla leptogenesis framework. The impact of charged lepton Yukawa equilibration during reheating has been studied extensively in the literature, for example in~\cite{Datta:2022jic,Datta:2023pav,Barman:2024jqh,Roshan:2025mwl}. In particular, Ref.~\cite{Datta:2023pav} considered an explicit inflaton--RHN coupling $\yphn$ and showed that the faster expansion rate during reheating, together with the modified temperature evolution, delays the equilibration of charged lepton Yukawa interactions. As a result, the generated asymmetry can differ significantly from the standard thermal leptogenesis scenario. However, those analyses still assume reheating temperatures well above the sphaleron freeze-out temperature. In this work, we leave a detailed study of flavour effects in low-reheating leptogenesis for future investigation. Our present goal is more modest: to examine how far the reheating temperature can be pushed below the sphaleron freeze-out temperature while still reproducing the observed baryon asymmetry, and to understand the associated limitations and trade-offs involved in doing so. 
\acknowledgments
The work of D.B. is supported by the Science and Engineering Research Board (SERB), Government of India grant CRG/2022/000603. BB would like to acknowledge SRM University-AP for support through the Internal Research Grant Policy.
\section*{{\bf Note added:}}
During the completion of this work, Ref.~\cite{Garcia:2026ulw} appeared, also studying leptogenesis in low-reheating scenarios. Although, our results largely agrees with~\cite{Garcia:2026ulw}, but our analysis differs in several aspects. Besides the fermionic and bosonic reheating scenarios considered in Ref.~\cite{Garcia:2026ulw}, we also investigate an RHN-dominated reheating scenario. We determine the RHN--SM couplings using the Casas--Ibarra parametrization, ensuring consistency with neutrino oscillation data, and explicitly track the sphaleron interaction rate throughout the reheating history. We further discuss the observational implications of these scenarios. Conversely, Ref.~\cite{Garcia:2026ulw} includes the effects of inflaton fragmentation, which have not considered in our analysis.
\appendix
\section{Details of inflationary potential}
\label{sec:inflation}
For the $\alpha$-attractor T-model of the form,
\begin{align}
& V(\phi )=\lambda\,M_P^4 \left[\tanh \left(\frac{\phi}{\sqrt{6\,\alpha}\,M_P}\right)\right]^n=\lambda\,M_P^4\,
\begin{dcases}
1\,, & \; \text{for}\; \phi \gg M_P,\\[10pt]
\left(\frac{\phi}{\sqrt{6\,\alpha}\,M_P}\right)^n\,, & \; \text{for}\; \phi\ll M_P\,,
\end{dcases}
\end{align}
we obtain the potential slow-roll parameters as,
\begin{align}
\epsilon_V(\phi)&=\frac{n^2}{3\alpha}
\,\mathrm{csch}^2\!\left(\sqrt{\frac{2}{3\alpha}}\frac{\phi}{M_P}\right),\\
\eta_V(\phi)&=\frac{n}{3\alpha}
\left[n-\cosh\!\left(\sqrt{\frac{2}{3\alpha}}\frac{\phi}{M_P}
\right)\right]\mathrm{csch}^2\!\left(\sqrt{\frac{2}{3\alpha}}\frac{\phi}{M_P}
\right)\,.
\end{align}
Inflation ends approximately when $\epsilon_V(\phi_{I})\simeq1$, giving
\begin{align}
\phi_{I}\simeq \sqrt{\frac{3\alpha}{2}}\,M_P\,
\sinh^{-1}\!\left(\frac{n}{\sqrt{3\alpha}}\right)\,.
\end{align}
The number of e-folds between horizon exit of the pivot scale $k_\star=0.05~{\rm Mpc}^{-1}$ and the end of inflation is
\begin{align}
N_\star=\frac{3\alpha}{2n}
\left[\cosh\!\left(\sqrt{\frac{2}{3\alpha}}\frac{\phi_\star}{M_P}
\right)-\cosh\!\left(\sqrt{\frac{2}{3\alpha}}\frac{\phi_{I}}{M_P}\right)\right]\,.
\end{align}
The spectral index and tensor-to-scalar ratio are,
\begin{align}
r &=16\epsilon_V(\phi_\star)\,,\\
n_s-1 &=2\eta_V(\phi_\star)-6\epsilon_V(\phi_\star)\,,
\end{align}
which yield
\begin{align}
r^T(\phi_\star)&=\frac{16n^2}{3\alpha}\,\mathrm{csch}^2\!\left(
\sqrt{\frac{2}{3\alpha}}\frac{\phi_\star}{M_P}
\right)\,,
\\
n_s^T(\phi_\star)-1&=-\frac{2n}{3\alpha}\,\mathrm{csch}^2\!\left(
\sqrt{\frac{2}{3\alpha}}\frac{\phi_\star}{M_P}
\right)\left[2n+\cosh\!\left(\sqrt{\frac{2}{3\alpha}}\frac{\phi_\star}{M_P}\right)\right]\,.
\label{eq:nS_T}
\end{align}
Inverting Eq.~\eqref{eq:nS_T}, one can obtain $\phi_\star$. We use the recent P-ACT-LB determination $n_s=0.9743\pm0.0034$~\cite{AtacamaCosmologyTelescope:2025blo}. The tensor-to-scalar ratio is defined by
\begin{align}
r\equiv\frac{\Delta_t^2(k_\star)}{\Delta_s^2(k_\star)}\,,
\end{align}
with
\begin{align}
\Delta_t^2=\frac{2}{\pi^2}\frac{\mathcal{H}_\star^2}{M_P^2}\,,
\qquad\Delta_s^2=\frac{1}{8\pi^2}
\frac{\mathcal{H}_\star^2}{M_P^2}
\frac{1}{\epsilon_\star}\,.
\end{align}
Using $\Delta_s^2(k_\star)=2.1\times10^{-9}$~\cite{Planck:2018jri} and the current BK18+Planck bound $r<0.036$, one finds
\begin{align}
\mathcal{H}_\star\simeq \mathcal{H}_{I}\lesssim
4.4\times10^{13}\ {\rm GeV}\,,
\end{align}
and consequently
\begin{align}
\rho_\phi(a_I)=3\,M_P^2\,\mathcal{H}_{I}^2
\lesssim3.4\times10^{64}\ {\rm GeV}^4\,.
\end{align}
The inflationary potential at horizon exit is
\begin{align}
V(\phi_\star)=\frac{3\pi^2}{2}\,
M_P^4\,\Delta_s^2\,r\,,
\end{align}
implying
\begin{align}
\lambda=\left(\frac{3\pi^2}{2}\,\Delta_s^2\,r\right)\,\tanh^{-n}\!\left(\frac{\phi_\star}{\sqrt{6\alpha}\,M_P}\right)\,.
\end{align} 
\bibliography{Bibliography}

@article{Ghoshal:2022kqp,
    author = "Ghoshal, Anish and Samanta, Rome and White, Graham",
    title = "{Bremsstrahlung High-frequency Gravitational Wave Signatures of High-scale Non-thermal Leptogenesis}",
    eprint = "2211.10433",
    archivePrefix = "arXiv",
    primaryClass = "hep-ph",
    month = "11",
    year = "2022"
}

@article{Kallosh:2013hoa,
    author = "Kallosh, Renata and Linde, Andrei",
    title = "{Universality Class in Conformal Inflation}",
    eprint = "1306.5220",
    archivePrefix = "arXiv",
    primaryClass = "hep-th",
    doi = "10.1088/1475-7516/2013/07/002",
    journal = "JCAP",
    volume = "07",
    pages = "002",
    year = "2013"
}

@article{Caprini:2018mtu,
    author = "Caprini, Chiara and Figueroa, Daniel G.",
    title = "{Cosmological Backgrounds of Gravitational Waves}",
    eprint = "1801.04268",
    archivePrefix = "arXiv",
    primaryClass = "astro-ph.CO",
    doi = "10.1088/1361-6382/aac608",
    journal = "Class. Quant. Grav.",
    volume = "35",
    number = "16",
    pages = "163001",
    year = "2018"
}

@article{Bernal:2022wck,
    author = "Bernal, Nicol\'as and Xu, Yong",
    title = "{WIMPs during reheating}",
    eprint = "2209.07546",
    archivePrefix = "arXiv",
    primaryClass = "hep-ph",
    reportNumber = "PI/UAN-2022-722FT",
    doi = "10.1088/1475-7516/2022/12/017",
    journal = "JCAP",
    volume = "12",
    pages = "017",
    year = "2022"
}

@article{Seto:2001qf,
    author = "Seto, Naoki and Kawamura, Seiji and Nakamura, Takashi",
    title = "{Possibility of direct measurement of the acceleration of the universe using 0.1-Hz band laser interferometer gravitational wave antenna in space}",
    eprint = "astro-ph/0108011",
    archivePrefix = "arXiv",
    doi = "10.1103/PhysRevLett.87.221103",
    journal = "Phys. Rev. Lett.",
    volume = "87",
    pages = "221103",
    year = "2001"
}

@article{Punturo:2010zz,
    author = "Punturo, M. and others",
    editor = "Ricci, Fulvio",
    title = "{The Einstein Telescope: A third-generation gravitational wave observatory}",
    doi = "10.1088/0264-9381/27/19/194002",
    journal = "Class. Quant. Grav.",
    volume = "27",
    pages = "194002",
    year = "2010"
}

@article{Kudoh:2005as,
    author = "Kudoh, Hideaki and Taruya, Atsushi and Hiramatsu, Takashi and Himemoto, Yoshiaki",
    title = "{Detecting a gravitational-wave background with next-generation space interferometers}",
    eprint = "gr-qc/0511145",
    archivePrefix = "arXiv",
    reportNumber = "UTAP-544, RESCEU-37-05",
    doi = "10.1103/PhysRevD.73.064006",
    journal = "Phys. Rev. D",
    volume = "73",
    pages = "064006",
    year = "2006"
}

@article{Kawasaki:2000en,
    author = "Kawasaki, M. and Kohri, Kazunori and Sugiyama, Naoshi",
    title = "{MeV scale reheating temperature and thermalization of neutrino background}",
    eprint = "astro-ph/0002127",
    archivePrefix = "arXiv",
    doi = "10.1103/PhysRevD.62.023506",
    journal = "Phys. Rev. D",
    volume = "62",
    pages = "023506",
    year = "2000"
}

@article{Herman:2022fau,
    author = "Herman, Nicolas and Lehoucq, L\'eonard and F\'{u}zfa, Andr\'e",
    title = "{Electromagnetic Antennas for the Resonant Detection of the Stochastic Gravitational Wave Background}",
    eprint = "2203.15668",
    archivePrefix = "arXiv",
    primaryClass = "gr-qc",
    month = "3",
    year = "2022"
}

@article{Crowder:2005nr,
    author = "Crowder, Jeff and Cornish, Neil J.",
    title = "{Beyond LISA: Exploring future gravitational wave missions}",
    eprint = "gr-qc/0506015",
    archivePrefix = "arXiv",
    doi = "10.1103/PhysRevD.72.083005",
    journal = "Phys. Rev. D",
    volume = "72",
    pages = "083005",
    year = "2005"
}

@article{Corbin:2005ny,
    author = "Corbin, Vincent and Cornish, Neil J.",
    title = "{Detecting the cosmic gravitational wave background with the big bang observer}",
    eprint = "gr-qc/0512039",
    archivePrefix = "arXiv",
    doi = "10.1088/0264-9381/23/7/014",
    journal = "Class. Quant. Grav.",
    volume = "23",
    pages = "2435--2446",
    year = "2006"
}

@article{Reitze:2019iox,
    author = "Reitze, David and others",
    title = "{Cosmic Explorer: The U.S. Contribution to Gravitational-Wave Astronomy beyond LIGO}",
    eprint = "1907.04833",
    archivePrefix = "arXiv",
    primaryClass = "astro-ph.IM",
    reportNumber = "LIGO-P1900316",
    journal = "Bull. Am. Astron. Soc.",
    volume = "51",
    number = "7",
    pages = "035",
    year = "2019"
}

@article{Maggiore:2019uih,
    author = "Maggiore, Michele and others",
    title = "{Science Case for the Einstein Telescope}",
    eprint = "1912.02622",
    archivePrefix = "arXiv",
    primaryClass = "astro-ph.CO",
    doi = "10.1088/1475-7516/2020/03/050",
    journal = "JCAP",
    volume = "03",
    pages = "050",
    year = "2020"
}

@article{Sathyaprakash:2012jk,
    author = "Sathyaprakash, B. and others",
    editor = "Hannam, Mark and Sutton, Patrick and Hild, Stefan and van den Broeck, Chris",
    title = "{Scientific Objectives of Einstein Telescope}",
    eprint = "1206.0331",
    archivePrefix = "arXiv",
    primaryClass = "gr-qc",
    doi = "10.1088/0264-9381/29/12/124013",
    journal = "Class. Quant. Grav.",
    volume = "29",
    pages = "124013",
    year = "2012",
    note = "[Erratum: Class.Quant.Grav. 30, 079501 (2013)]"
}

@article{Maggiore:1999vm,
    author = "Maggiore, Michele",
    title = "{Gravitational wave experiments and early universe cosmology}",
    eprint = "gr-qc/9909001",
    archivePrefix = "arXiv",
    reportNumber = "IFUP-TH-20-99",
    doi = "10.1016/S0370-1573(99)00102-7",
    journal = "Phys. Rept.",
    volume = "331",
    pages = "283--367",
    year = "2000"
}

@article{Garcia:2020wiy,
    author = "Garcia, Marcos A. G. and Kaneta, Kunio and Mambrini, Yann and Olive, Keith A.",
    title = "{Inflaton Oscillations and Post-Inflationary Reheating}",
    eprint = "2012.10756",
    archivePrefix = "arXiv",
    primaryClass = "hep-ph",
    reportNumber = "UMN-TH-4006/20, FTPI-MINN-20/37, IFT-UAM/CSIC-20-185, KIAS-P20071",
    doi = "10.1088/1475-7516/2021/04/012",
    journal = "JCAP",
    volume = "04",
    pages = "012",
    year = "2021"
}

@article{Kallosh:2013yoa,
    author = "Kallosh, Renata and Linde, Andrei and Roest, Diederik",
    title = "{Superconformal Inflationary $\alpha$-Attractors}",
    eprint = "1311.0472",
    archivePrefix = "arXiv",
    primaryClass = "hep-th",
    doi = "10.1007/JHEP11(2013)198",
    journal = "JHEP",
    volume = "11",
    pages = "198",
    year = "2013"
}

@article{Starobinsky:1980te,
    author = "Starobinsky, Alexei A.",
    editor = "Khalatnikov, I. M. and Mineev, V. P.",
    title = "{A New Type of Isotropic Cosmological Models Without Singularity}",
    doi = "10.1016/0370-2693(80)90670-X",
    journal = "Phys. Lett. B",
    volume = "91",
    pages = "99--102",
    year = "1980"
}

@inproceedings{Starobinsky:1981vz,
    author = "Starobinsky, Alexei A.",
    title = "{Nonsingular Model of the Universe with the Quantum Gravitational de Sitter Stage and its Observational Consequences}",
    booktitle = "{Second Seminar on Quantum Gravity}",
    year = "1981"
}

@article{Kofman:1985aw,
    author = "Kofman, L. A. and Linde, Andrei D. and Starobinsky, Alexei A.",
    title = "{Inflationary Universe Generated by the Combined Action of a Scalar Field and Gravitational Vacuum Polarization}",
    doi = "10.1016/0370-2693(85)90381-8",
    journal = "Phys. Lett. B",
    volume = "157",
    pages = "361--367",
    year = "1985"
}

@article{Starobinsky:1983zz,
    author = "Starobinsky, A. A.",
    title = "{The Perturbation Spectrum Evolving from a Nonsingular Initially De-Sitter Cosmology and the Microwave Background Anisotropy}",
    journal = "Sov. Astron. Lett.",
    volume = "9",
    pages = "302",
    year = "1983"
}

@article{Turner:1983he,
    author = "Turner, Michael S.",
    title = "{Coherent Scalar Field Oscillations in an Expanding Universe}",
    reportNumber = "EFI-83-29-CHICAGO",
    doi = "10.1103/PhysRevD.28.1243",
    journal = "Phys. Rev. D",
    volume = "28",
    pages = "1243",
    year = "1983"
}

@article{Peloso:2000hy,
    author = "Peloso, Marco and Sorbo, Lorenzo",
    title = "{Preheating of massive fermions after inflation: Analytical results}",
    eprint = "hep-ph/0003045",
    archivePrefix = "arXiv",
    doi = "10.1088/1126-6708/2000/05/016",
    journal = "JHEP",
    volume = "05",
    pages = "016",
    year = "2000"
}

@article{Opferkuch:2019zbd,
    author = "Opferkuch, Toby and Schwaller, Pedro and Stefanek, Ben A.",
    title = "{Ricci Reheating}",
    eprint = "1905.06823",
    archivePrefix = "arXiv",
    primaryClass = "gr-qc",
    reportNumber = "CERN-TH-2019-063, MITP/19-032",
    doi = "10.1088/1475-7516/2019/07/016",
    journal = "JCAP",
    volume = "07",
    pages = "016",
    year = "2019"
}

@article{Kallosh:2013maa,
    author = "Kallosh, Renata and Linde, Andrei",
    title = "{Non-minimal Inflationary Attractors}",
    eprint = "1307.7938",
    archivePrefix = "arXiv",
    primaryClass = "hep-th",
    doi = "10.1088/1475-7516/2013/10/033",
    journal = "JCAP",
    volume = "10",
    pages = "033",
    year = "2013"
}

@article{Buchmuller:2004nz,
    author = "Buchmuller, W. and Di Bari, P. and Plumacher, M.",
    title = "{Leptogenesis for pedestrians}",
    eprint = "hep-ph/0401240",
    archivePrefix = "arXiv",
    reportNumber = "DESY-03-100, UAB-FT-551, CERN-TH-2003-199",
    doi = "10.1016/j.aop.2004.02.003",
    journal = "Annals Phys.",
    volume = "315",
    pages = "305--351",
    year = "2005"
}

@article{LISA:2017pwj,
    author = "Amaro-Seoane, Pau and others",
    collaboration = "LISA",
    title = "{Laser Interferometer Space Antenna}",
    eprint = "1702.00786",
    archivePrefix = "arXiv",
    primaryClass = "astro-ph.IM",
    month = "2",
    year = "2017"
}

@article{Mohapatra:1979ia,
    author = "Mohapatra, Rabindra N. and Senjanovic, Goran",
    title = "{Neutrino Mass and Spontaneous Parity Nonconservation}",
    reportNumber = "MDDP-TR-80-060, MDDP-PP-80-105, CCNY-HEP-79-10",
    doi = "10.1103/PhysRevLett.44.912",
    journal = "Phys. Rev. Lett.",
    volume = "44",
    pages = "912",
    year = "1980"
}

@article{ParticleDataGroup:2020ssz,
    author = "Zyla, P. A. and others",
    collaboration = "Particle Data Group",
    title = "{Review of Particle Physics}",
    doi = "10.1093/ptep/ptaa104",
    journal = "PTEP",
    volume = "2020",
    number = "8",
    pages = "083C01",
    year = "2020"
}

@article{Amin:2010dc,
    author = "Amin, Mustafa A. and Easther, Richard and Finkel, Hal",
    title = "{Inflaton Fragmentation and Oscillon Formation in Three Dimensions}",
    eprint = "1009.2505",
    archivePrefix = "arXiv",
    primaryClass = "astro-ph.CO",
    doi = "10.1088/1475-7516/2010/12/001",
    journal = "JCAP",
    volume = "12",
    pages = "001",
    year = "2010"
}

@article{Lozanov:2019jxc,
    author = "Lozanov, Kaloian D.",
    title = "{Lectures on Reheating after Inflation}",
    eprint = "1907.04402",
    archivePrefix = "arXiv",
    primaryClass = "astro-ph.CO",
    month = "7",
    year = "2019"
}

@article{Garcia:2023dyf,
    author = "Garcia, Marcos A. G. and Gross, Mathieu and Mambrini, Yann and Olive, Keith A. and Pierre, Mathias and Yoon, Jong-Hyun",
    title = "{Effects of fragmentation on post-inflationary reheating}",
    eprint = "2308.16231",
    archivePrefix = "arXiv",
    primaryClass = "hep-ph",
    reportNumber = "UMN--TH--4223/23, FTPI--MINN--23/15, DESY-23-122",
    doi = "10.1088/1475-7516/2023/12/028",
    journal = "JCAP",
    volume = "12",
    pages = "028",
    year = "2023"
}

@article{Fukugita:1986hr,
    author = "Fukugita, M. and Yanagida, T.",
    title = "{Baryogenesis Without Grand Unification}",
    reportNumber = "RIFP-641",
    doi = "10.1016/0370-2693(86)91126-3",
    journal = "Phys. Lett. B",
    volume = "174",
    pages = "45--47",
    year = "1986"
}

@article{Kuzmin:1985mm,
    author = "Kuzmin, V. A. and Rubakov, V. A. and Shaposhnikov, M. E.",
    title = "{On the Anomalous Electroweak Baryon Number Nonconservation in the Early Universe}",
    reportNumber = "IC/85/8",
    doi = "10.1016/0370-2693(85)91028-7",
    journal = "Phys. Lett. B",
    volume = "155",
    pages = "36",
    year = "1985"
}

@article{Sakharov:1967dj,
    author = "Sakharov, A. D.",
    title = "{Violation of CP Invariance, C asymmetry, and baryon asymmetry of the universe}",
    doi = "10.1070/PU1991v034n05ABEH002497",
    journal = "Pisma Zh. Eksp. Teor. Fiz.",
    volume = "5",
    pages = "32--35",
    year = "1967"
}

@article{deSalas:2015glj,
    author = "de Salas, P.F. and Lattanzi, M. and Mangano, G. and Miele, G. and Pastor, S. and Pisanti, O.",
    title = "{Bounds on very low reheating scenarios after Planck}",
    eprint = "1511.00672",
    archivePrefix = "arXiv",
    primaryClass = "astro-ph.CO",
    reportNumber = "IFIC-15-70",
    doi = "10.1103/PhysRevD.92.123534",
    journal = "Phys. Rev. D",
    volume = "92",
    number = "12",
    pages = "123534",
    year = "2015"
}

@article{Kofman:1997yn,
    author = "Kofman, Lev and Linde, Andrei D. and Starobinsky, Alexei A.",
    title = "{Towards the theory of reheating after inflation}",
    eprint = "hep-ph/9704452",
    archivePrefix = "arXiv",
    reportNumber = "IFA-97-28, SU-ITP-97-18",
    doi = "10.1103/PhysRevD.56.3258",
    journal = "Phys. Rev. D",
    volume = "56",
    pages = "3258--3295",
    year = "1997"
}

@article{Bernal:2020ywq,
    author = "Bernal, Nicol\'as and Ghoshal, Anish and Hajkarim, Fazlollah and Lambiase, Gaetano",
    title = "{Primordial Gravitational Wave Signals in Modified Cosmologies}",
    eprint = "2008.04959",
    archivePrefix = "arXiv",
    primaryClass = "gr-qc",
    month = "8",
    year = "2020"
}

@article{Samanta:2020cdk,
    author = "Samanta, Rome and Datta, Satyabrata",
    title = "{Gravitational wave complementarity and impact of NANOGrav data on gravitational leptogenesis: cosmic strings}",
    eprint = "2009.13452",
    archivePrefix = "arXiv",
    primaryClass = "hep-ph",
    month = "9",
    year = "2020"
}

@article{Zyla:2020zbs,
    author = "Zyla, P. A. and others",
    collaboration = "Particle Data Group",
    title = "{Review of Particle Physics}",
    doi = "10.1093/ptep/ptaa104",
    journal = "PTEP",
    volume = "2020",
    number = "8",
    pages = "083C01",
    year = "2020"
}

@article{Schechter:1980gr,
      author         = "Schechter, J. and Valle, J. W. F.",
      title          = "{Neutrino Masses in SU(2) x U(1) Theories}",
      journal        = "Phys. Rev.",
      volume         = "D22",
      year           = "1980",
      pages          = "2227",
      doi            = "10.1103/PhysRevD.22.2227",
      reportNumber   = "SU-4217-167, COO-3533-167",
      SLACcitation   = "%%CITATION = PHRVA,D22,2227;%%"
}

@article{Schechter:1981cv,
    author = "Schechter, J. and Valle, J. W. F.",
    title = "{Neutrino Decay and Spontaneous Violation of Lepton Number}",
    reportNumber = "SU-4217-203, COO-3533-203",
    doi = "10.1103/PhysRevD.25.774",
    journal = "Phys. Rev. D",
    volume = "25",
    pages = "774",
    year = "1982"
}

@article{Yanagida:1979as,
      author         = "Yanagida, Tsutomu",
      title          = "{HORIZONTAL SYMMETRY AND MASSES OF NEUTRINOS}",
      booktitle      = "{Proceedings: Workshop on the Unified Theories and the
                        Baryon Number in the Universe: Tsukuba, Japan, February
                        13-14, 1979}",
      journal        = "Conf. Proc.",
      volume         = "C7902131",
      year           = "1979",
      pages          = "95-99",
      reportNumber   = "KEK-79-18-95",
      SLACcitation   = "%%CITATION = CONFP,C7902131,95;%%"
}

@article{GellMann:1980vs,
      author         = "Gell-Mann, Murray and Ramond, Pierre and Slansky,
                        Richard",
      title          = "{Complex Spinors and Unified Theories}",
      booktitle      = "{Supergravity Workshop Stony Brook, New York, September
                        27-28, 1979}",
      journal        = "Conf. Proc.",
      volume         = "C790927",
      year           = "1979",
      pages          = "315-321",
      eprint         = "1306.4669",
      archivePrefix  = "arXiv",
      primaryClass   = "hep-th",
      reportNumber   = "PRINT-80-0576",
      SLACcitation   = "%%CITATION = ARXIV:1306.4669;%%"
}

@article{Davidson:2002qv,
      author         = "Davidson, Sacha and Ibarra, Alejandro",
      title          = "{A Lower bound on the right-handed neutrino mass from
                        leptogenesis}",
      journal        = "Phys. Lett.",
      volume         = "B535",
      year           = "2002",
      pages          = "25-32",
      doi            = "10.1016/S0370-2693(02)01735-5",
      eprint         = "hep-ph/0202239",
      archivePrefix  = "arXiv",
      primaryClass   = "hep-ph",
      reportNumber   = "OUTP-02-10P, IPPP-02-16, DCPT-02-32",
      SLACcitation   = "%%CITATION = HEP-PH/0202239;%%"
}

@article{Pilaftsis:2003gt,
      author         = "Pilaftsis, Apostolos and Underwood, Thomas E. J.",
      title          = "{Resonant leptogenesis}",
      journal        = "Nucl. Phys.",
      volume         = "B692",
      year           = "2004",
      pages          = "303-345",
      doi            = "10.1016/j.nuclphysb.2004.05.029",
      eprint         = "hep-ph/0309342",
      archivePrefix  = "arXiv",
      primaryClass   = "hep-ph",
      reportNumber   = "MC-TH-2003-09",
      SLACcitation   = "%%CITATION = HEP-PH/0309342;%%"
}

@article{Lazarides:1991wu,
    author = "Lazarides, George and Shafi, Q.",
    title = "{Origin of matter in the inflationary cosmology}",
    reportNumber = "BA-90-78",
    doi = "10.1016/0370-2693(91)91090-I",
    journal = "Phys. Lett. B",
    volume = "258",
    pages = "305--309",
    year = "1991"
}

@article{Murayama:1992ua,
    author = "Murayama, H. and Suzuki, Hiroshi and Yanagida, T. and Yokoyama, Jun'ichi",
    title = "{Chaotic inflation and baryogenesis by right-handed sneutrinos}",
    reportNumber = "TU-423, YITP-U-92-28",
    doi = "10.1103/PhysRevLett.70.1912",
    journal = "Phys. Rev. Lett.",
    volume = "70",
    pages = "1912--1915",
    year = "1993"
}

@article{Kolb:1996jt,
    author = "Kolb, Edward W. and Linde, Andrei D. and Riotto, Antonio",
    title = "{GUT baryogenesis after preheating}",
    eprint = "hep-ph/9606260",
    archivePrefix = "arXiv",
    reportNumber = "FERMILAB-PUB-96-133-A, SU-ITP-96-22",
    doi = "10.1103/PhysRevLett.77.4290",
    journal = "Phys. Rev. Lett.",
    volume = "77",
    pages = "4290--4293",
    year = "1996"
}

@article{Giudice:1999fb,
    author = "Giudice, G.F. and Peloso, M. and Riotto, A. and Tkachev, I.",
    title = "{Production of massive fermions at preheating and leptogenesis}",
    eprint = "hep-ph/9905242",
    archivePrefix = "arXiv",
    reportNumber = "CERN-TH-99-117",
    doi = "10.1088/1126-6708/1999/08/014",
    journal = "JHEP",
    volume = "08",
    pages = "014",
    year = "1999"
}

@article{Asaka:1999yd,
    author = "Asaka, T. and Hamaguchi, Koichi and Kawasaki, M. and Yanagida, T.",
    title = "{Leptogenesis in inflaton decay}",
    eprint = "hep-ph/9906366",
    archivePrefix = "arXiv",
    reportNumber = "UT-853, RESCEU-15-99",
    doi = "10.1016/S0370-2693(99)01020-5",
    journal = "Phys. Lett. B",
    volume = "464",
    pages = "12--18",
    year = "1999"
}

@article{Asaka:1999jb,
    author = "Asaka, T. and Hamaguchi, Koichi and Kawasaki, M. and Yanagida, T.",
    title = "{Leptogenesis in inflationary universe}",
    eprint = "hep-ph/9907559",
    archivePrefix = "arXiv",
    reportNumber = "UT-855, RESCEU-28-99",
    doi = "10.1103/PhysRevD.61.083512",
    journal = "Phys. Rev. D",
    volume = "61",
    pages = "083512",
    year = "2000"
}

@article{Hamaguchi:2001gw,
    author = "Hamaguchi, Koichi and Murayama, Hitoshi and Yanagida, T.",
    title = "{Leptogenesis from N dominated early universe}",
    eprint = "hep-ph/0109030",
    archivePrefix = "arXiv",
    reportNumber = "UT-957, LBNL-48679, UCB-PTH-01-30",
    doi = "10.1103/PhysRevD.65.043512",
    journal = "Phys. Rev. D",
    volume = "65",
    pages = "043512",
    year = "2002"
}

@article{Casas:2001sr,
    author = "Casas, J. A. and Ibarra, A.",
    title = "{Oscillating neutrinos and $\mu \to e, \gamma$}",
    eprint = "hep-ph/0103065",
    archivePrefix = "arXiv",
    reportNumber = "IEM-FT-211-01, OUTP-01-11P, IFT-UAM-CSIC-01-08",
    doi = "10.1016/S0550-3213(01)00475-8",
    journal = "Nucl. Phys. B",
    volume = "618",
    pages = "171--204",
    year = "2001"
}

@article{Davidson:2008bu,
    author = "Davidson, Sacha and Nardi, Enrico and Nir, Yosef",
    title = "{Leptogenesis}",
    eprint = "0802.2962",
    archivePrefix = "arXiv",
    primaryClass = "hep-ph",
    doi = "10.1016/j.physrep.2008.06.002",
    journal = "Phys. Rept.",
    volume = "466",
    pages = "105--177",
    year = "2008"
}

@article{Planck:2018jri,
    author = "Akrami, Y. and others",
    collaboration = "Planck",
    title = "{Planck 2018 results. X. Constraints on inflation}",
    eprint = "1807.06211",
    archivePrefix = "arXiv",
    primaryClass = "astro-ph.CO",
    doi = "10.1051/0004-6361/201833887",
    journal = "Astron. Astrophys.",
    volume = "641",
    pages = "A10",
    year = "2020"
}

@article{Barman:2021tgt,
    author = "Barman, Basabendu and Borah, Debasish and Roshan, Rishav",
    title = "{Nonthermal leptogenesis and UV freeze-in of dark matter: Impact of inflationary reheating}",
    eprint = "2103.01675",
    archivePrefix = "arXiv",
    primaryClass = "hep-ph",
    doi = "10.1103/PhysRevD.104.035022",
    journal = "Phys. Rev. D",
    volume = "104",
    number = "3",
    pages = "035022",
    year = "2021"
}

@article{Hahn-Woernle:2008tsk,
    author = "Hahn-Woernle, F. and Plumacher, M.",
    title = "{Effects of reheating on leptogenesis}",
    eprint = "0801.3972",
    archivePrefix = "arXiv",
    primaryClass = "hep-ph",
    doi = "10.1016/j.nuclphysb.2008.07.032",
    journal = "Nucl. Phys. B",
    volume = "806",
    pages = "68--83",
    year = "2009"
}

@article{Barman:2021ost,
    author = "Barman, Basabendu and Borah, Debasish and Das, Suruj Jyoti and Roshan, Rishav",
    title = "{Non-thermal origin of asymmetric dark matter from inflaton and primordial black holes}",
    eprint = "2111.08034",
    archivePrefix = "arXiv",
    primaryClass = "hep-ph",
    reportNumber = "PI/UAN-2021-705FT",
    doi = "10.1088/1475-7516/2022/03/031",
    journal = "JCAP",
    volume = "03",
    number = "03",
    pages = "031",
    year = "2022"
}

@article{Dror:2019syi,
    author = "Dror, Jeff A. and Hiramatsu, Takashi and Kohri, Kazunori and Murayama, Hitoshi and White, Graham",
    title = "{Testing the Seesaw Mechanism and Leptogenesis with Gravitational Waves}",
    eprint = "1908.03227",
    archivePrefix = "arXiv",
    primaryClass = "hep-ph",
    reportNumber = "IPMU19-0108, DESY-19-138, DESY 19-138, KEK-TH-2147, KEK-Cosmo-241",
    doi = "10.1103/PhysRevLett.124.041804",
    journal = "Phys. Rev. Lett.",
    volume = "124",
    number = "4",
    pages = "041804",
    year = "2020"
}

@article{Ichikawa:2008ne,
    author = "Ichikawa, Kazuhide and Suyama, Teruaki and Takahashi, Tomo and Yamaguchi, Masahide",
    title = "{Primordial Curvature Fluctuation and Its Non-Gaussianity in Models with Modulated Reheating}",
    eprint = "0807.3988",
    archivePrefix = "arXiv",
    primaryClass = "astro-ph",
    doi = "10.1103/PhysRevD.78.063545",
    journal = "Phys. Rev. D",
    volume = "78",
    pages = "063545",
    year = "2008"
}

@article{Sesana:2019vho,
    author = "Sesana, Alberto and others",
    title = "{Unveiling the gravitational universe at $\mu$-Hz frequencies}",
    eprint = "1908.11391",
    archivePrefix = "arXiv",
    primaryClass = "astro-ph.IM",
    doi = "10.1007/s10686-021-09709-9",
    journal = "Exper. Astron.",
    volume = "51",
    number = "3",
    pages = "1333--1383",
    year = "2021"
}

@article{Drewes:2017fmn,
    author = "Drewes, Marco and Kang, Jin U and Mun, Ui Ri",
    title = "{CMB constraints on the inflaton couplings and reheating temperature in $\alpha$-attractor inflation}",
    eprint = "1708.01197",
    archivePrefix = "arXiv",
    primaryClass = "astro-ph.CO",
    reportNumber = "TUM-HEP-1093-17",
    doi = "10.1007/JHEP11(2017)072",
    journal = "JHEP",
    volume = "11",
    pages = "072",
    year = "2017"
}

@article{Drewes:2019rxn,
    author = "Drewes, Marco",
    title = "{Measuring the inflaton coupling in~the~CMB}",
    eprint = "1903.09599",
    archivePrefix = "arXiv",
    primaryClass = "astro-ph.CO",
    reportNumber = "CP3-19-12",
    doi = "10.1088/1475-7516/2022/09/069",
    journal = "JCAP",
    volume = "09",
    pages = "069",
    year = "2022"
}

@article{Giovannini:1998bp,
    author = "Giovannini, Massimo",
    title = "{Gravitational waves constraints on postinflationary phases stiffer than radiation}",
    eprint = "hep-ph/9806329",
    archivePrefix = "arXiv",
    doi = "10.1103/PhysRevD.58.083504",
    journal = "Phys. Rev. D",
    volume = "58",
    pages = "083504",
    year = "1998"
}

@article{Hild:2010id,
    author = "Hild, S. and others",
    title = "{Sensitivity Studies for Third-Generation Gravitational Wave Observatories}",
    eprint = "1012.0908",
    archivePrefix = "arXiv",
    primaryClass = "gr-qc",
    doi = "10.1088/0264-9381/28/9/094013",
    journal = "Class. Quant. Grav.",
    volume = "28",
    pages = "094013",
    year = "2011"
}

@article{Sarkar:1995dd,
    author = "Sarkar, Subir",
    title = "{Big bang nucleosynthesis and physics beyond the standard model}",
    eprint = "hep-ph/9602260",
    archivePrefix = "arXiv",
    reportNumber = "OUTP-95-16-P",
    doi = "10.1088/0034-4885/59/12/001",
    journal = "Rept. Prog. Phys.",
    volume = "59",
    pages = "1493--1610",
    year = "1996"
}

@article{Hannestad:2004px,
    author = "Hannestad, Steen",
    title = "{What is the lowest possible reheating temperature?}",
    eprint = "astro-ph/0403291",
    archivePrefix = "arXiv",
    doi = "10.1103/PhysRevD.70.043506",
    journal = "Phys. Rev. D",
    volume = "70",
    pages = "043506",
    year = "2004"
}

@article{DeBernardis:2008zz,
    author = "De Bernardis, Francesco and Pagano, Luca and Melchiorri, Alessandro",
    title = "{New constraints on the reheating temperature of the universe after WMAP-5}",
    doi = "10.1016/j.astropartphys.2008.09.005",
    journal = "Astropart. Phys.",
    volume = "30",
    pages = "192--195",
    year = "2008"
}

@article{Planck:2018vyg,
    author = "Aghanim, N. and others",
    collaboration = "Planck",
    title = "{Planck 2018 results. VI. Cosmological parameters}",
    eprint = "1807.06209",
    archivePrefix = "arXiv",
    primaryClass = "astro-ph.CO",
    doi = "10.1051/0004-6361/201833910",
    journal = "Astron. Astrophys.",
    volume = "641",
    pages = "A6",
    year = "2020",
    note = "[Erratum: Astron.Astrophys. 652, C4 (2021)]"
}

@article{Giovannini:1999bh,
    author = "Giovannini, Massimo",
    title = "{Production and detection of relic gravitons in quintessential inflationary models}",
    eprint = "astro-ph/9903004",
    archivePrefix = "arXiv",
    reportNumber = "TUPT-01-99",
    doi = "10.1103/PhysRevD.60.123511",
    journal = "Phys. Rev. D",
    volume = "60",
    pages = "123511",
    year = "1999"
}

@article{Riazuelo:2000fc,
    author = "Riazuelo, Alain and Uzan, Jean-Philippe",
    title = "{Quintessence and gravitational waves}",
    eprint = "astro-ph/0004156",
    archivePrefix = "arXiv",
    doi = "10.1103/PhysRevD.62.083506",
    journal = "Phys. Rev. D",
    volume = "62",
    pages = "083506",
    year = "2000"
}

@article{Artymowski:2017pua,
    author = "Artymowski, Michal and Czerwinska, Olga and Lalak, Zygmunt and Lewicki, Marek",
    title = "{Gravitational wave signals and cosmological consequences of gravitational reheating}",
    eprint = "1711.08473",
    archivePrefix = "arXiv",
    primaryClass = "astro-ph.CO",
    reportNumber = "KCL-PH-TH-2017-59",
    doi = "10.1088/1475-7516/2018/04/046",
    journal = "JCAP",
    volume = "04",
    pages = "046",
    year = "2018"
}

@article{Boyle:2007zx,
    author = "Boyle, Latham A. and Buonanno, Alessandra",
    title = "{Relating gravitational wave constraints from primordial nucleosynthesis, pulsar timing, laser interferometers, and the CMB: Implications for the early Universe}",
    eprint = "0708.2279",
    archivePrefix = "arXiv",
    primaryClass = "astro-ph",
    doi = "10.1103/PhysRevD.78.043531",
    journal = "Phys. Rev. D",
    volume = "78",
    pages = "043531",
    year = "2008"
}

@article{Stewart:2007fu,
    author = "Stewart, Andrew and Brandenberger, Robert",
    title = "{Observational Constraints on Theories with a Blue Spectrum of Tensor Modes}",
    eprint = "0711.4602",
    archivePrefix = "arXiv",
    primaryClass = "astro-ph",
    doi = "10.1088/1475-7516/2008/08/012",
    journal = "JCAP",
    volume = "08",
    pages = "012",
    year = "2008"
}

@article{Figueroa:2019paj,
    author = "Figueroa, Daniel G. and Tanin, Erwin H.",
    title = "{Ability of LIGO and LISA to probe the equation of state of the early Universe}",
    eprint = "1905.11960",
    archivePrefix = "arXiv",
    primaryClass = "astro-ph.CO",
    doi = "10.1088/1475-7516/2019/08/011",
    journal = "JCAP",
    volume = "08",
    pages = "011",
    year = "2019"
}

@article{Ghoshal:2022ruy,
    author = "Ghoshal, Anish and Heurtier, Lucien and Paul, Arnab",
    title = "{Signatures of non-thermal dark matter with kination and early matter domination. Gravitational waves versus laboratory searches}",
    eprint = "2208.01670",
    archivePrefix = "arXiv",
    primaryClass = "hep-ph",
    reportNumber = "IPPP/22/54",
    doi = "10.1007/JHEP12(2022)105",
    journal = "JHEP",
    volume = "12",
    pages = "105",
    year = "2022"
}

@article{Caldwell:2022qsj,
    author = "Caldwell, Robert and others",
    title = "{Detection of early-universe gravitational-wave signatures and fundamental physics}",
    eprint = "2203.07972",
    archivePrefix = "arXiv",
    primaryClass = "gr-qc",
    doi = "10.1007/s10714-022-03027-x",
    journal = "Gen. Rel. Grav.",
    volume = "54",
    number = "12",
    pages = "156",
    year = "2022"
}

@article{Gouttenoire:2021jhk,
    author = "Gouttenoire, Yann and Servant, Geraldine and Simakachorn, Peera",
    title = "{Kination cosmology from scalar fields and gravitational-wave signatures}",
    eprint = "2111.01150",
    archivePrefix = "arXiv",
    primaryClass = "hep-ph",
    reportNumber = "DESY 21-134",
    month = "11",
    year = "2021"
}

@article{Seto:2003kc,
    author = "Seto, Naoki and Yokoyama, Jun'Ichi",
    title = "{Probing the equation of state of the early universe with a space laser interferometer}",
    eprint = "gr-qc/0305096",
    archivePrefix = "arXiv",
    reportNumber = "OU-TAP-206",
    doi = "10.1143/JPSJ.72.3082",
    journal = "J. Phys. Soc. Jap.",
    volume = "72",
    pages = "3082--3086",
    year = "2003"
}

@article{Li:2021htg,
    author = "Li, Bohua and Shapiro, Paul R.",
    title = "{Precision cosmology and the stiff-amplified gravitational-wave background from inflation: NANOGrav, Advanced LIGO-Virgo and the Hubble tension}",
    eprint = "2107.12229",
    archivePrefix = "arXiv",
    primaryClass = "astro-ph.CO",
    doi = "10.1088/1475-7516/2021/10/024",
    journal = "JCAP",
    volume = "10",
    pages = "024",
    year = "2021"
}

@article{Shtanov:1994ce,
    author = "Shtanov, Y. and Traschen, Jennie H. and Brandenberger, Robert H.",
    title = "{Universe reheating after inflation}",
    eprint = "hep-ph/9407247",
    archivePrefix = "arXiv",
    reportNumber = "BROWN-HET-957",
    doi = "10.1103/PhysRevD.51.5438",
    journal = "Phys. Rev. D",
    volume = "51",
    pages = "5438--5455",
    year = "1995"
}

@article{Bettoni:2018pbl,
    author = "Bettoni, Dario and Dom\`enech, Guillem and Rubio, Javier",
    title = "{Gravitational waves from global cosmic strings in quintessential inflation}",
    eprint = "1810.11117",
    archivePrefix = "arXiv",
    primaryClass = "astro-ph.CO",
    reportNumber = "HIP-2018-21/TH",
    doi = "10.1088/1475-7516/2019/02/034",
    journal = "JCAP",
    volume = "02",
    pages = "034",
    year = "2019"
}

@article{Hasegawa:2019jsa,
    author = "Hasegawa, Takuya and Hiroshima, Nagisa and Kohri, Kazunori and Hansen, Rasmus S. L. and Tram, Thomas and Hannestad, Steen",
    title = "{MeV-scale reheating temperature and thermalization of oscillating neutrinos by radiative and hadronic decays of massive particles}",
    eprint = "1908.10189",
    archivePrefix = "arXiv",
    primaryClass = "hep-ph",
    reportNumber = "KEK-TH-2149, KEK-Cosmo-242, RIKEN-iTHEMS-Report-19, IPMU19-0120",
    doi = "10.1088/1475-7516/2019/12/012",
    journal = "JCAP",
    volume = "12",
    pages = "012",
    year = "2019"
}

@article{Garcia:2023eol,
    author = "Garcia, Marcos A. G. and Pierre, Mathias",
    title = "{Reheating after Inflaton Fragmentation}",
    eprint = "2306.08038",
    archivePrefix = "arXiv",
    primaryClass = "hep-ph",
    reportNumber = "DESY-23-075",
    month = "6",
    year = "2023"
}

@article{Haque:2021dha,
    author = "Haque, Md Riajul and Maity, Debaprasad and Paul, Tanmoy and Sriramkumar, L.",
    title = "{Decoding the phases of early and late time reheating through imprints on primordial gravitational waves}",
    eprint = "2105.09242",
    archivePrefix = "arXiv",
    primaryClass = "astro-ph.CO",
    doi = "10.1103/PhysRevD.104.063513",
    journal = "Phys. Rev. D",
    volume = "104",
    number = "6",
    pages = "063513",
    year = "2021"
}

@article{Datta:2022jic,
    author = "Datta, Arghyajit and Roshan, Rishav and Sil, Arunansu",
    title = "{Effects of Reheating on Charged Lepton Yukawa Equilibration and Leptogenesis}",
    eprint = "2206.10650",
    archivePrefix = "arXiv",
    primaryClass = "hep-ph",
    doi = "10.1103/PhysRevLett.132.061802",
    journal = "Phys. Rev. Lett.",
    volume = "132",
    number = "6",
    pages = "061802",
    year = "2024"
}

@article{Datta:2023pav,
    author = "Datta, Arghyajit and Roshan, Rishav and Sil, Arunansu",
    title = "{Flavor leptogenesis during the reheating era}",
    eprint = "2301.10791",
    archivePrefix = "arXiv",
    primaryClass = "hep-ph",
    doi = "10.1103/PhysRevD.108.035029",
    journal = "Phys. Rev. D",
    volume = "108",
    number = "3",
    pages = "035029",
    year = "2023"
}

@article{Barman:2024slw,
    author = "Barman, Basabendu and Jyoti Das, Suruj and Haque, Md Riajul and Mambrini, Yann",
    title = "{Leptogenesis, primordial gravitational waves, and PBH-induced reheating}",
    eprint = "2403.05626",
    archivePrefix = "arXiv",
    primaryClass = "hep-ph",
    reportNumber = "CTPU-PTC-24-07",
    doi = "10.1103/PhysRevD.110.043528",
    journal = "Phys. Rev. D",
    volume = "110",
    number = "4",
    pages = "043528",
    year = "2024"
}

@article{Barman:2022gjo,
    author = "Barman, Basabendu and Borah, Debasish and Das Jyoti, Suruj and Roshan, Rishav",
    title = "{Cogenesis of Baryon Asymmetry and Gravitational Dark Matter from PBH}",
    eprint = "2204.10339",
    archivePrefix = "arXiv",
    primaryClass = "hep-ph",
    reportNumber = "PI/UAN-2022-714FT",
    month = "4",
    year = "2022"
}

@article{Lazarides:2022ezc,
    author = "Lazarides, George and Maji, Rinku and Roshan, Rishav and Shafi, Qaisar",
    title = "{A predictive SO(10) model}",
    eprint = "2210.03710",
    archivePrefix = "arXiv",
    primaryClass = "hep-ph",
    doi = "10.1088/1475-7516/2022/12/009",
    journal = "JCAP",
    volume = "12",
    pages = "009",
    year = "2022"
}

@article{Minkowski:1977sc,
    author = "Minkowski, Peter",
    title = "{$\mu \to e\gamma$ at a Rate of One Out of $10^{9}$ Muon Decays?}",
    reportNumber = "Print-77-0182 (BERN)",
    doi = "10.1016/0370-2693(77)90435-X",
    journal = "Phys. Lett. B",
    volume = "67",
    pages = "421--428",
    year = "1977"
}

@article{Barman:2024jqh,
    author = "Barman, Basabendu and Datta, Arghyajit and Haque, Md Riajul",
    title = "{Is leptogenesis during gravitational reheating flavourful?}",
    eprint = "2410.16381",
    archivePrefix = "arXiv",
    primaryClass = "hep-ph",
    month = "10",
    year = "2024"
}

@article{Domcke:2020quw,
    author = "Domcke, Valerie and Kamada, Kohei and Mukaida, Kyohei and Schmitz, Kai and Yamada, Masaki",
    title = "{Wash-In Leptogenesis}",
    eprint = "2011.09347",
    archivePrefix = "arXiv",
    primaryClass = "hep-ph",
    reportNumber = "CERN-TH-2020-196, RESCEU-22/20, DESY-20-202, TU-1112",
    doi = "10.1103/PhysRevLett.126.201802",
    journal = "Phys. Rev. Lett.",
    volume = "126",
    number = "20",
    pages = "201802",
    year = "2021"
}

@article{Maleknejad:2020pec,
    author = "Maleknejad, Azadeh",
    title = "{Chiral anomaly in SU(2)$_{R}$-axion inflation and the new prediction for particle cosmology}",
    eprint = "2103.14611",
    archivePrefix = "arXiv",
    primaryClass = "hep-ph",
    reportNumber = "CERN-TH-2021-034",
    doi = "10.1007/JHEP06(2021)113",
    journal = "JHEP",
    volume = "21",
    pages = "113",
    year = "2020"
}

@article{Hamada:2018epb,
    author = "Hamada, Yuta and Kitano, Ryuichiro and Yin, Wen",
    title = "{Leptogenesis via Neutrino Oscillation Magic}",
    eprint = "1807.06582",
    archivePrefix = "arXiv",
    primaryClass = "hep-ph",
    reportNumber = "KEK-TH-2064",
    doi = "10.1007/JHEP10(2018)178",
    journal = "JHEP",
    volume = "10",
    pages = "178",
    year = "2018"
}

@article{Eijima:2019hey,
    author = "Eijima, Shintaro and Kitano, Ryuichiro and Yin, Wen",
    title = "{Throwing away antimatter via neutrino oscillations during the reheating era}",
    eprint = "1908.11864",
    archivePrefix = "arXiv",
    primaryClass = "hep-ph",
    reportNumber = "KEK-TH-2153",
    doi = "10.1088/1475-7516/2020/03/048",
    journal = "JCAP",
    volume = "03",
    pages = "048",
    year = "2020"
}

@article{Kallosh:2015lwa,
    author = "Kallosh, Renata and Linde, Andrei",
    title = "{Planck, LHC, and $\alpha$-attractors}",
    eprint = "1502.07733",
    archivePrefix = "arXiv",
    primaryClass = "astro-ph.CO",
    doi = "10.1103/PhysRevD.91.083528",
    journal = "Phys. Rev. D",
    volume = "91",
    pages = "083528",
    year = "2015"
}

@article{DOnofrio:2014rug,
    author = "D'Onofrio, Michela and Rummukainen, Kari and Tranberg, Anders",
    title = "{Sphaleron Rate in the Minimal Standard Model}",
    eprint = "1404.3565",
    archivePrefix = "arXiv",
    primaryClass = "hep-ph",
    doi = "10.1103/PhysRevLett.113.141602",
    journal = "Phys. Rev. Lett.",
    volume = "113",
    number = "14",
    pages = "141602",
    year = "2014"
}

@article{Maity:2024cpq,
    author = "Maity, Suvashis and Haque, Md Riajul",
    title = "{Probing the early universe with future GW observatories}",
    eprint = "2407.18246",
    archivePrefix = "arXiv",
    primaryClass = "astro-ph.CO",
    doi = "10.1088/1475-7516/2025/04/091",
    journal = "JCAP",
    volume = "04",
    pages = "091",
    year = "2025"
}

@article{Barman:2024ujh,
    author = "Barman, Basabendu and Basu, Arindam and Borah, Debasish and Chakraborty, Amit and Roshan, Rishav",
    title = "{Testing leptogenesis and dark matter production during reheating with primordial gravitational waves}",
    eprint = "2410.19048",
    archivePrefix = "arXiv",
    primaryClass = "hep-ph",
    doi = "10.1103/PhysRevD.111.055016",
    journal = "Phys. Rev. D",
    volume = "111",
    number = "5",
    pages = "055016",
    year = "2025"
}

@article{Barman:2022yos,
    author = "Barman, Basabendu and Borah, Debasish and Dasgupta, Arnab and Ghoshal, Anish",
    title = "{Probing high scale Dirac leptogenesis via gravitational waves from domain walls}",
    eprint = "2205.03422",
    archivePrefix = "arXiv",
    primaryClass = "hep-ph",
    doi = "10.1103/PhysRevD.106.015007",
    journal = "Phys. Rev. D",
    volume = "106",
    number = "1",
    pages = "015007",
    year = "2022"
}

@article{Borboruah:2024eha,
    author = "Borboruah, Zafri A. and Ghoshal, Anish and Malhotra, Lekhika and Yajnik, Urjit",
    title = "{Inflationary gravitational wave spectral shapes as a test for low-scale leptogenesis}",
    eprint = "2405.06603",
    archivePrefix = "arXiv",
    primaryClass = "hep-ph",
    doi = "10.1103/hbd6-4qp7",
    journal = "Phys. Rev. D",
    volume = "112",
    number = "9",
    pages = "095017",
    year = "2025"
}

@article{Borboruah:2025hai,
    author = "Borboruah, Zafri A. and Deppisch, Frank F. and Ghoshal, Anish and Malhotra, Lekhika",
    title = "{Inflationary gravitational waves and laboratory searches as complementary probes of right-handed neutrinos}",
    eprint = "2504.15374",
    archivePrefix = "arXiv",
    primaryClass = "hep-ph",
    doi = "10.1103/qyld-mf33",
    journal = "Phys. Rev. D",
    volume = "112",
    number = "5",
    pages = "056003",
    year = "2025"
}

@article{Berbig:2023yyy,
    author = "Berbig, Maximilian and Ghoshal, Anish",
    title = "{Impact of high-scale Seesaw and Leptogenesis on inflationary tensor perturbations as detectable gravitational waves}",
    eprint = "2301.05672",
    archivePrefix = "arXiv",
    primaryClass = "hep-ph",
    doi = "10.1007/JHEP05(2023)172",
    journal = "JHEP",
    volume = "05",
    pages = "172",
    year = "2023"
}

@article{Datta:2025vyu,
    author = "Datta, Satyabrata and Ghoshal, Anish and Spalding, Angus and White, Graham",
    title = "{Gravitational wave spectral shapes as a probe of long lived right-handed neutrinos, leptogenesis and dark matter. Global versus local B {\ensuremath{-}} L cosmic strings}",
    eprint = "2511.01779",
    archivePrefix = "arXiv",
    primaryClass = "astro-ph.CO",
    doi = "10.1007/JHEP03(2026)245",
    journal = "JHEP",
    volume = "03",
    pages = "245",
    year = "2026"
}

@article{Datta:2024tne,
    author = "Datta, Arghyajit and Sil, Arunansu",
    title = "{Probing Leptogenesis through Gravitational Waves}",
    eprint = "2410.01900",
    archivePrefix = "arXiv",
    primaryClass = "hep-ph",
    month = "10",
    year = "2024"
}

@article{Chianese:2024gee,
    author = "Chianese, Marco and Datta, Satyabrata and Miele, Gennaro and Samanta, Rome and Saviano, Ninetta",
    title = "{Probing flavored regimes of leptogenesis with gravitational waves from cosmic strings}",
    eprint = "2406.01231",
    archivePrefix = "arXiv",
    primaryClass = "hep-ph",
    doi = "10.1103/PhysRevD.111.L041305",
    journal = "Phys. Rev. D",
    volume = "111",
    number = "4",
    pages = "L041305",
    year = "2025"
}

@article{Blasi:2020wpy,
    author = "Blasi, Simone and Brdar, Vedran and Schmitz, Kai",
    title = "{Fingerprint of low-scale leptogenesis in the primordial gravitational-wave spectrum}",
    eprint = "2004.02889",
    archivePrefix = "arXiv",
    primaryClass = "hep-ph",
    reportNumber = "CERN-TH-2020-055",
    doi = "10.1103/PhysRevResearch.2.043321",
    journal = "Phys. Rev. Res.",
    volume = "2",
    number = "4",
    pages = "043321",
    year = "2020"
}

@article{Borah:2025bfa,
    author = "Borah, Debasish and Saha, Indrajit",
    title = "{Gravitational waves from seesaw assisted collapsing domain walls}",
    eprint = "2512.22339",
    archivePrefix = "arXiv",
    primaryClass = "hep-ph",
    month = "12",
    year = "2025"
}

@article{Greene:1998nh,
    author = "Greene, Patrick B. and Kofman, Lev",
    title = "{Preheating of fermions}",
    eprint = "hep-ph/9807339",
    archivePrefix = "arXiv",
    reportNumber = "UH-IFA-98-44",
    doi = "10.1016/S0370-2693(99)00020-9",
    journal = "Phys. Lett. B",
    volume = "448",
    pages = "6--12",
    year = "1999"
}

@article{Kanemura:2025rct,
    author = "Kanemura, Shinya and Kaneta, Kunio and Nanda, Dibyendu",
    title = "{Gravitational waves from supermassive right-handed neutrinos produced at preheating}",
    eprint = "2508.00315",
    archivePrefix = "arXiv",
    primaryClass = "hep-ph",
    reportNumber = "OU-HET 1281",
    doi = "10.1103/rr5v-3jvg",
    journal = "Phys. Rev. D",
    volume = "113",
    number = "5",
    pages = "055046",
    year = "2026"
}

@article{Nardi:2005hs,
    author = "Nardi, Enrico and Nir, Yosef and Racker, Juan and Roulet, Esteban",
    title = "{On Higgs and sphaleron effects during the leptogenesis era}",
    eprint = "hep-ph/0512052",
    archivePrefix = "arXiv",
    doi = "10.1088/1126-6708/2006/01/068",
    journal = "JHEP",
    volume = "01",
    pages = "068",
    year = "2006"
}

@article{Nardi:2006fx,
    author = "Nardi, Enrico and Nir, Yosef and Roulet, Esteban and Racker, Juan",
    title = "{The Importance of flavor in leptogenesis}",
    eprint = "hep-ph/0601084",
    archivePrefix = "arXiv",
    doi = "10.1088/1126-6708/2006/01/164",
    journal = "JHEP",
    volume = "01",
    pages = "164",
    year = "2006"
}

@article{DeSimone:2006nrs,
    author = "De Simone, Andrea and Riotto, Antonio",
    title = "{On the impact of flavour oscillations in leptogenesis}",
    eprint = "hep-ph/0611357",
    archivePrefix = "arXiv",
    reportNumber = "CERN-PH-TH-2006-242, MIT-CTP-3789",
    doi = "10.1088/1475-7516/2007/02/005",
    journal = "JCAP",
    volume = "02",
    pages = "005",
    year = "2007"
}

@article{Blanchet:2011xq,
    author = "Blanchet, Steve and Di Bari, Pasquale and Jones, David A. and Marzola, Luca",
    title = "{Leptogenesis with heavy neutrino flavours: from density matrix to Boltzmann equations}",
    eprint = "1112.4528",
    archivePrefix = "arXiv",
    primaryClass = "hep-ph",
    doi = "10.1088/1475-7516/2013/01/041",
    journal = "JCAP",
    volume = "01",
    pages = "041",
    year = "2013"
}

@article{Moffat:2018wke,
    author = "Moffat, K. and Pascoli, S. and Petcov, S. T. and Schulz, H. and Turner, J.",
    title = "{Three-flavored nonresonant leptogenesis at intermediate scales}",
    eprint = "1804.05066",
    archivePrefix = "arXiv",
    primaryClass = "hep-ph",
    reportNumber = "IPPP-18-25, SISSA-17-2018-FISI, IPPP/18/25, FERMILAB-PUB-18-100-T, IPMU18-0062, SISSA 17/2018/FISI",
    doi = "10.1103/PhysRevD.98.015036",
    journal = "Phys. Rev. D",
    volume = "98",
    number = "1",
    pages = "015036",
    year = "2018"
}

@article{Granelli:2021fyc,
    author = "Granelli, A. and Moffat, K. and Petcov, S. T.",
    title = "{Aspects of high scale leptogenesis with low-energy leptonic CP violation}",
    eprint = "2107.02079",
    archivePrefix = "arXiv",
    primaryClass = "hep-ph",
    doi = "10.1007/JHEP11(2021)149",
    journal = "JHEP",
    volume = "11",
    pages = "149",
    year = "2021"
}

@article{Blanchet:2006be,
    author = "Blanchet, Steve and Di Bari, Pasquale",
    title = "{Flavor effects on leptogenesis predictions}",
    eprint = "hep-ph/0607330",
    archivePrefix = "arXiv",
    doi = "10.1088/1475-7516/2007/03/018",
    journal = "JCAP",
    volume = "03",
    pages = "018",
    year = "2007"
}

@article{Blanchet:2006ch,
    author = "Blanchet, S. and Di Bari, P. and Raffelt, G. G.",
    title = "{Quantum Zeno effect and the impact of flavor in leptogenesis}",
    eprint = "hep-ph/0611337",
    archivePrefix = "arXiv",
    reportNumber = "MPP-2006-155",
    doi = "10.1088/1475-7516/2007/03/012",
    journal = "JCAP",
    volume = "03",
    pages = "012",
    year = "2007"
}

@inproceedings{Gell-Mann:1979ijt,
    author = "Gell{\nobreakdash-}Mann, Murray and Ramond, Pierre and Slansky, Richard",
    title = "{The Family Group in Grand Unified Theories}",
    booktitle = "{International Symposium on Fundamentals of Quantum Theory and Quantum Field Theory}",
    eprint = "hep-ph/9809459",
    archivePrefix = "arXiv",
    reportNumber = "CALT-68-709",
    month = "2",
    year = "1979"
}

@article{Chowdhury:2026zox,
    author = "Chowdhury, Tammi and Jenks, Leah and Kolb, Edward W. and Long, Andrew J. and McDonough, Evan",
    title = "{Nonthermal leptogenesis via cosmological gravitational particle production is tested by inflationary gravitational waves}",
    eprint = "2605.05304",
    archivePrefix = "arXiv",
    primaryClass = "hep-ph",
    month = "5",
    year = "2026"
}

@article{Bhandari:2025ufp,
    author = "Bhandari, Dipendu and Sil, Arunansu",
    title = "{Exploring leptogenesis in the era of first order electroweak phase transition}",
    eprint = "2504.03837",
    archivePrefix = "arXiv",
    primaryClass = "hep-ph",
    doi = "10.1103/lp9n-97ns",
    journal = "Phys. Rev. D",
    volume = "113",
    number = "9",
    pages = "L091702",
    year = "2026"
}

@article{AtacamaCosmologyTelescope:2025blo,
    author = "Louis, Thibaut and others",
    collaboration = "Atacama Cosmology Telescope",
    title = "{The Atacama Cosmology Telescope: DR6 power spectra, likelihoods and {\ensuremath{\Lambda}}CDM parameters}",
    eprint = "2503.14452",
    archivePrefix = "arXiv",
    primaryClass = "astro-ph.CO",
    reportNumber = "FERMILAB-PUB-25-0071-PPD",
    doi = "10.1088/1475-7516/2025/11/062",
    journal = "JCAP",
    volume = "11",
    pages = "062",
    year = "2025"
}

@article{Garcia:2026ulw,
    author = "Garcia, Marcos A. G. and Henrich, Stephen E. and Ke, Wenqi and Olive, Keith A.",
    title = "{Leptogenesis and Low Reheating Temperatures}",
    eprint = "2607.08663",
    archivePrefix = "arXiv",
    primaryClass = "hep-ph",
    reportNumber = "UMN--TH--4534/26, FTPI--MINN--26/14",
    month = "7",
    year = "2026"
}

@inproceedings{Allen:1996vm,
    author = "Allen, Bruce",
    title = "{The Stochastic gravity wave background: Sources and detection}",
    booktitle = "{Les Houches School of Physics: Astrophysical Sources of Gravitational Radiation}",
    eprint = "gr-qc/9604033",
    archivePrefix = "arXiv",
    reportNumber = "WISC-MILW-96-TH-22",
    pages = "373--417",
    month = "4",
    year = "1996"
}

@article{Allen:1997ad,
    author = "Allen, Bruce and Romano, Joseph D.",
    title = "{Detecting a stochastic background of gravitational radiation: Signal processing strategies and sensitivities}",
    eprint = "gr-qc/9710117",
    archivePrefix = "arXiv",
    reportNumber = "WISC-MILW-97-TH-14",
    doi = "10.1103/PhysRevD.59.102001",
    journal = "Phys. Rev. D",
    volume = "59",
    pages = "102001",
    year = "1999"
}

@article{Haque:2023zhb,
    author = "Haque, Md Riajul and Maity, Debaprasad and Mondal, Rajesh",
    title = "{Gravitational neutrino reheating}",
    eprint = "2311.07684",
    archivePrefix = "arXiv",
    primaryClass = "hep-ph",
    doi = "10.1103/PhysRevD.109.063543",
    journal = "Phys. Rev. D",
    volume = "109",
    number = "6",
    pages = "063543",
    year = "2024"
}

@article{Haque:2024zdq,
    author = "Haque, Md Riajul and Maity, Debaprasad and Mondal, Rajesh",
    title = "{Thermal and nonthermal dark matters with gravitational neutrino reheating}",
    eprint = "2408.12450",
    archivePrefix = "arXiv",
    primaryClass = "hep-ph",
    doi = "10.1103/PhysRevD.111.063546",
    journal = "Phys. Rev. D",
    volume = "111",
    number = "6",
    pages = "063546",
    year = "2025"
}

@article{Roshan:2025mwl,
    author = "Roshan, Rishav and Show, Sudipta",
    title = "{Delayed Charged Lepton Yukawa Equilibration in Minimal Seesaw}",
    eprint = "2510.23712",
    archivePrefix = "arXiv",
    primaryClass = "hep-ph",
    doi = "10.1103/nfkx-q6z7",
    journal = "Phys. Rev. Lett.",
    volume = "137",
    number = "11",
    pages = "111802",
    year = "2026"
}

@article{Bandyopadhyay:2026gci,
    author = "Bandyopadhyay, Disha and Borah, Debasish and Das, Nayan",
    title = "{QCD axion misalignment during reheating}",
    eprint = "2610.00458",
    archivePrefix = "arXiv",
    primaryClass = "hep-ph",
    month = "9",
    year = "2026"
}

@article{Bernal:2026lpr,
    author = "Bernal, Nicol{\'a}s and Fong, Chee Sheng and Zapata, {\'O}scar",
    title = "{Seesaw Cosmology}",
    eprint = "2607.27325",
    archivePrefix = "arXiv",
    primaryClass = "hep-ph",
    month = "7",
    year = "2026"
}
\bibliographystyle{JHEP}
\end{document}